%% file: amestudy.tex
\documentclass[twocolumn,traditabstract,longauth,a4]{aa}
\usepackage{graphicx,amsmath}
\usepackage{epsf}
\usepackage{txfonts}
\usepackage{url}
\usepackage{longtable}
\usepackage{lscape}
\usepackage[hyperindex,breaklinks=true, colorlinks, citecolor=blue]{hyperref}  
\usepackage{breakurl}
\usepackage{comment}
\usepackage{natbib}
\usepackage{upgreek}
\usepackage{fixltx2e}

\bibpunct{(}{)}{;}{a}{}{,} 

\newcommand{\planck}{\Planck}  
\newcommand{\IRAS}{\textit{IRAS\/}}
\newcommand{\WMAP}{\textit{WMAP\/}}
\newcommand{\COBE}{\textit{COBE\/}}
\newcommand{\Spitzer}{\textit{Spitzer\/}}
\newcommand{\healpix}{{\tt HEALPix}}
\newcommand{\sextractor}{{\tt SExtractor}}
\newcommand{\hii}{\ion{H}{ii}}

\include{Planck}

\begin{document}

\title{\planck\ intermediate results. XV. A study of anomalous microwave emission in Galactic clouds}
\titlerunning{A study of AME in Galactic clouds}

\input{AuthorList_PIP_77_Proj_7_10_Peel_authors_and_institutes.tex}


\abstract{Anomalous microwave emission (AME) is believed to be due to electric dipole radiation from small spinning dust grains. The aim of this paper is a statistical study of the basic properties of AME regions and the environment in which they emit.  We used {\it WMAP} and \Planck\ maps, combined with ancillary radio and IR data, to construct a sample of 98~candidate AME sources, assembling SEDs for each source using aperture photometry on $1\deg$-smoothed maps from 0.408\,GHz up to 3000\,GHz.  Each spectrum is fitted with a simple model of free-free, synchrotron (where necessary), cosmic microwave background (CMB), thermal dust, and spinning dust components.  We find that 42 of the 98 sources have significant ($>5\sigma$) excess emission at frequencies between  20 and 60\,GHz.  An analysis of the potential contribution of optically thick free-free emission from ultra-compact \hii\ regions, using IR colour criteria, reduces the significant AME sample to 27 regions. The spectrum of the AME is consistent with model spectra of spinning dust.  Peak frequencies are in the range 20--35\,GHz except for the California Nebula (NGC1499), which appears to have a high spinning dust peak frequency of ($50\pm17$)\,GHz. The AME regions tend to be more spatially extended than regions with little or no AME. The AME intensity is strongly correlated with the sub-millimetre/IR flux densities and comparable to previous AME detections in the literature.  AME emissivity, defined as the ratio of AME to dust optical depth, varies by an order of magnitude for the AME regions.  The AME regions tend to be associated with cooler dust in the range 14--20\,K and an average emissivity index, $\beta_{\rm d}$, of +1.8, while the non-AME regions are typically warmer, at 20--27\,K.  In agreement with previous studies, the AME emissivity appears to decrease with increasing column density.  This supports the idea of AME originating from small grains that are known to be depleted in dense regions, probably due to coagulation onto larger grains.  We also find a correlation between the AME emissivity (and to a lesser degree the spinning dust peak frequency) and the intensity of the interstellar radiation field, $G_0$.  Modelling of this trend suggests that both radiative and collisional excitation are important for the spinning dust emission.  The most significant AME regions tend to have relatively less ionized gas (free-free emission), although this could be a selection effect. The infrared excess, a measure of the heating of dust associated with \hii\ regions, is typically $>4$ for AME sources, indicating that the dust is not primarily heated by hot OB stars. The AME regions are associated with known dark nebulae and have higher $12\micron/25\micron$ ratios. The emerging picture is that the bulk of the AME is coming from the polycyclic aromatic hydrocarbons and small dust grains from the colder neutral interstellar medium phase.
}


\maketitle   
\allearlypapers

%


\section{Introduction} 
\label{sec:introduction}

Anomalous microwave emission (AME) has been observed in a few directions of the Galaxy and is an important foreground for the cosmic microwave background (CMB) \citep{Kogut1996, Leitch1997, Finkbeiner2002, Finkbeiner2004,deOliveira-Costa2004, Dobler2008, Miville-Deschenes2008, Gold2011}. There is strong evidence, particularly in the Perseus and $\rho$~Ophiuchi clouds \citep{Watson2005,Casassus2008,planck2011-7.2}, that AME is due to electric dipole radiation from small spinning dust grains.  Along these sight lines, there is highly significant excess emission above free-free, synchrotron, CMB, and thermal dust in the frequency range $10$--$100$\,GHz. The spectral energy distributions (SEDs) are peaked at about $30$\,GHz, and can be fitted by physically-motivated theoretical models of spinning dust \citep{Draine1998b, Ali-Hamoud2009, Hoang2010,Hoang2011}. AME has been detected in \hii\ regions \citep{Dickinson2006,Dickinson2007,Dickinson2009a,Todorovic2010}, dust clouds \citep{Casassus2006,Casassus2008,Scaife2009}, a supernova remnant \citep{Scaife2007}, and in one external galaxy \citep{Murphy2010,Scaife2010a}. There is also evidence for AME in the diffuse emission at high Galactic latitudes \citep{Peel2012,Macellari2011,Ghosh2012}.  

Definitive evidence for spinning dust was provided by \citet{planck2011-7.2}. Accurate SEDs of the Perseus and $\rho$~Ophiuchi clouds were easily fitted by a physically motivated model for the clouds, including spinning dust components associated with the atomic and molecular phases of the interstellar medium (ISM). The model was found to be an excellent fit with physical parameters that were reasonable for these regions. \cite{planck2011-7.3} applied an inversion technique to separate the various contributions of the ISM in Galactocentric rings along the Galactic plane and found that $25\pm 5\,\%$ of the 30\,GHz emission comes from AME and was consistent with spinning dust associated with atomic and molecular gas but not with the ionized phase. Component separation of the diffuse emission at intermediate latitudes in the southern Gould Belt region \citep{planck2013-XII} revealed an AME component consistent with spinning dust emitting at a peak frequency of ($25.5\pm1.5$)\,GHz (in flux density units), compatible with plausible values for the local density and radiation field.

To date there has been no detailed study of AME in a reasonable sample of sources. \cite{Dickinson2007} observed six southern \hii\ regions with the Cosmic Background Imager at 31\,GHz and found tentative evidence for excess emission from the RCW49 complex. \cite{Scaife2008} observed a sample of 16 compact \hii\ regions at 15\,GHz with the Arcminute Microkelvin Imager (AMI) and found no evidence for excess emission; the spectrum was consistent with optically thin free-free emission from warm ionized gas. \cite{Todorovic2010} surveyed the Galactic plane at longitudes $27\deg\le l\le46\deg$ with the Very Small Array (VSA) at 33\,GHz and found statistical evidence for AME in nine regions, but with an emissivity relative to $100\micron$ brightness that was 30--50\,\% of the average high latitude value.

In this paper, we have assembled a sample of 98 Galactic clouds selected at \Planck\footnote{\Planck\ (\url{http://www.esa.int/Planck}) is a project of the European Space Agency (ESA) with instruments provided by two scientific consortia funded by ESA member states (in particular the lead countries France and Italy), with contributions from NASA (USA) and telescope reflectors provided by a collaboration between ESA and a scientific consortium led and funded by Denmark.} frequencies to investigate their SEDs and constrain the contribution of AME. Due to the large beam size of the lowest \WMAP/\Planck\ channels and the low frequency radio data, there is sometimes a mix of sources within the beam. Many of the sources can be classed as diffuse \hii\ regions, although we have found a few AME sources with no obvious associated \hii\ region and very weak free-free emission. Many of the regions are in large star-forming complexes, which at $1\deg$ resolution contain many individual sources. These are often located in the vicinity of molecular clouds, which produce strong thermal dust emission. Nevertheless, combining \Planck\ data with ancillary radio and far-infrared data we assemble their SEDs from 0.408\,GHz to 5000\,GHz. We fit the SEDs with a simple model of free-free, synchrotron (where appropriate), thermal dust, CMB, and AME (spinning dust) components to determine whether there is evidence for AME at frequencies $20$--$60$\,GHz and if so, if it agrees with spinning dust models. For the most significant ($\geq 5\,\sigma$) AME detections, we investigate the observational properties of these regions and compare them with each other and with regions that do not show strong AME. In particular, we would like to distinguish AME and ``non-AME'' regions using observational and physical properties. This is the first statistical study of AME regions to date.

In Sect.~\ref{sec:data} we describe the \Planck\ and ancillary data used in our analysis. Section~\ref{sec:sample} describes the sample selection, aperture photometry, and model-fitting. Section~\ref{sec:ame_regions} presents the results of the quantification of AME in these sources. Section~\ref{sec:statistical_study} investigates the correlation of AME with source properties.  Section~\ref{sec:conclusions} gives a brief discussion and conclusions.


\begin{table*}[tmb]
\begingroup
\newdimen\tblskip \tblskip=5pt
\caption{Sources of the datasets used in this paper, as well as centre frequencies, angular resolutions, and references.}
\label{tab:data}
\nointerlineskip
\vskip -3mm
\footnotesize
\setbox\tablebox=\vbox{
   \newdimen\digitwidth 
   \setbox0=\hbox{\rm 0} 
   \digitwidth=\wd0 
   \catcode`*=\active 
   \def*{\kern\digitwidth}
   \newdimen\signwidth 
   \setbox0=\hbox{+} 
   \signwidth=\wd0 
   \catcode`!=\active 
   \def!{\kern\signwidth}
    \halign{\hbox to 2.15in{#\leaderfil}\tabskip 1.0em&
            \hfil#\hfil&
            \hfil#\hfil&
            \hfil#\hfil&
            #\hfil\tabskip 0pt\cr
    \noalign{\doubleline\vskip 2pt}
    \omit&Frequency&&&\cr
    \omit\hfil Telescope/Survey\hfil&[GHz]&Resolution&Coverage&\omit\hfil Reference\hfil\cr
\noalign{\vskip 4pt\hrule\vskip 6pt}
Haslam&         0.408& 51\parcm0& Full sky& \citet{Haslam1982}\cr
Reich&          *1.42& 35\parcm4& Full sky& \citet{Reich1982,Reich1986,Reich2001}\cr
Jonas&          **2.3& 20\parcm0& Southern sky& \citet{Jonas1998}\cr
\WMAP~9-year&   *22.8& \,\,51\parcm3\rlap{$^{\rm a}$}& Full sky& \citet{Bennett2013}\cr
\Planck&        *28.4& 32\parcm3& Full sky& \cite{planck2013-p01}\cr
\WMAP~9-year&   *33.0& \,\,39\parcm1\rlap{$^{\rm a}$}& Full sky& \citet{Bennett2013}\cr
\WMAP~9-year& *  40.7& \,\,30\parcm8\rlap{$^{\rm a}$}& Full sky& \citet{Bennett2013}\cr
\Planck&        *44.1& 27\parcm1& Full sky& \cite{planck2013-p01}\cr
\WMAP~9-year&   *60.7& \,\,21\parcm1\rlap{$^{\rm a}$}& Full sky& \citet{Bennett2013}\cr
\Planck&        *70.4& 13\parcm3& Full sky& \cite{planck2013-p01}\cr
\WMAP~9-year&   *93.5& \,\,14\parcm8\rlap{$^{\rm a}$}& Full sky& \citet{Bennett2013}\cr
\Planck&        **100& *9\parcm7& Full sky& \cite{planck2013-p01}\cr
\Planck&        **143& *7\parcm3& Full sky& \cite{planck2013-p01}\cr
\Planck&        **217& *5\parcm0& Full sky& \cite{planck2013-p01}\cr
\Planck&        **353& *4\parcm8& Full sky& \cite{planck2013-p01}\cr
\Planck&        **545& *4\parcm7& Full sky& \cite{planck2013-p01}\cr
\Planck&        **857& *4\parcm3& Full sky& \cite{planck2013-p01}\cr
\COBE-DIRBE&    *1249& 37\parcm1& Full sky& \citet{Hauser1998}\cr
\COBE-DIRBE&    *2141& 38\parcm0& Full sky& \citet{Hauser1998}\cr
\COBE-DIRBE&    *2997& 38\parcm6& Full sky& \citet{Hauser1998}\cr
\IRAS~(IRIS) Band 4 ($100\micron$)& *3000& *4\parcm7& Near-full sky& \citet{Miville-Deschenes2005}\cr 
\IRAS~(IRIS) Band 3 ($60\micron$)&  *5000& *3\parcm6& Near-full sky& \citet{Miville-Deschenes2005}\cr 
\IRAS~(IRIS) Band 2 ($25\micron$)&  12000& *3\parcm5& Near-full sky& \citet{Miville-Deschenes2005}\cr 
\IRAS~(IRIS) Band 1 ($12\micron$)&  25000& *3\parcm5& Near-full sky& \citet{Miville-Deschenes2005}\cr 
\Spitzer~IRAC/MIPS& $8$, $24\micron$& $2\arcsec$, $6\arcsec$& Partial& \cite{Fazio2004,Rieke2004}\cr
\noalign{\vskip 3pt\hrule\vskip 4pt}
}}
\endPlancktablewide
\tablenote {{\rm a}} We use the symmeterized, 1\deg-smoothed version.\par
\endgroup
\end{table*}

\section{Data}
\label{sec:data}

\subsection{Planck data} 
\label{sec:planck}

\Planck\ \citep{tauber2010a, planck2011-1.1} is the third generation space mission to measure the anisotropy of the CMB.  It observes the sky in nine frequency bands covering 30--857\,GHz with high sensitivity and angular resolution from 31\arcm\ to 5\arcm.  The Low Frequency Instrument (LFI; \citealt{Mandolesi2010, Bersanelli2010, planck2011-1.4}) covers the 30, 44, and 70\,GHz bands with amplifiers cooled to 20\,\hbox{K}.  The High Frequency Instrument (HFI; \citealt{Lamarre2010, planck2011-1.5}) covers the 100, 143, 217, 353, 545, and 857\,GHz bands with bolometers cooled to 0.1\,\hbox{K}.  Polarization is measured in all but the highest two bands \citep{Leahy2010, Rosset2010}.  A combination of radiative cooling and three mechanical coolers produces the temperatures needed for the detectors and optics \citep{planck2011-1.3}.  Two data processing centers (DPCs) check and calibrate the data and make maps of the sky \citep{planck2011-1.7, planck2011-1.6}.  \Planck's sensitivity, angular resolution, and frequency coverage make it a powerful instrument for Galactic and extragalactic astrophysics as well as cosmology.  Early astrophysics results are given in Planck Collaboration VIII--XXVI 2011, based on data taken between 13~August 2009 and 7~June 2010.  Intermediate astrophysics results are now being presented in a series of papers based on data taken between 13~August 2009 and 27~November 2010. 

In this paper we use {\it Planck} data from the 2013 distribution of released products \citep{planck2013-p01}, based on data acquired during the ``nominal'' operations period from 13~August 2009 to 27 November 2010, and available from the \Planck\ Legacy Archive\footnote{\url{http://www.sciops.esa.int/index.php?project=planck\&page=Planck\_Legacy\_Archive}}. Specifically, we use the nine temperature maps summarized in Table~\ref{tab:data}. We also use a CMB-subtracted version for testing the robustness of the detections, using the {\tt SMICA} CMB map \citep{planck2013-p06}. We use the standard conversion factors from CMB to Rayleigh-Jeans (RJ) units and updated colour corrections described in \cite{planck2013-p01}. The \Planck~bands centred at 100 and 217\,GHz are known to be contaminated by CO lines. We corrected these channels using the \cite{Dame2001} integrated CO map smoothed to $1\deg$ resolution and scaled with the conversion factors described in \cite{planck2013-p03a}; however, for some sources, we still see  discrepancies with the spectral model at the $>10\,\%$ level. We therefore did not include these two channels in our fitting of the spectral model. The CO contamination in the 353\,GHz channel is small, typically $<1\,\%$ \citep{planck2013-p03a}, and we do not see significant deviations in our SEDs. Therefore, no correction was made for CO lines in the 353\,GHz band.

Although we limit ourselves to bright Galactic regions with typical flux densities at 30\,GHz far greater than 10\,Jy, at $1\deg$ angular scales the integrated flux density of CMB fluctuations can be $10$~Jy or more at 100\,GHz, a significant fraction of the total flux density of some of the sources in our sample.  CMB-subtracted maps would, in principle, be most appropriate for our analysis.  However, in bright regions near the Galactic plane, significant foreground residuals remain in the CMB maps produced by the \Planck\ component separation codes in 2013 \citep{planck2013-p06}, which used only \Planck\ data and frequencies for separation.  These regions can be masked for cosmological work, but they are precisely the regions that we need here.  Investigations comparing CMB-subtracted with non-CMB-subtracted maps revealed biases in the plane at the level of $10$--$15\,\%$. Furthermore, incorrect subtraction, particularly at frequencies near $100$\,GHz, resulted in high $\chi^2$ values for some SEDs, and poorly fitted thermal dust components.  We therefore use the CMB-subtracted maps only for finding regions of AME, and use non-CMB-subtracted maps for the photometric analysis, where we fit for a CMB component in the spectrum of each source, using the full data available in Table~\ref{tab:data} (see Sect.~\ref{sec:ancillary}). In this way we do not bias the flux densities (due to the component separation process), and more importantly, we can characterize and propagate the uncertainty due to the CMB fluctuation. The AME amplitudes from both datasets agree within a fraction of the uncertainty for the majority of sources.  In the future, \Planck\ component separation will also make use of many of the external datasets listed in Table~\ref{tab:data}, and it may be possible to subtract the CMB directly.

\subsection{Ancillary data} 
\label{sec:ancillary}

We use a range of ancillary data to allow the SEDs to be determined from radio (around 1\,GHz) to far-infrared (around 3000\,GHz). All ancillary data are summarized, along with the \Planck\ data, in Table~\ref{tab:data}. These data have been smoothed to a common resolution of $1\deg$ since some of the maps have only slightly higher resolution than this. The smoothing also reduces the effects of any residual beam asymmetry in some cases, e.g., \WMAP\ and \Planck, where non-circular beams vary across the map. 

We analysed the northern sky survey at 12--18\,GHz from the COSMOSOMAS experiments \citep{Gallegos2001}; however, due to the filtering of emission on large angular scales and large intrinsic beam width, the majority of the sources were strongly affected by negative filtering artefacts from neighbouring bright sources. The exceptions were G160.26$-$18.62 and G173.6+2.8, which were previously reported by \cite{planck2011-7.2}. We therefore did not consider further the COSMOSOMAS data in our analysis.

In the following sections, we describe the ancillary data in more detail.

\subsubsection{Radio surveys} \label{sec:radio}
Data at low frequencies (around 1\,GHz) are important for excluding regions with synchrotron emission, and for estimating the level of free-free emission. Ideally, we would have several frequency channels in the range $1$--$10$\,GHz; however, no large area surveys exist above 2.3\,GHz, except for higher resolution surveys that do not retain large-angular-scale information. We therefore use the three well-known surveys at 0.408, 1.42, and 2.326\,GHz.

The all-sky survey of \cite{Haslam1982} at 0.408\,GHz is widely used as a tracer of synchrotron emission at high Galactic latitudes; however, it also contains strong free-free radiation from the Galactic plane and from \hii\ regions, where the free-free typically dominates over synchrotron emission even at these lower frequencies. 

A number of different versions of the 0.408\,GHz map are available. The most widely used is the NCSA\footnote{National Center for Supercomputing Applications (NCSA), located at the University of Illinois at Urbana-Champaign; \url{http://www.ncsa.illinois.edu}} destriped and desourced version available on the LAMBDA website\footnote{\url{http://lambda.gsfc.nasa.gov/}} at an angular resolution of $1\deg$. This map has been Fourier filtered to remove large-scale striations, and bright sources have been subtracted, including many of the bright \hii\ regions. Since we want to retain all the sources for this work, we use a less-processed version of the map\footnote {Available from the Bonn Survey Sampler webpage \url{http://www.mpifr-bonn.mpg.de/survey.html}} at $51\arcmin$ resolution that was originally sampled in a 2-D Cartesian projection with $0\pdeg33\times0\pdeg33$ square pixels and B1950 coordinate frame. This version retains all the bright compact sources, although striations are much more visible by eye.  However, at low latitudes and in bright regions, the striations are negligible compared to the sky signal. This map was regridded into the \healpix\ format \citep{Gorski2005} using a procedure that computes the surface intersection between individual pixels of the survey with the intersecting \healpix\ pixels (see Appendix A of \citealt{Paradis2012b}). After smoothing the resulting map with a $31\parcm6$ FWHM Gaussian kernel to bring it to $1\deg$ resolution, this new map gave results more consistent with the 1.42 and 2.326\,GHz maps. 

The Reich et al.~full-sky 1.42\,GHz map \citep{Reich1982,Reich1986,Reich2001} has $36\arcmin$ resolution, and the \cite{Jonas1998} 2.326\,GHz map of the southern hemisphere has $20\arcmin$ resolution. These have been destriped but not source-subtracted.  Although the 2.326\,GHz map covers up to +15\deg, we do not use declinations $>+10\deg$ because the smoothing operation affects the edges of the map.

The 0.408\,GHz map is formally calibrated on angular scales of $5\deg$ by comparison with the 404\,MHz survey of \cite{Pauliny-Toth1962}, while the 1.42\,GHz and 2.326\,GHz maps are tied to absolute sky horn measurements by \cite{Webster1974} and \cite{Bersanelli1994}, respectively. Our study is at $1\deg$ resolution, with some regions being extended to 2--3\deg. Therefore one would expect the brightness temperature (and thus flux density) to be under-estimated for many of our sources. The maximum correction factor is given by the full-beam to main-beam ratio, which quantifies the power in the full beam (including sidelobes) compared to the main beam. The largest correction factor we applied is 1.55 for the Reich et al. 1.42\,GHz survey, based on comparisons with bright calibrator sources.  We did not make any corrections to the 0.408 and 2.326\,GHz maps, since they were found to be consistent to within $10\,\%$ of the 1.4\,GHz data for the majority of the sources in our sample and for bright extragalactic sources. We also note that the positional accuracy of these maps, particularly the 0.408\,GHz map, is not particularly good. Visual inspection of the maps suggests inconsistencies of bright sources at the level of up to $15\arcmin$ at 0.408\,GHz.  For our analysis, however, this is not likely to be a major source of error, since our integration aperture has a diameter of 2\deg. 

We assumed a $10\,\%$ uncertainty in the radio data at all three frequencies. For the 408\,MHz map, which has striations, we added an additional 3.8\,Jy uncertainty corresponding to the baseline uncertainty of $\,\pm3$\,K \citep{Haslam1982} at $1\deg$ angular scales. This is required to bring the $\chi^2$ value to within acceptable levels for some sources.  This additional uncertainty is not always required for sources in our sample, and we find, in fact, that we overestimated our uncertainties in many cases (see Sect.~\ref{sec:robustness}).

\subsubsection{WMAP} \label{sec:wmap}
{\it WMAP} 9-year data are included in our analysis \citep{Bennett2013}. The data span 23 to 94\,GHz and thus complement \Planck\ data, particularly the K-band (22.8\,GHz) channel. The $1\degr$-smoothed maps available from the LAMBDA website are used. We apply colour corrections to the central frequencies using the recipe described by \cite{Bennett2013}; the local spectral index across each band is calculated using the best-fitting model (see Sect.~\ref{sec:model_fitting}). This does not exactly take into account curvature of the spectrum, but is a good approximation given that the colour corrections are typically a few percent. For the majority of sources studied in this paper we are not limited by instrumental noise and we assume a $3\,\%$ overall calibration uncertainty.

\subsubsection{Submm/infrared data} \label{sec:submm}

To sample the peak of the blackbody curve for temperatures greater than 15\,K, we include the {\it COBE}-DIRBE data at $240\micron$ (1249\,GHz), $140\micron$ (2141\,GHz), and $100\micron$ (2997\,GHz). The DIRBE data are the Zodi-Subtracted Mission Average (ZSMA) maps \citep{Hauser1998} regridded into the \healpix\ format using the same procedure as used for the 408\,MHz map described in Sect.~\ref{sec:radio}. Colour corrections are applied as described in the DIRBE explanatory supplement version 2.3. Data at higher frequencies are not included in the spectral fits, since they are dominated by transiently heated grains not in thermal equilibrium with the interstellar radiation field and therefore not easily modelled by a single modified blackbody curve. Furthermore, at wavelengths $\lesssim40\micron$ the spectrum contains many emission/absorption lines, which complicates the modelling. For the statistical comparison, we also include the shorter wavelengths of DIRBE band 7 (4995\,GHz) and the \IRAS\ 12\micron\ (25000\,GHz) and 25\micron\ (12000\,GHz) bands. We use the IRIS maps of \cite{Miville-Deschenes2005}, which have had bright sources and a model of zodiacal light removed. Residuals from zodiacal-light subtraction are known to be an issue at  wavelengths shorter than about 25\micron, but are not expected to be significant for the bright regions in this study because the zodiacal light is relatively smooth spatially except for a narrow band at low ecliptic latitudes. We test this assumption by comparing the flux densities from improved zodiacal-light-subtracted maps (Marc-Antoine Miville-Deschenes, private comm.) where the residuals are clearly much smaller. We obtained consistent results within a fraction of the errors; the scatter is less than 5\,\% at the worst band ($12\micron$). Sources were not removed for $|b|>5\deg$ and therefore do not affect the majority of the sources in our sample.

We use \textit{Spitzer} data where available at 8 and 24\micron\ as a dust diagnostic for the polycyclic aromatic hydrocarbons (PAHs) and very small grains (VSGs), respectively. The \textit{Spitzer} data are obtained from the \textit{Spitzer} data archive,\footnote{\url{http://sha.ipac.caltech.edu/applications/Spitzer/SHA/}} and are reprocessed for  the purposes of this paper in order to mitigate possible systematics. An extended emission correction is applied to the 8\micron\ data, and the zodiacal light contribution is subtracted from both the 8 and 24\micron\ data. Bright point sources are extracted from both bands to enable us to investigate the extended emission, and an overlap correction is applied to ensure a consistent background level. Finally, all the reprocessed data are combined to produce the final maps used in this analysis; see \cite{Tibbs2011} for more details. We are able to measure flux densities for 24 regions.





\section{Sample selection and SED fitting} \label{sec:sample}

In this section we cover the methods we use to create the sample of sources. Section~\ref{sec:hii_sample} describes the source detection method that forms the main sample. Section~\ref{sec:create} describes the component subtraction method for detecting potential AME regions. Section~\ref{sec:final_sample} summarizes the final sample of 98 sources. Section~\ref{sec:aperflux} describes the aperture photometry method used to extract the flux densities of the sources. Section~\ref{sec:model_fitting} describes the model-fitting that is adopted to quantify the various components and to assess the contribution of AME. Section~\ref{sec:example_seds} presents example SEDs\footnote{Strictly speaking, the SED is frequency multiplied by the flux density (with  units W\,m$^{-2}$). Here we use the term for the flux density spectrum (units W\,m$^{-2}$\,Hz$^{-1}$).} and a summary of what is observed in our sample.

\subsection{Detection of bright sources} 
\label{sec:hii_sample}

At high radio frequencies ($30$--$70\,$GHz), synchrotron and thermal dust emission are expected to be relatively faint. The dominant emission mechanism is thought to be optically thin free-free emission ($\alpha \approx\! -0.14$, where $S \propto \nu^{\alpha}$), with a possible contribution from AME. Free-free emission is expected to be particularly strong near the Galactic plane due to the presence of \hii\ regions and ionized gas near OB stars. This allows \hii\ regions to be detected by simply searching for bright sources in individual frequency maps.  However, in this paper we are mainly interested in constructing accurate SEDs across the radio/submillimetre/far-infrared wavelength range, which requires the detection of the brightest clouds at {\it all\/} {\it WMAP}/\Planck\ frequencies. We used the \sextractor\ software \citep{Bertin1996}, which was used in the ``Sextra'' pipeline for the Planck Early Release Compact Source Catalogue \citep{planck2011-1.10}, to detect bright sources at each \Planck\ frequency of the CMB-subtracted maps. 

We begin with a \sextractor\ catalogue of 1194 sources detected at $70\,$GHz. To increase reliability and to ensure the region is bright at all \Planck\ frequencies, this catalogue is further cross-matched with the 28.4 and $100\,$GHz catalogues, using a matching radius of the largest beam FWHM ($16\parcm38$). This results in 462 sources that are well-detected across the $30$--$100$\,GHz range. We remove extragalactic sources by searching the NASA Extragalactic Database (NED\footnote{\url{http://ned.ipac.caltech.edu/}}) for radio galaxies. Approximately half of all detected sources, and a majority at $|b| \gtrsim 5\deg$, are found to be extragalactic, most of which are likely blazars. We also remove a small number of sources associated with known bright supernova remnants \citep{Green2009} and planetary nebulae \citep{Acker1992}. The SIMBAD\footnote{\url{http://simbad.u-strasbg.fr/simbad/}} database is found to be useful for confirming that a region is dominated by Galactic emission and that many of our sources are in fact large \hii\ complexes or parts of molecular clouds. These regions often contain several individual sources.

The final stage of catalogue trimming is made by visual inspection of the maps and preliminary SEDs made by aperture photometry (Sects~\ref{sec:aperflux}, \ref{sec:model_fitting}, and \ref{sec:example_seds}). We make visual inspection at this resolution, since the final SEDs are to be constructed using $1\deg$-smoothed maps (to ensure that the response to diffuse emission is the same at all frequencies). To ensure a robust sample, sources that are not well-defined after smoothing to $1\deg$ (i.e., do not show a definite peak of emission on scales of $\lesssim 2\deg$), or are relatively faint ($\ll10\,$Jy at a frequency of $30\,$GHz), are discarded, except for a few cases at several degrees distance from the Galactic plane.  We find a few sources whose positions are not exactly centred on the peak of the emission at frequencies of $20$--$60$\,GHz, with offsets as large as 10--20\arcmin. This can occur because of the complexity of the Galactic plane, which after filtering can produce multiple peaks in close proximity to each other. In these cases, we manually shift the position to the approximate centre of the hotspot. Since we are using a large $1\deg$~radius aperture (see Sect.~\ref{sec:aperflux}), this makes little difference to the SEDs. We identify 94 candidate AME sources using this technique.

\subsection{Detection of AME regions by component subtraction} 
\label{sec:create}

We use a simple CMB/foreground subtraction method to isolate AME from the other diffuse components. This method is essentially the same as was used by \citet{planck2011-7.2}, where potential AME regions were located by a simple subtraction of the non-AME components from the 28.4\,GHz \Planck~CMB-subtracted map. The one difference is that here we only use the 0.408\,GHz map to trace the synchrotron emission, which is extrapolated with a single power law and a spectral index $\beta=-3.0$ ($T \propto \nu^{\,\beta}$). This is a typical value of the slope between 408\,MHz and {\it WMAP}/\Planck\ frequencies \citep{Davies2006,Gold2011}. The combination of the 1.4\,GHz and 2.3\,GHz maps is not used, as it creates large-scale artefacts. Although there is some evidence of flattening ($\beta \approx\! -2.7$) of the synchrotron index at low Galactic latitudes (e.g., \citealp{Gold2009}), we use the typical high latitude value. For most sources on the Galactic plane, the synchrotron emission is a minor component at frequencies above $23$\,GHz. For the free-free component we use the dust-corrected H$\alpha$ map of \cite{Dickinson2003}.  For thermal dust, we use model 8 of \cite{Finkbeiner1999}.  Both are calculated at a frequency of 28.4\,GHz.

We smooth the \Planck~CMB-subtracted maps to a resolution of 1\deg\ and subtract the non-AME components from the \Planck~28.4\,GHz map to create a map of residuals. A 5\deg-smoothed version is also created and subtracted from the 1\deg~map to remove large-scale emission and highlight the compact regions most suited for this analysis. The diffuse emission removed here will be the focus of future papers.

\begin{figure}[tb]
\begin{center}
\includegraphics[width=0.30\textwidth,angle=90]{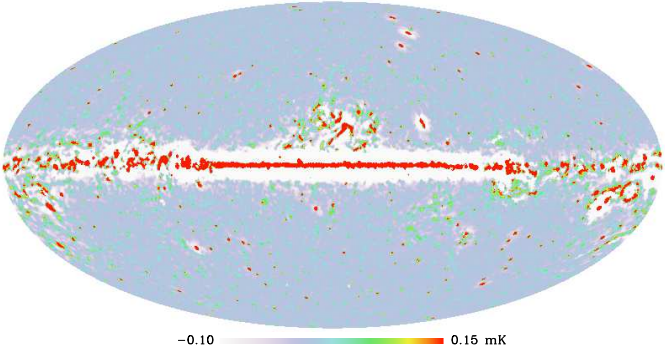}\\
\caption{Map of residuals at 28.4\,GHz after subtracting off synchrotron, free-free, thermal dust, and CMB components (see text), in mK (R-J) units. A $5\deg$-smoothed version of the map is subtracted to remove extended diffuse emission to more easily identify bright, relatively compact sources. This map is shown in the Mollweide projection, with $l=0\deg$ in the centre and increasing to the left.}
\label{fig:residuals}
\end{center}
\end{figure}

The resulting map of residuals at 28.4\,GHz is shown in Fig.~\ref{fig:residuals}. The large-scale features, including negative artefacts, are not of concern here. Instead, we used this map as a ``finding chart'' to identify new regions that emit detectable levels of AME. Approximately 100 bright well-defined sources are located by eye and a spectrum is produced for each one using aperture photometry (see Sect.~\ref{sec:aperflux}). The well-known AME regions in Ophiuchus and Perseus stand out in this map. Lots of free-free emission (usually because it can be self-absorbed at lower frequencies) and synchrotron point sources (with a flatter spectral index than $\beta=-3.0$, and hence not removed completely by extrapolating the synchrotron map assuming a steep spectrum) can be found in this residual map. Most of the 100 AME candidates are \hii\ regions; 20 sources show evidence for excess emission at $30$\,GHz based on an initial spectral fit, out of which 16 have already been identified using the source-detection method (Sect.~\ref{sec:hii_sample}). The four additional sources found using this technique are G037.79$-$00.11, G293.35$-$24.47, G317.51$-$00.11, and G344.75+23.97. 

\begin{figure*}[tb]
\begin{center}
\includegraphics[width=0.25\textwidth,angle=90]{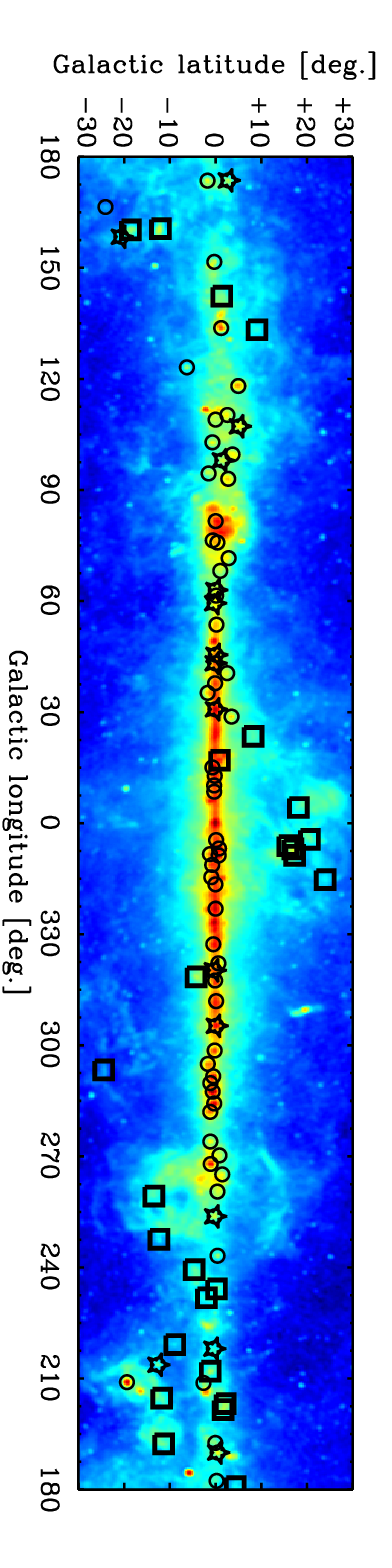}\\
\caption{CMB-subtracted \Planck\ 28.4\,GHz map covering the entire Galactic plane and latitudes $|b|<30\deg$. The colour scale has a logarithmic stretch. Regions with the most significant AME are highlighted as thick squares while the rest of the sample are shown as circles. Regions with significant excess emission but with a potential UC\hii\ contribution ($f_{\rm max}^{\rm UC\hii}>0.25$) are shown as star symbols (see Sect.~\ref{sec:uchii}).}
\label{fig:dr3map}
\end{center}
\end{figure*}

\subsection{Final sample}
\label{sec:final_sample}

The final sample contains 98 sources, listed in Table~\ref{tab:main_list}. The superscript letter after the name indicates which method the source is chosen from. Most of the sources are located using the \sextractor\ detection technique, with a few of the AME-dominated sources being detected using the component subtraction method. We also indicate if a source is already known from previous AME studies. A few previously identified AME candidates are not on this list because they are not detected at high significance in the \Planck\ data, mostly due to the limited angular resolution of this study. These include RCW175 \citep{Dickinson2009a}, LDN1621 \citep{Dickinson2010}, M78 \citep{Castellanos2011}, LDN1780 \citep{Vidal2011}, and LDN1111/675/1246 \citep{Scaife2009,Scaife2010b}. Associations with known objects are listed in the notes column of Table~\ref{tab:main_list}.

The \Planck\ CMB-subtracted map with the locations of the sources is shown in Fig.~\ref{fig:dr3map}. Most of the sources lie within a few degrees of the Galactic plane. A few sources are in the well-known regions of Ophiuchus ($l=0\deg$), Perseus ($l=160\deg$), Orion ($l=200\deg$), and the Gum nebula $(l=260\deg$). The most significant ($\sigma_{\rm AME}>5$ and $f_{\rm max}^{\rm UC\hii}<0.25$; see Sect.~\ref{sec:ame_regions}) AME sources are shown as thick squares; sources that have excess emission ($\sigma_{\rm AME}>5$) but have a potentially large contribution of optically thick free-free emission from ultra-compact \hii\ (UC\hii) regions ($f_{\rm max}^{\rm UC\hii}>0.25$) are shown as stars. It is interesting to see that these AME-bright sources appear to cluster in certain regions, particularly along the local Gould Belt region \citep{planck2013-XII}. There seem to be no bright AME regions along the lines-of-sight to the local spiral arm at $l=90\deg$ and $l=270\deg$. In general, few of the most significant AME sources lie on the plane. This is partly explained by the removal of AME sources that have a potential  UC\hii\ contribution, based on infrared sources  (see Sect.~\ref{sec:uchii}), which preferentially lie in the Galactic plane. In addition, there is a selection effect, since the high free-free brightness temperatures and overall confusion in the plane make it more difficult to identify individual AME-bright objects. It may also be that these sight-lines contain a strong component of free-free emission from warm ionized gas, which is thought to exhibit less AME than cold neutral medium (CNM) or molecular clouds \citep{planck2011-7.3}. With our incomplete sample, such claims cannot be confirmed in this study.

\subsection{Aperture photometry} 
\label{sec:aperflux}

We use the \healpix\ aperture photometry code developed for \citet{planck2011-7.2} to extract the flux densities of the regions from the maps. This software has also been used to investigate at the polarization of AME from $\rho$~Ophiuchi in \citet{Dickinson2011}. After converting from CMB thermodynamic units (K$_{\rm CMB}$) to RJ units (K$_{\rm RJ}$) at the central frequency, the maps are converted to units of Jy\,pixel\mo\ using $S=2kT_{\rm RJ}\Omega \nu^2 / c^2$, where $\Omega$ is the \healpix\ pixel solid angle. The pixels are then summed in a circular aperture of $60\arcmin$ to obtain an integrated flux density. An estimate of the background is subtracted using a median estimator of pixels lying at radii between $80\arcmin$ and $100\arcmin$. By using Monte Carlo injection of sources, we find that this choice of aperture and annulus size provides the least scatter in recovered flux densities, and is a reasonable balance for obtaining an appropriate background level without subtracting appreciable flux density from the source itself. 

The flux density uncertainties are estimated from the rms of the values in the background annulus and added in quadrature to the absolute calibration uncertainties for each map (see Sect.~\ref{sec:ancillary}). Simulations of injected point-like sources show that the flux density estimates are unbiased and that the uncertainties are reasonable; however, the exact value of flux density uncertainty for each source is difficult to quantify, since it depends very strongly not only on the brightness of the source and background, but also on the morphology of the emission in the vicinity of the source. This will be discussed further in Sect.~\ref{sec:robustness}. Colour corrections, based on the local spectral index across each band, are applied during the model-fitting, as described in the next section.

\subsection{Model fitting} 
\label{sec:model_fitting}

We take the flux density $S$ for each source from the aperture photometry and fit a simple model of free-free, synchrotron (where appropriate), CMB, thermal dust, and spinning dust components:

\begin{equation}
S = S_{\rm ff} + S_{\rm sync} + S_{\rm td} + S_{\rm CMB} + S_{\rm sp}~.
\end{equation}
The free-free flux density $S_{\rm ff}$ is calculated from the brightness temperature $T_{\rm ff}$, based on the optical depth $\tau_{\rm ff}$, using the standard formula
\begin{equation}
S_{\rm ff} = \frac{2 \, k \,T_{\rm ff} \, \Omega \, \nu^2}{c^2}~,
\end{equation} 
where $k$ is the Boltzmann constant, $\Omega$ is the solid angle of the aperture, and $\nu$ is the frequency, with 
\begin{equation} \label{eq:freefree}
T_{\rm ff} = T_\mathrm{e} \, (1-e^{-\tau_{\rm ff}})~,
\end{equation}
and the optical depth $\tau_{\rm ff}$ is given by
\begin{equation}
\tau_{\rm ff} = 5.468 \times 10^{-2} \, T_\mathrm{e}^{-1.5} \, \nu^{-2} \,  {\rm EM} \, g_{\rm ff}~,
\end{equation}
in which the Gaunt factor can be approximated\footnote{Here we use the approximation given by \cite{Draine_book}, which is accurate to better than 1\,\% even up to frequencies of 100\,GHz and higher.} by 
\begin{equation}
\label{eq:gaunt}
g_{\rm ff} = {\rm ln} \left( {\rm exp} \left[5.960 - \frac{\sqrt{3}}{\pi} {\rm ln} (Z_i \,\nu_9 \,T_4^{-3/2}) \right] + 2.71828 \right)~.
\end{equation}
For the analysis of AME, we assume a fixed electron temperature of 8000\,K for $T_\mathrm{e}$ for all regions, fitting only for the emission measure (EM). Note that this is not the true EM, but an effective EM over the $1\deg$ radius aperture. For compact sources, the quoted EM will be underestimated.

For six sources, we also include a synchrotron component modelled as a power law with amplitude $A_{\rm sync}$ and variable flux density spectral index $\alpha$,
\begin{equation}
S_{\rm sync} = A_{\rm sync} \left( \frac{\nu}{{\rm GHz}}\right)^{\alpha}~.
\end{equation}

The thermal dust is fitted using a modified blackbody model,
\begin{equation}
S_{\rm td} =  2 \, h \, \frac{\nu^3}{c^2} \frac{1}{e^{h\nu/kT_{\rm d}}-1} \, \tau_{250} \, (\nu/1.2\,\textrm{THz})^{\,\beta_\mathrm{d}} \,\Omega~,
\end{equation}
fitting for the optical depth $\tau_{250}$, the dust temperature $T_\mathrm{d}$, and the emissivity index $\beta_\mathrm{d}$.
The CMB is fitted using the differential of a blackbody at $T_{\rm CMB}=2.7255$\,K \citep{Fixsen2009}
\begin{equation}
S_{\rm CMB} = \left(\frac{2 \, k \, \Omega \, \nu^2}{c^2}\right) \Delta T_{\rm CMB}~.
\end{equation}
Here $\Delta T_{\rm CMB}$ is the CMB fluctuation temperature in thermodynamic units. The spinning dust is fitted using
\begin{equation}
S_{\rm sp} = A_{\rm sp} \, j_{(\nu+\nu_{\rm shift})} \,\Omega ~,
\end{equation}
where we use a model for $j_{\nu}$ calculated using the {\tt SPDUST} (v2) code \citep{Ali-Hamoud2009,Silsbee2011}. We choose a model corresponding to the warm ionized medium (WIM) with a peak at 28.1\,GHz to give the generic shape, and allow for a shift of this model with frequency. We therefore fit for two parameters corresponding to the AME amplitude $A_{\rm sp}$, and a frequency shift $\nu_{\rm shift}$. Note that the units of $A_{\rm sp}$ are formally of column density (cm$^{-2}$). If the spinning dust model was appropriate for the line-of-sight, and no frequency shift was applied, then this would indeed be the column density $N_{\rm H}$; however, since this quantity is model-dependent and there is potentially a shift in frequency, we do not take this to be a reliable estimate of $N_{\rm H}$. Similarly, in this paper we do not attempt to fit specific spinning dust models to each source, hence the derived column density is not necessarily physical; $A_{\rm sp}$ is essentially the flux density at the peak normalized to the spinning dust model. Given the large uncertainties and difficulty in separating the various spectral components, we have not attempted to look for deviations from the basic spinning dust model \citep{Hoang2011}.

The least-squares fit is calculated using the {\tt MPFIT}\footnote{\url{http://purl.com/net/mpfit}} \citep{Markwardt2009} package written in IDL, with starting values estimated from the data and with amplitude parameters constrained to be positive except for the CMB, which is allowed to go negative. {\tt MPFIT} also provides estimates of the $1\sigma$ uncertainties for each parameter, taken as the square root of the diagonal elements of the parameter covariance matrix. We note four special cases in Table~\ref{tab:main_list} (G068.16+01.02, G076.38--00.62, G118.09+04.96 and G289.80--01.15) where the fitting returned $A_{\rm sp}=0.0\pm0.0$. These could be mitigated by removing the positivity prior, with best-fitting negative values still being consistent with zero. Instead, for these special cases, we fixed $A_{\rm sp}$ to zero to make the fits more physically meaningful, since the spinning dust spectrum should not go negative.

\begin{figure*}[tb]
\begin{center}
\includegraphics[scale=0.33]{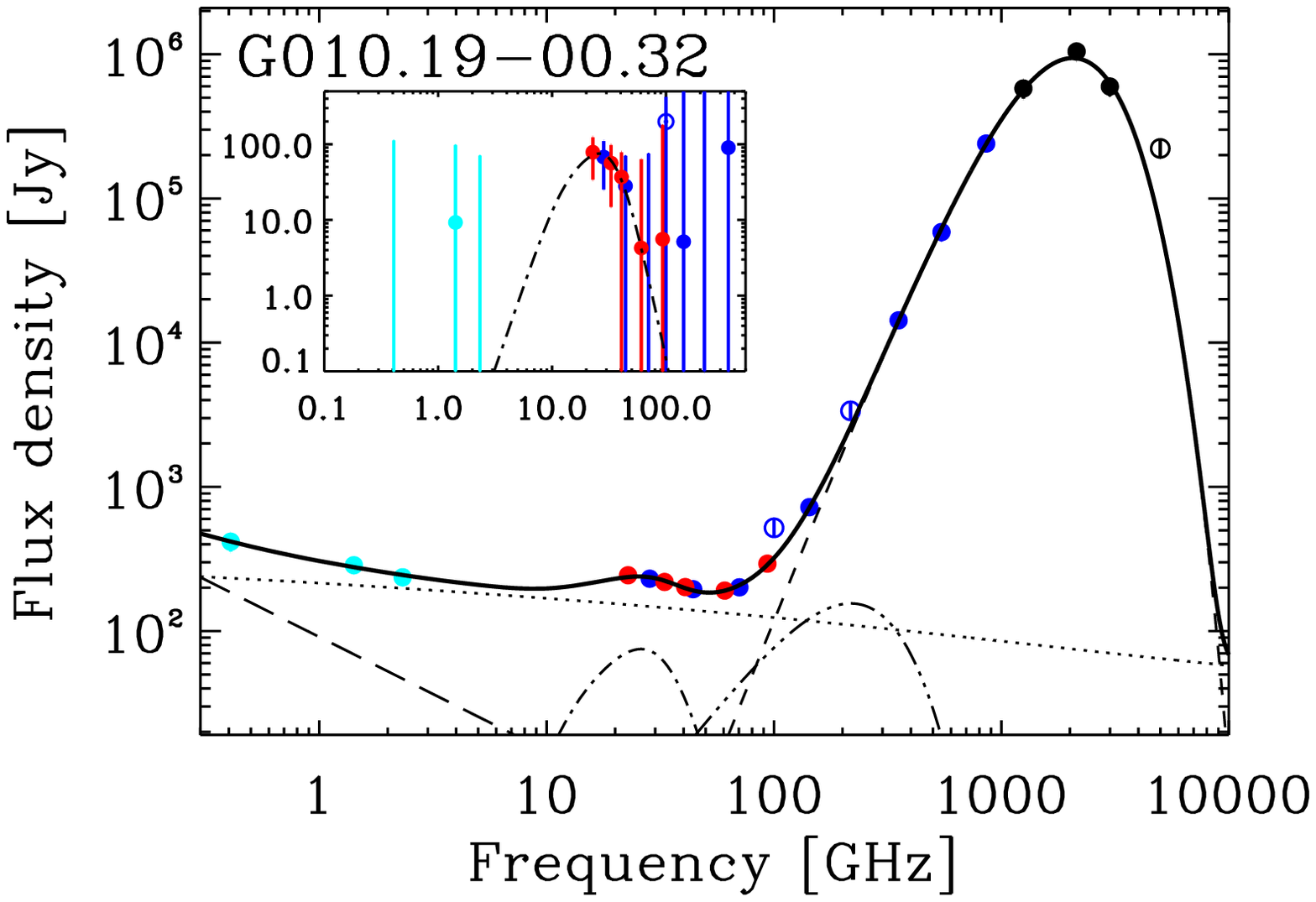}
\includegraphics[scale=0.33]{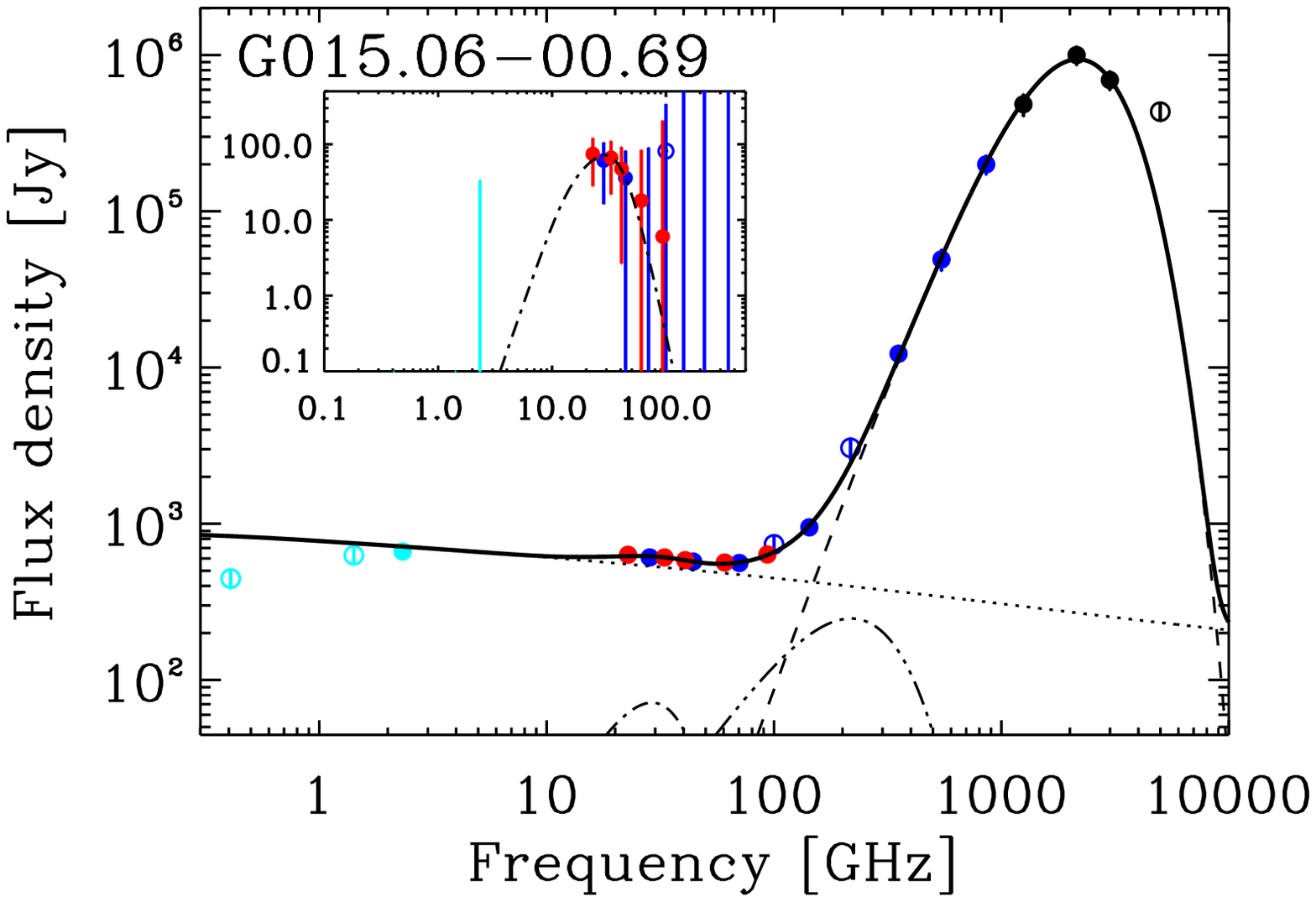}
\includegraphics[scale=0.33]{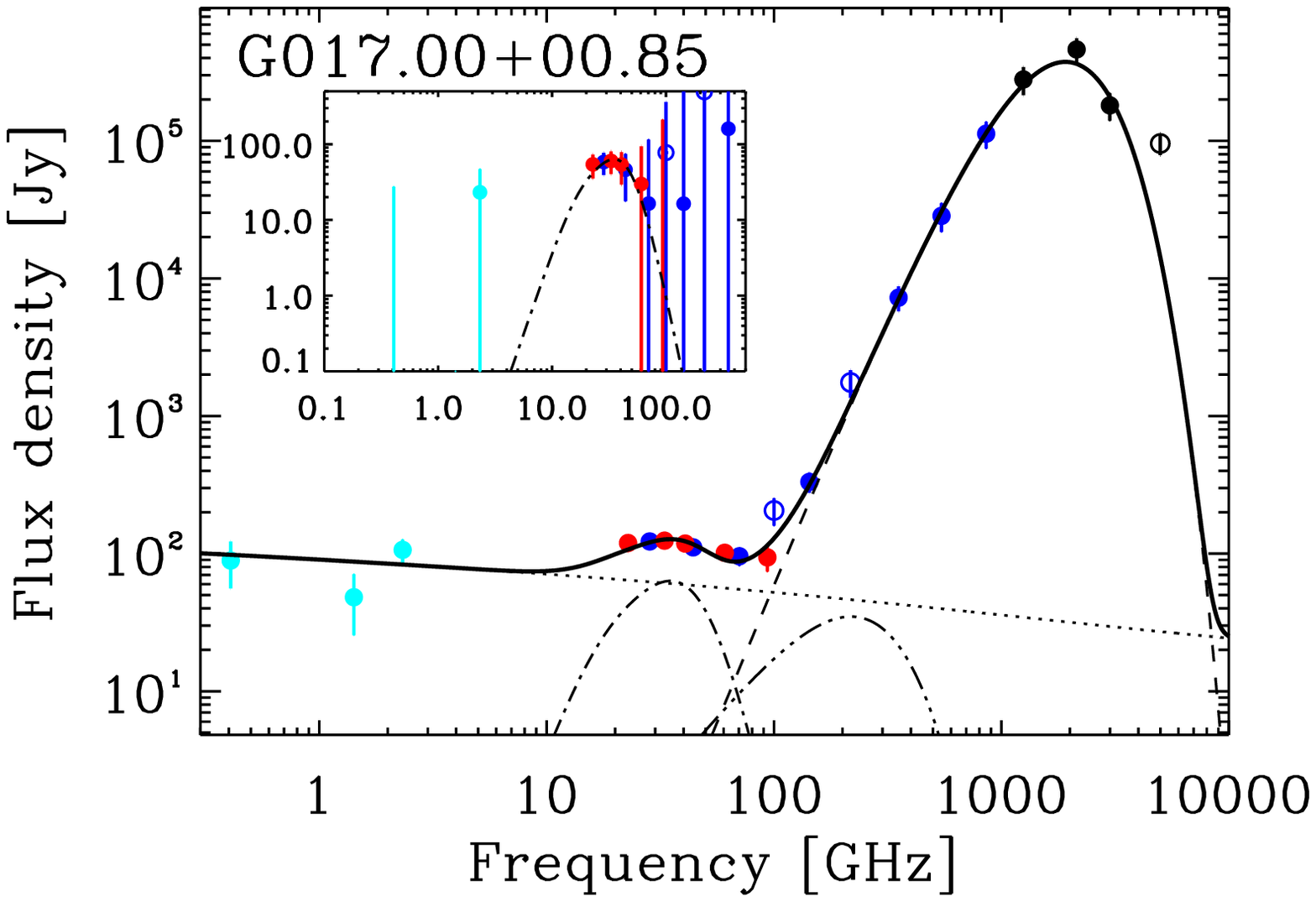}
\includegraphics[scale=0.33]{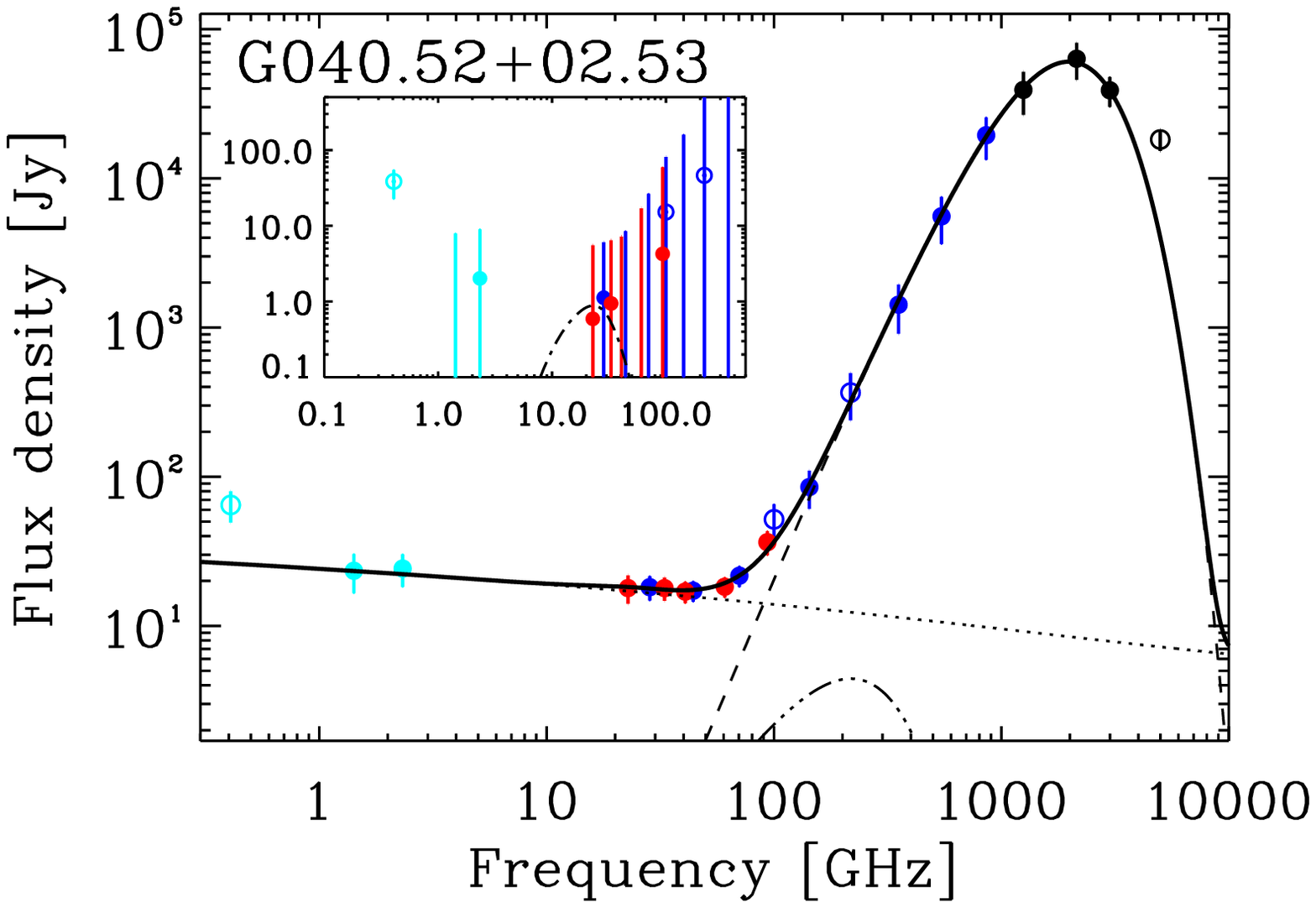}
\includegraphics[scale=0.33]{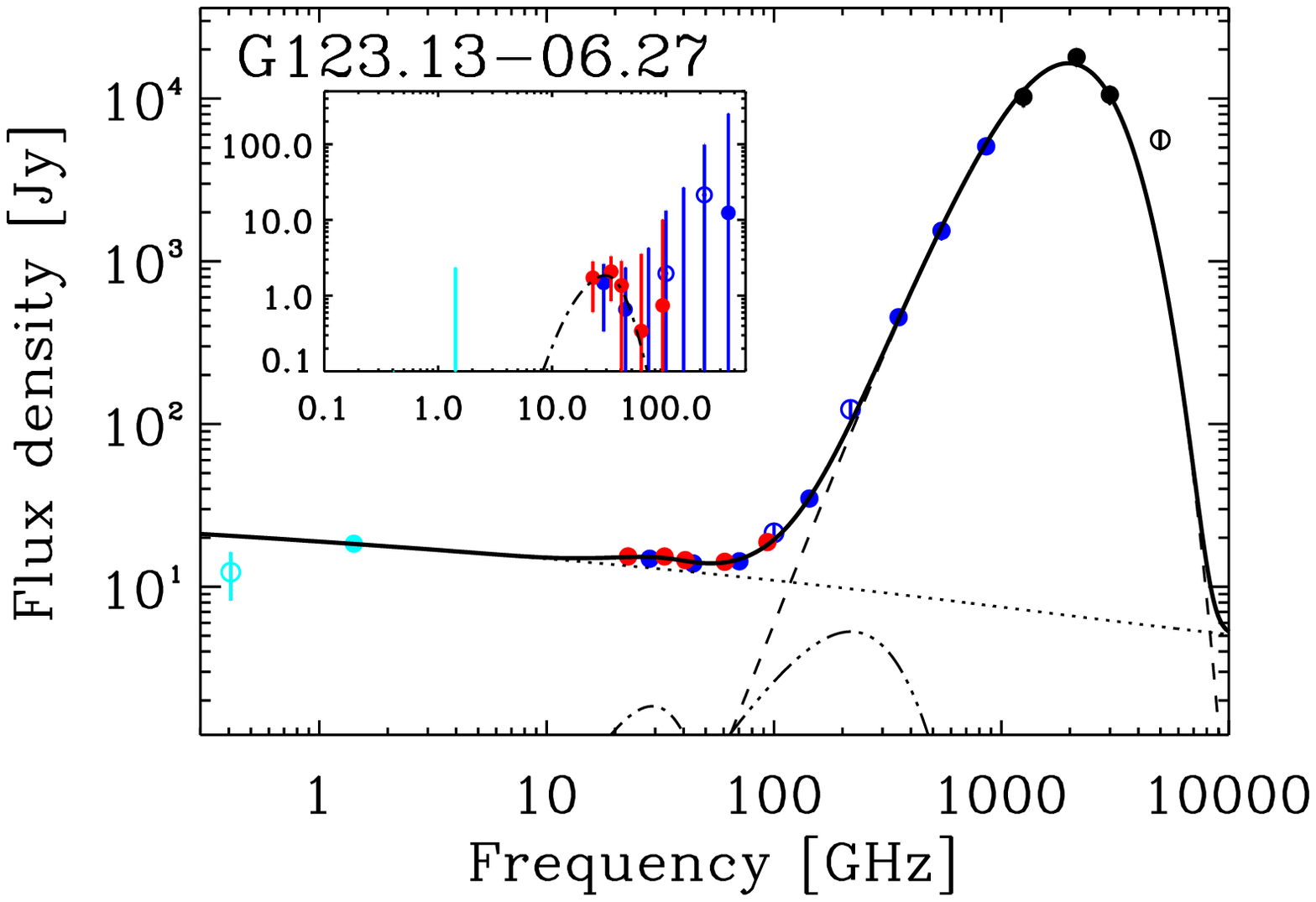}
\includegraphics[scale=0.33]{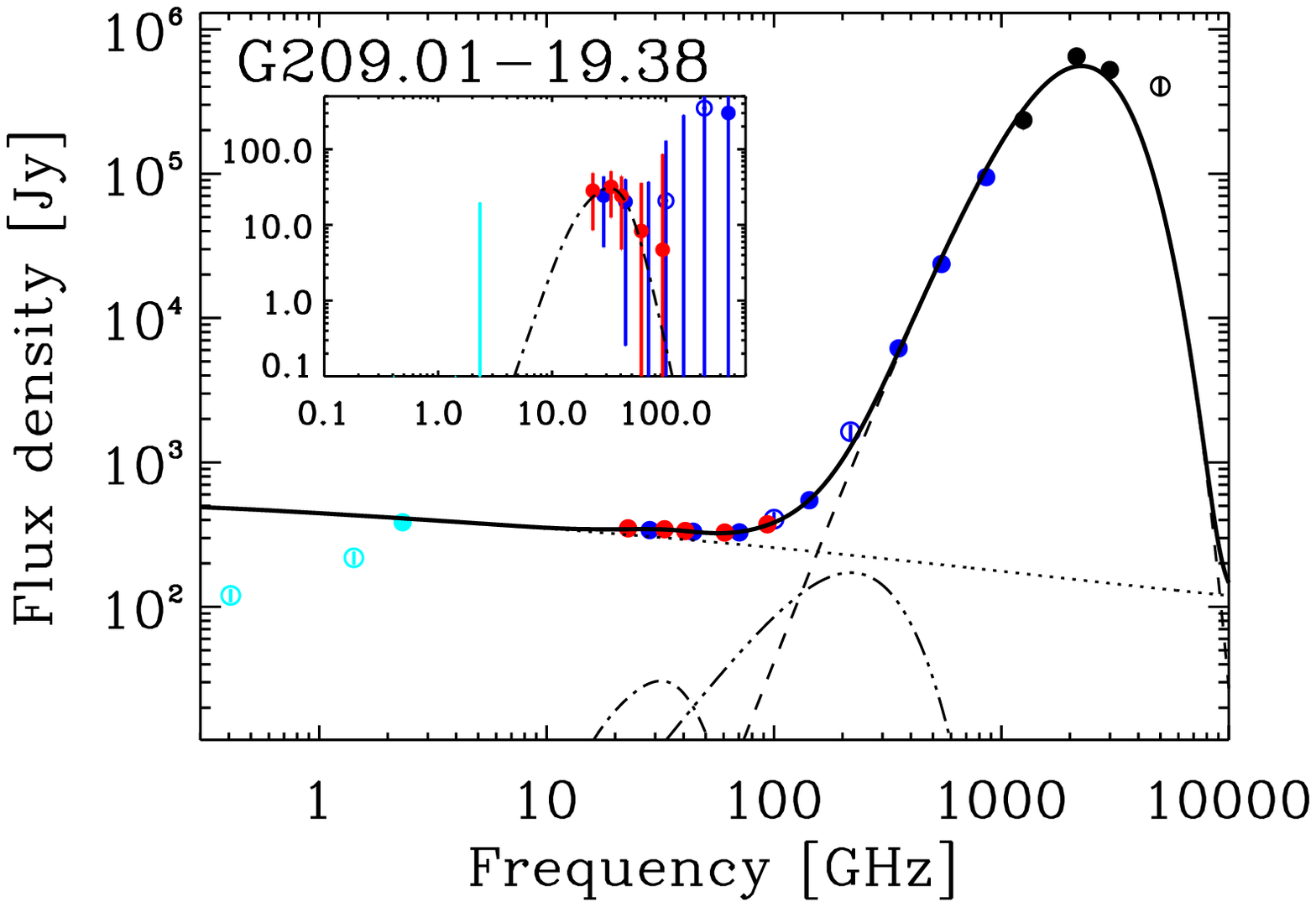}
\includegraphics[scale=0.33]{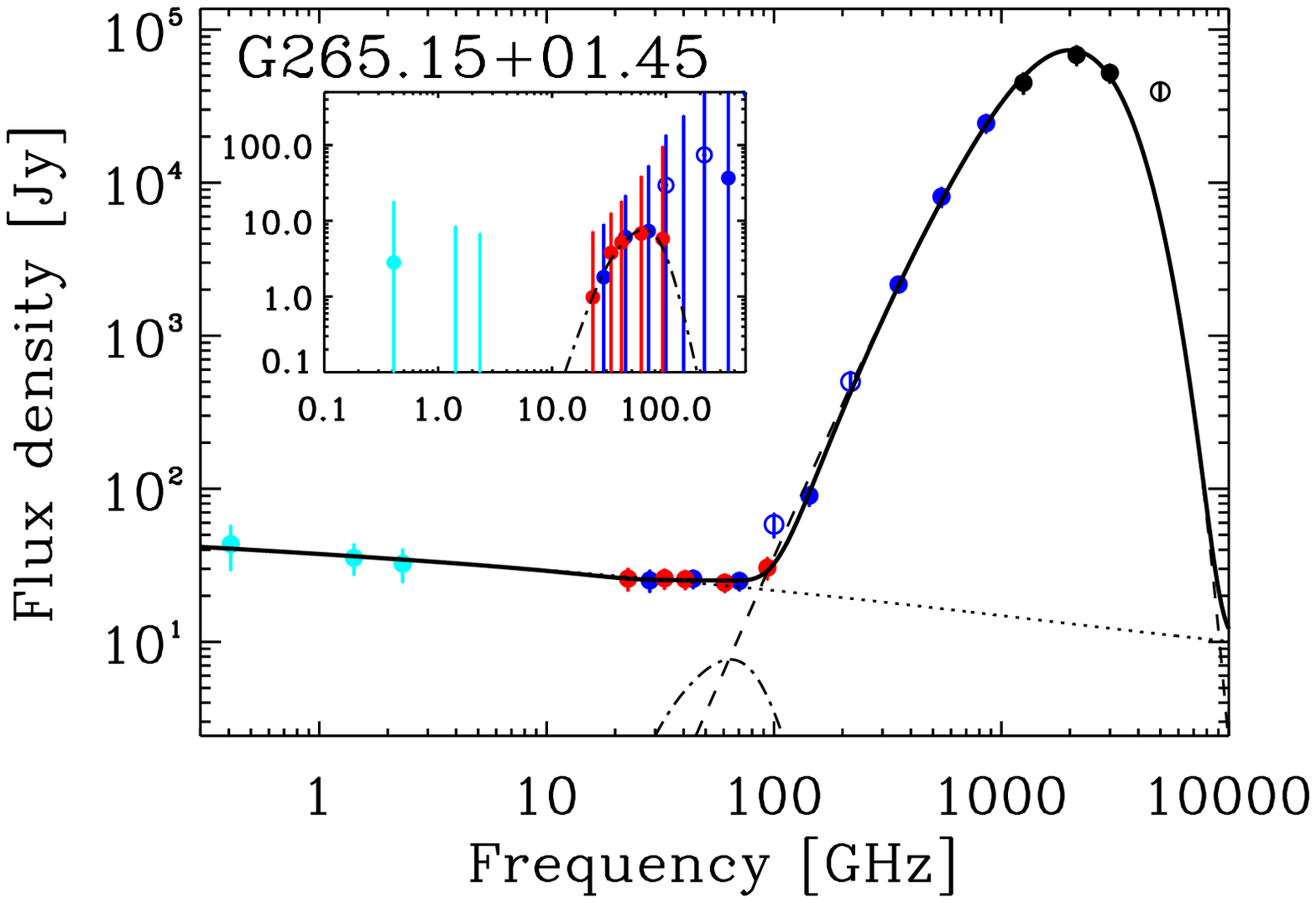}
\includegraphics[scale=0.33]{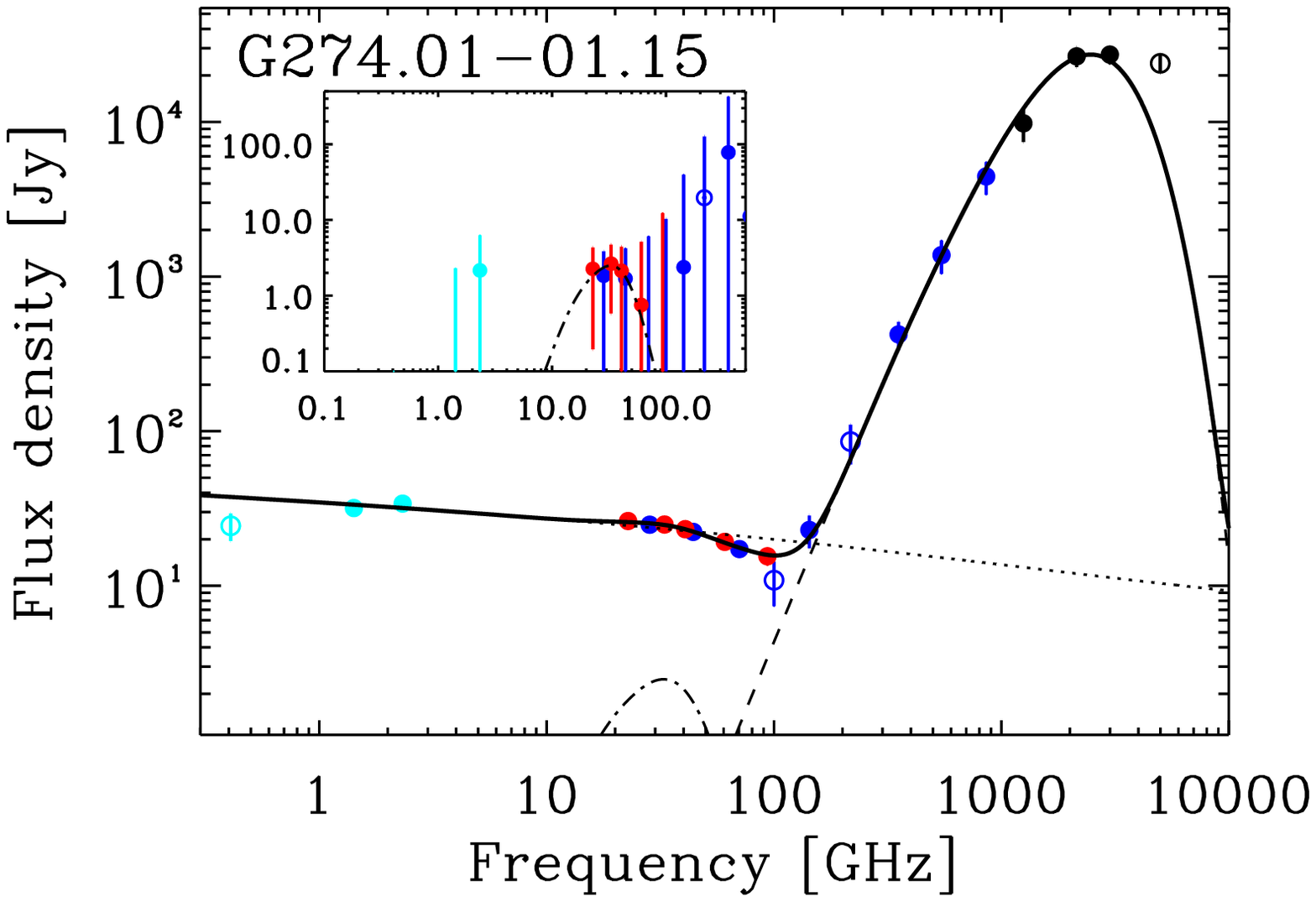}
\includegraphics[scale=0.33]{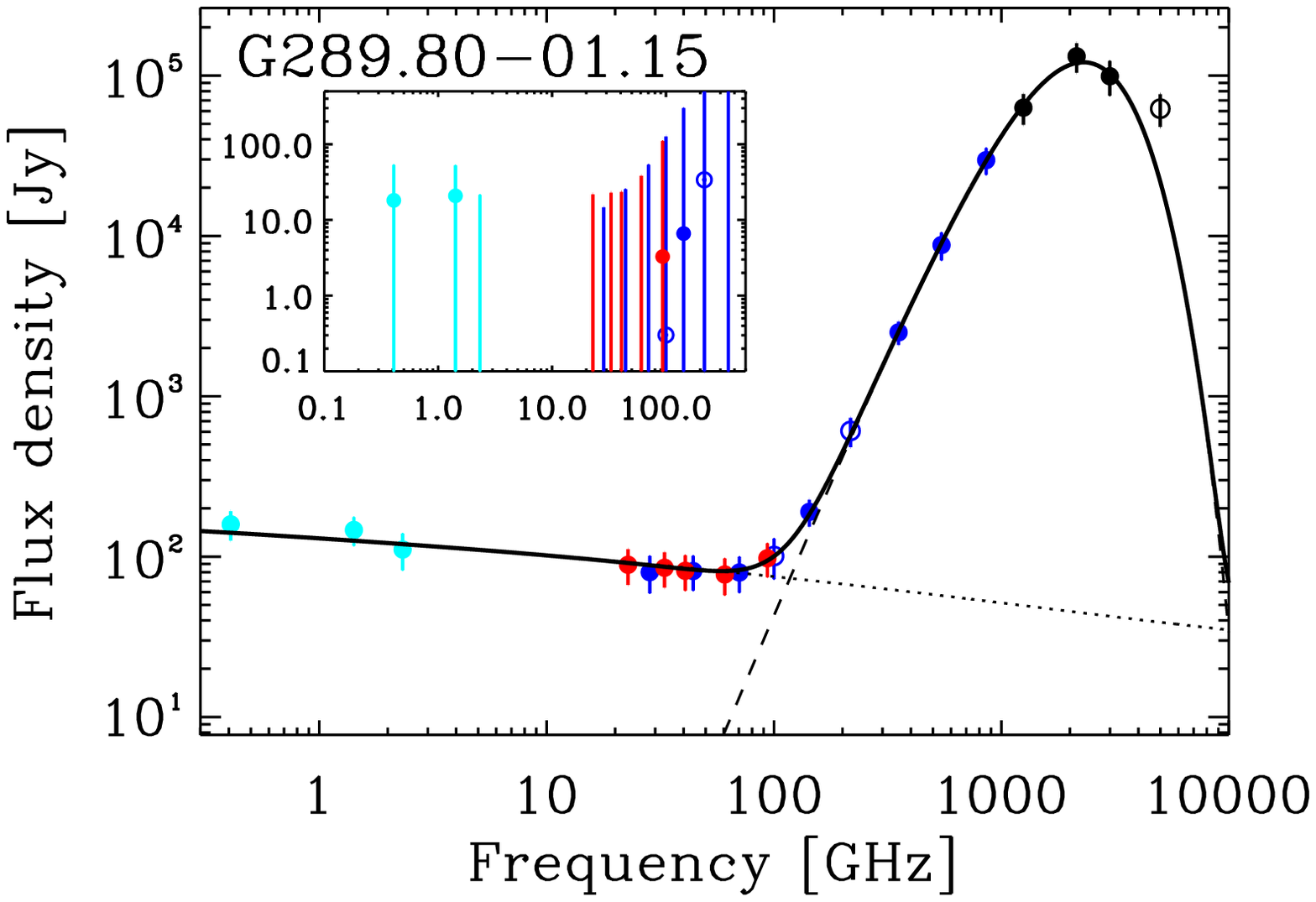}
\caption{Example SEDs (see text for description of individual SEDs) of sources with little or no AME (see Fig.~\ref{fig:ame_seds} for SEDs with significant AME). Data points are shown as circles with errors and are colour-coded for radio data (cyan), {\it WMAP} (red), \Planck\ (blue), and DIRBE/ \IRAS\ (black). The best-fitting model of free-free (dotted line), synchrotron (long dashed line), thermal dust (short dashed line), CMB (triple-dot dashed line), and spinning dust (dot-dashed line) are shown. Data included in the fit are shown as filled circles, while the other data are unfilled. The residual spectrum, after subtraction of free-free, synchrotron, CMB and thermal dust components, is shown as the insert.}
\label{fig:example_seds}
\end{center}
\end{figure*}

\subsection{Example SEDs}
\label{sec:example_seds}

Some example SEDs for regions with weak AME are shown in Fig.~\ref{fig:example_seds}; see Sect.~\ref{sec:maps_seds} and Fig.~\ref{fig:ame_seds} for SEDs with significant AME. Filled circles are used for data included in the fit, and unfilled circles are for display purposes only. We begin by including data from 0.408\,GHz up to 3000\,GHz and make a least-squares fit to the data. In general, the SEDs are well-fitted by our simple model, although the uncertainties appear to be over-estimated. This can be seen in some of the example SEDs in Fig.~\ref{fig:example_seds} and in the reduced $\chi^2$ values in Table~\ref{tab:main_list}; the mean value for the entire sample is $\bar{\chi^2}=0.59$.  However, our uncertainties are justified for some sources where the scatter is consistent with our assigned uncertainties. An example of this is G017.00+00.85, where there is considerable scatter at low frequencies. 

All sources show a strong thermal dust component peaking at about $2000$--$3000$\,GHz, indicative of dust grains at $T_{\rm d}\approx\!20$\,K. The one-component modified blackbody function reproduces the spectrum above 100\,GHz remarkably well for the majority of our sources; however, the 100/217\,GHz data points are often inconsistent with the model due to the CO line contamination within the {\Planck~bands}. For this reason, as previously explained, we exclude the 100/217\,GHz data from all our fits. 

Another effect seen in our SEDs is that of the fluctuations in the CMB. Although the CMB fluctuations are faint (with an rms of $70\,\mu$K at $1\deg$ scales), the large aperture that we integrate over results in a typical integrated CMB flux density of $7$\,Jy at $100$\,GHz, based on the standard deviation of flux densities from  Monte Carlo simulations of a CMB-only sky, assuming the {\it WMAP} 7-year power spectrum \citep{larson2010}. It is important to note that these fluctuations are about the mean CMB temperature, and thus can be negative or positive. Figure~\ref{fig:example_seds} shows examples of both; G209.01--19.38 contains a large positive CMB fluctuation ($\Delta T_{\rm CMB} = 371 \pm 102\,\mu$K), and G274.01$-$01.15, showing a strong negative fluctuation ($\Delta T_{\rm CMB}=-37\pm10\,\mu$K). The negative CMB flux densities cause a dip in the spectrum at frequencies near $100$\,GHz, which could be misinterpreted as spinning dust at lower frequencies. 

Similarly, over-fitting by a strong positive CMB fluctuation could affect the AME intensity. This could happen when there is a flattening of the thermal dust spectral index at frequencies below 353\,GHz \citep{PIP96}, which can be accounted for by the CMB component. A clear example of this is G015.06$-$00.69, shown in Fig.~\ref{fig:example_seds}. There is an apparent flattening of the thermal dust spectral index, which appears as an excess at frequencies $\approx\! 100$--353\,GHz relative to the one component dust model, an effect that has been observed before \citep{Paradis2009,Paradis2012}. In this case, the fitted CMB temperature, $\Delta T_{\rm CMB}=(533\pm251)\,\mu$K, is larger than what could realistically be attributed to a pure CMB fluctuation ($\gtrsim 150\,\mu$K). Fortunately, because the uncertainties are large and the CMB does not contribute strongly at frequencies where AME is dominant ($10$--60\,GHz), this does not have a major impact on the AME results. This will be discussed further in Sect.~\ref{sec:robustness}.

At frequencies below $100$\,GHz, optically thin free-free emission is seen in many sources and is sometimes consistent with the low frequency radio data at $\approx\!1$\,GHz and {\it WMAP}/{\it Planck} data at 20--100\,GHz (e.g., G265.15+01.45 and G289.80$-$01.15 in Fig.~\ref{fig:example_seds}). These sources justify our use of the 0.408, 1.42, and 2.326\,GHz data, and show that the overall calibration factors are within the uncertainties assumed in this study. Where there is evidence of absorption at low frequencies, or if there is a discrepancy between 0.408\,GHz and the other low frequency data at 1.4/2.3\,GHz, we omit the 0.408\,GHz data point (and occasionally the 1.42\,GHz data point) in the fit (e.g., G123.13$-$06.27, G209.01$-$19.38, and G274.01$-$01.15 in Fig.~\ref{fig:example_seds}). For some sources (e.g., G015.06$-$00.69), we choose not to include the 0.408 and 1.42\,GHz data, since they both show evidence of absorption. At 2.3\,GHz, the data are consistent with optically thin free-free emission and are a good match to the {\it WMAP}/\Planck\ data. This is acceptable since the free-free component usually contains only one free parameter (i.e., EM). Sometimes this is necessary because the maps show considerably higher background relative to the source itself, due to the high levels of synchrotron emission at frequencies $\lesssim 1$\,GHz. This can affect the estimated flux density both inside the aperture and also in the background annulus, resulting in a bias that can be either high or low and may account for data points that are discrepant with the other low frequency data, particularly at 0.408\,GHz.

Figure~\ref{fig:example_seds} shows examples of other situations.  Synchrotron-dominated sources are omitted in our sample except for six sources where the low frequency data are seen to be a good fit to a power-law (amplitude and spectral index) by visual inspection. These sources are G010.19$-$00.32 (Fig.~\ref{fig:example_seds}), G008.51$-$00.31, G012.80$-$00.19, G037.79$-$0.11, G344.75+23.97, and G355.44+0.11. No strong supernova remnants are included in our sample; however, weak supernova remnants (SNR) are identified (see the Notes column of Table~\ref{tab:main_list}) in some regions. We obtain flux density spectral indices that are in the expected range ($-0.7$ to $-1.2$) for supernova remnants, with a mean value of $-0.9$.



\section{Regions of AME}
\label{sec:ame_regions}

\subsection{Significance of AME detections}
\label{sec:ame_significance}

Visual inspection of the SEDs suggests that a large fraction (at least half) of the regions chosen for this study may exhibit excess emission at frequencies in the range $20$--$60$\,GHz.  All sources have a bright thermal dust component that peaks at $2000$--3000\,GHz and becomes subdominant at frequencies below 100\,GHz; most have a contribution of free-free emission. Approximately half of the 98 sources appear to contain more emission at 20--60\,GHz compared to a simple extrapolation of optically thin free-free and thermal dust components. 

To quantify the level of AME and its significance, we use the spinning dust amplitude $A_{\rm sp}$ and its uncertainty $\sigma_{A_{\rm sp}}$, directly from the SED fitting. These are listed in Table~\ref{tab:main_list}, along with the significance level of the AME detection, $\sigma_{\rm AME} \equiv A_{\rm sp}/\sigma_{{A_{\rm sp}}}$. We also tried subtracting the non-AME components in the SED, thus leaving the AME residual, which gave similar results. 

We focus on the AME detections that are at $>5\,\sigma$. Of the 98 sources, 42 initially show highly significant ($\sigma_{\rm AME} >5$) evidence for AME while 29 do not exhibit strong AME ($\sigma_{\rm AME}<2$). Hereafter, we will refer to these as ``AME regions'' (shown in boldface in Table~\ref{tab:main_list}) and ``non-AME regions'', respectively. This leaves 42 sources that show some evidence ($\sigma_{\rm AME}=2$--$5$) of excess emission attributable to AME, which will be referred to as ``semi-significant AME regions''. Note that we reclassify 15 of the AME sources into the semi-significant category due to the potential contamination from UC\hii\ regions and this will be discussed in Sect.~\ref{sec:uchii}. This leaves 27 that we classify as strong AME detections.

\begin{figure}[tb]
\begin{center}
\includegraphics[scale=0.5]{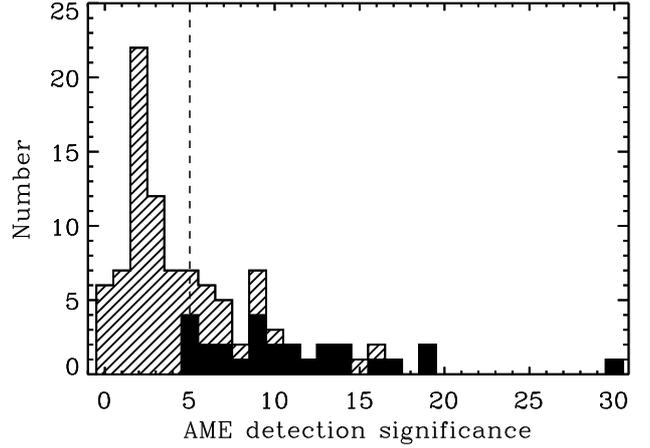}
\caption{Histogram of the AME significance values, $\sigma_{\rm AME}$, for the sample of 98 sources. The $5\,\sigma$ limit is shown as a dashed line. Sources that are significant and have a maximum contribution from UC\hii\ regions ($f_{\rm max}^{\rm UC\hii}<0.25$) are shown as the filled histogram. }
\label{fig:hist_sdnh_signif}
\end{center}
\end{figure}

Figure~\ref{fig:hist_sdnh_signif} shows a histogram of the AME significances, $\sigma_{\rm AME}$.   As discussed in Sect.~\ref{sec:robustness}, although there are some concerns for a few sources regarding possible modelling and systematic errors that could be contaminating our results, there is strong evidence for AME in a number of sources.

Two of the strongest detections are the well-known sources within the Perseus (G160.26--18.62) and $\rho$~Ophiuchi (G353.05+16.90) clouds, studied by \cite{planck2011-7.2}. These are the most easily detectable AME-dominated sources in the sky, exhibiting $70$--$80\,\%$ of AME at $30$\,GHz, and are detected at a level of $30\,\sigma$ and $17\,\sigma$, respectively. We detect them at higher significance in this paper due to using the spinning dust fit amplitude directly (rather than subtracting the best-fitting model and determining the significance from the residuals over a restricted frequency range). For $\rho$~Ophiuchi the best-fitting dust temperature and spectral index are consistent with the values from \citet{planck2011-7.2}, while for Perseus the spectral index is somewhat different; we attribute this to the filtering that was applied in that paper to allow the inclusion of the COSMOSOMAS data. The thermal dust optical depth $\tau_{250}$ was higher in the analysis from the early paper. In the early paper, the quoted optical depth was the value calculated from the modelling of the spinning dust, assuming a given PAH abundance, rather than from the thermal dust component of the SED. For these sources, the bulk of the AME was modelled as originating from the denser molecular component, which has a higher optical depth associated with it.

The two new sources detected by \cite{planck2011-7.2},  G107.20+05.20 and G173.62+02.79, are also high in the significance list at $9.9\sigma$ and $5.6\sigma$, respectively. Note that the details of the SEDs are not identical to those presented in \cite{planck2011-7.2} because the size of the background annulus has changed and the \Planck\ maps have been updated; however, the differences are small and within the stated uncertainties.

\subsection{Ultra-Compact \hii\ regions (UC\hii)}
\label{sec:uchii}

\begin{figure}[tb]
\begin{center}
\includegraphics[scale=0.5]{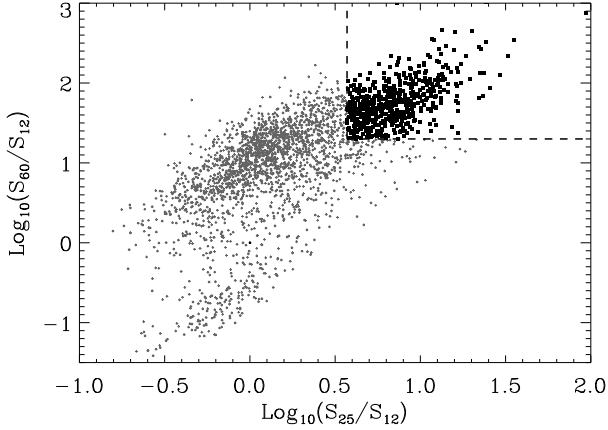}
\caption{Colour-colour plot of \IRAS\ PSC sources (grey plus symbols) that lie within the apertures of all 98 sources in our sample. UC\hii\ candidates (solid black squares) tend to have ratios ${\rm log}_{10}(S_{60}/S_{12}) \geq 1.30$ and ${\rm log}_{10}(S_{25}/S_{12})\geq 0.57$, corresponding to the top-right hand corner of this plot (marked with a dashed line). }
\label{fig:uchii}
\end{center}
\end{figure}

Optically thick free-free emission at frequencies above a few GHz can come from UC\hii\ regions with ${\rm EM}\gtrsim 10^{7}$\,cm$^{-6}$\,pc, which can be optically thick up to $10$\,GHz or higher \citep{Kurtz2002,Kurtz2005}. A nearby UC\hii\ region with typical parameters ($T_e=8000$\,K, angular size $1\arcsec$) could have a flux density of up to $10$\,Jy at LFI frequencies, while most are at $\ll 1$\,Jy \citep{Wood1989}.  However, in our low resolution analysis there could be numerous sources within the aperture that may be contributing at frequencies $\gtrsim 10$\,GHz. Since no high resolution radio surveys exist at frequencies above a few gigahertz\footnote{The CORNISH 5\,GHz source catalogue \citep{Purcell2013} covering $l=10^{\circ}$--$65^{\circ}$ has recently been published and provides useful additional information. The two significant AME sources that are covered by CORNISH (G017.00+00.85 and G062.98+00.05) do not contain any bright ($>1$\,Jy) sources at 5\,GHz.}, we use the \IRAS\ Point Source Catalogue (PSC; \citealt{Beichman1988}) to identify UC\hii\ candidates within each aperture. \cite{Wood1989b} found that due to the warmer dust temperatures in the vicinity of OB stars, UC\hii\ regions tend to have \IRAS\ colour ratios of ${\rm log}_{10}(S_{60}/S_{12}) \geq 1.30$ and ${\rm log}_{10}(S_{25}/S_{12})\geq 0.57$. 

\begin{figure}[tb]
\begin{center}
\includegraphics[scale=0.5]{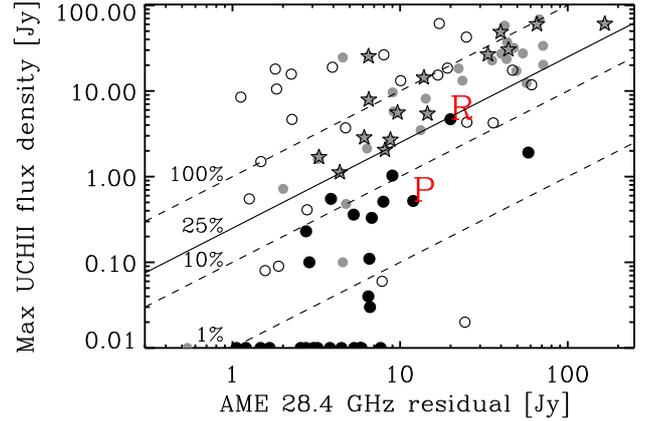}
\caption{Estimated maximum contribution from UC\hii\ regions plotted against 28.4\,GHz AME residual flux density. The most significant AME sources ($\sigma_{\rm AME}>5$ and $f_{\rm max}^{\rm UC\hii}<0.25$) are shown as filled black circles, while non-AME regions ($\sigma_{\rm AME}<2$) are shown as unfilled circles. Semi-significant AME sources ($\sigma_{\rm AME}=2$--5) are shown as filled grey circles. Significant AME regions that have a potentially large contribution from UC\hii\ ($f_{\rm max}^{\rm UC\hii}>0.25$) are re-classed as semi-significant (grey circles) and are highlighted by star symbols in this plot. Regions with no matched UC\hii\ regions are set to 0.01 and lie on the bottom of the plot. The dashed lines correspond to different maximum fractions of UC\hii\ flux density: 1, 10, 25 (solid line), and 100\,\% of the 28.4\,GHz residual flux density. The Perseus (P) and $\rho$~Ophiuchi (R) clouds are indicated.}
\label{fig:uchii_max2cm}
\end{center}
\end{figure}

Figure~\ref{fig:uchii} shows the distribution of these colour ratios for all sources that lie within the apertures of our sample. It can be seen that about $20\,\%$ of sources lie in this region of colour-colour space. To identify UC\hii\ regions we choose sources that lie within this colour-colour range, but exclude sources that are identified as extragalactic (IRAS ${\rm IDFLAG}=0$) or are upper limits at either $25\micron$ or $60\micron$. Candidate UC\hii\ regions are marked as solid black squares in Fig.~\ref{fig:uchii}. For each of our 98 sources, we typically find 10--50 matched \IRAS\ PSC sources and a few UC\hii\ candidates within the aperture. Some apertures contain no apparent UC\hii\ regions, while a few have 10--20; the median value is three UC\hii\ candidates for each aperture.

To quantify how much of the radio flux density at frequencies $\gtrsim 10$\,GHz could be attributed to UC\hii, we use the ${100\,\mu{\rm m}}$ flux densities of UC\hii\ candidates to predict the {\it maximum} flux density at a wavelength of 2\,cm (15\,GHz). \cite{Kurtz1994} measured the ratio of 100\micron\ to 2\,cm flux densities, and found it to lie in the range of 1000 to 400\,000, with no UC\hii\ regions having $S_{100}/S_{2}<1000$.  We can therefore sum up  $S_{100}$ for all UC\hii\ candidates within each aperture and use the factor of 1000 to give a maximum 15\,GHz flux density from UC\hii\ regions, $S_{\rm max}^{\rm UC\hii}$ \citep{Dickinson2013}. Strictly speaking, this is an estimate at 15\,GHz, but most \hii\ regions are in the optically thin regime (i.e., have turned over) at frequencies $\gtrsim 15$\,GHz \citep{Kurtz1994}. This is therefore very likely to be an over-estimate in many cases. First, some of the \IRAS\ PSC sources will not be UC\hii, as discussed by \cite{Wood1989b}. \cite{Ramesh1997} suggested that only about one quarter of these candidates were indeed UC\hii\ regions, due to contamination by cloud cores with lower mass stars, while \cite{Bourke2005} showed that the contamination by low-mass (i.e., non-ionizing) protostars is at the $10\,\%$ level or greater. On the other hand, such sources are typically weak and will therefore limit such a bias. Second, the average value of  $S_{100}$ is $\gtrsim 3000$, with a tail to higher values \citep{Kurtz1994}. Therefore, our maximum UC\hii\ flux density determination is very conservative and is likely to be a factor of a few above the true value.

We define the maximum fractional contribution at 28.4\,GHz, $f_{\rm max}^{\rm UC\hii}=S_{\rm max}^{\rm UC\hii}/S_{\rm resid}^{28.4}$, where 1.0 represents a possible 100\,\% contribution of UC\hii\ to the AME amplitude. Figure~\ref{fig:uchii_max2cm} plots $S_{\rm max}^{\rm UC\hii}$ against the 28.4\,GHz AME residual flux density; the values of $f_{\rm max}^{\rm UC\hii}$ for 1,10, 25, and 100\,\% are shown as straight lines. A wide range of values is seen, with some regions having no UC\hii\ matches and therefore  no contribution (they are set to 0.01\,Jy in Fig.~\ref{fig:uchii_max2cm}), while many other sources have a potentially significant contribution from UC\hii. Out of the 42 regions that have $\sigma_{\rm AME}>5$, four sources have $f_{\rm max}^{\rm UC\hii}>1$, nine have $f_{\rm max}^{\rm UC\hii}>0.5$, fifteen have $f_{\rm max}^{\rm UC\hii}>0.25$, and eighteen have $f_{\rm max}^{\rm UC\hii}>0.1$ (Table~\ref{tab:main_list}).

We choose to remove the fifteen sources with $f_{\rm max}^{\rm UC\hii}>0.25$ from the significant AME list and re-classify them as semi-significant hereafter. Therefore they appear as grey circles in subsequent plots, and in Fig.~\ref{fig:uchii_max2cm} highlighted with star symbols. This cut-off value is chosen to keep G353.05+16.90 ($\rho$~Ophiuchi) in the significant AME sample, since there is good evidence for it not to harbour UC\hii\ regions \citep{Casassus2008}. Inspection of the SEDs of those sources with high $f_{\rm max}^{\rm UC\hii}$ reveals cases where the spectrum at frequencies between 20 and 60\,GHz is indeed very flat (as expected from optically thin free-free emission from \hii\ regions). The most prominent example of this is the source G213.71$-$12.60, which has the highest $f_{\rm max}^{\rm UC\hii}$ value of 3.9, and a high apparent excess significance of $15\,\sigma$. This is likely to be mostly dominated by UC\hii\ at these frequencies.

We note in this section that the two new AME regions reported by \cite{planck2011-7.2}, G107.20+05.20 and G173.62+02.79, have among the highest values of $f_{\rm max}^{\rm UC\hii}\approx\! 1$ (Table~\ref{tab:main_list}). A preliminary analysis of high resolution (2\arcm) follow-up observations at 15\,GHz with the AMI \citep{Perrott2013} indicates that for both these sources, the bulk of the {\it Planck} flux density is diffuse (on scales larger than 10~arcmin), making it unlikely for UC\hii\ to explain the excess. They found that while G107.2+5.2 may harbour a hyper-compact \hii\ (HC\hii) region with a rising spectrum at 15\,GHz, but with a flux density $\ll1$\,Jy, a spinning dust model is preferred. 

To quantify possible UC\hii\ regions (or other compact sources) at high latitudes, we searched the NED to look for compact radio sources and, where available, the GB6/PMN 4.85\,GHz maps\footnote{Maps downloaded from the SkyView: \url{http://skyview.gsfc.nasa.gov}}  \citep{Condon1991,Condon1993,Condon1994}. Within $1\deg$ of the central position there are typically two to five compact sources with flux densities in the range 0.03 to 0.1\,Jy. Most of these are already identified as extragalactic sources (therefore not UC\hii\ regions) and typically have flat or falling spectra. For example, near G023.47+8.19 lies the radio source PMN J0805$-$0438 with a flux density 0.193\,Jy at 4.85\,GHz. Comparison with measurements at 1.4\,GHz yields a spectral index of $-0.8\pm0.2$ for this source, implying a negligible contribution to the flux density at 23\,GHz. The most significant case is the flat-spectrum radio source PKS 1552$-$033 near G005.4+36.50, which has a flux density of 0.26\,Jy at 20\,GHz \citep{Murphy2010-ATCA}. This source contributes $15\,\%$ of the AME flux density of $1.7\pm0.3$\,Jy, but is within the uncertainty of the residual AME.

\begin{figure*}
\begin{center}
\includegraphics[scale=0.65]{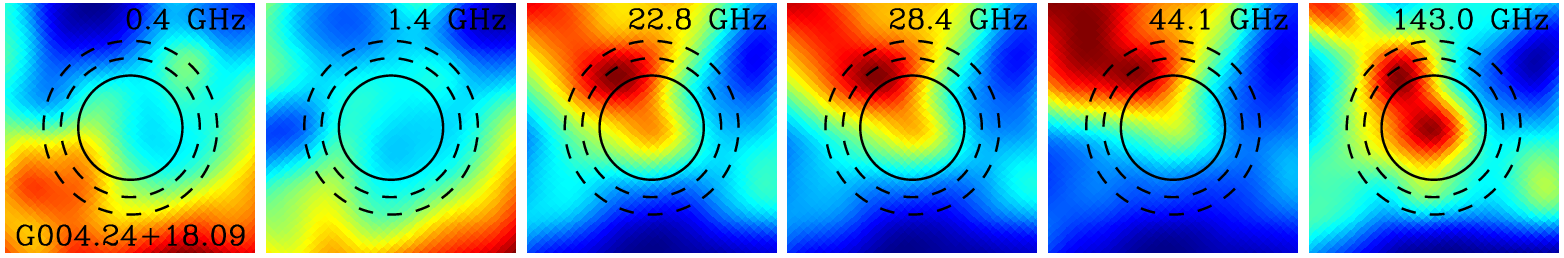}
\includegraphics[scale=0.65]{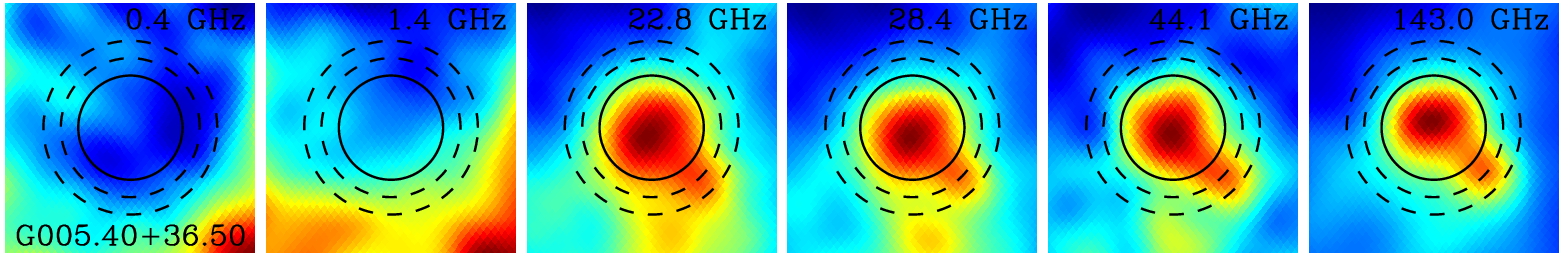}
\includegraphics[scale=0.65]{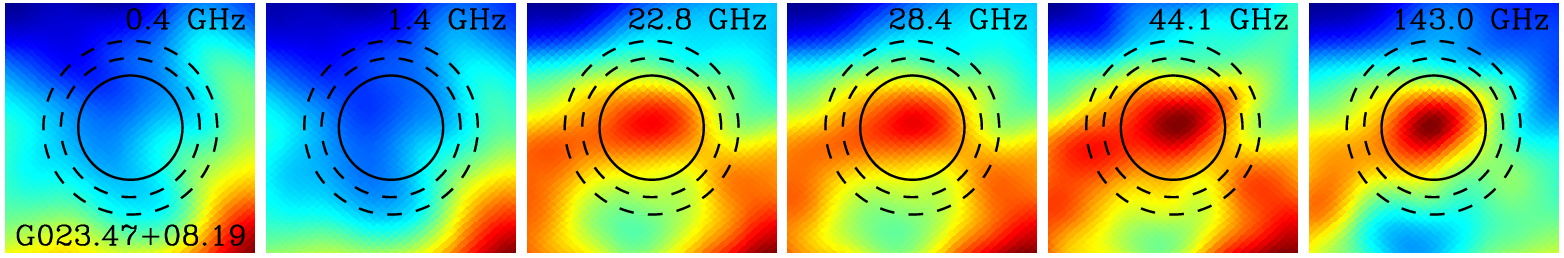}
\includegraphics[scale=0.65]{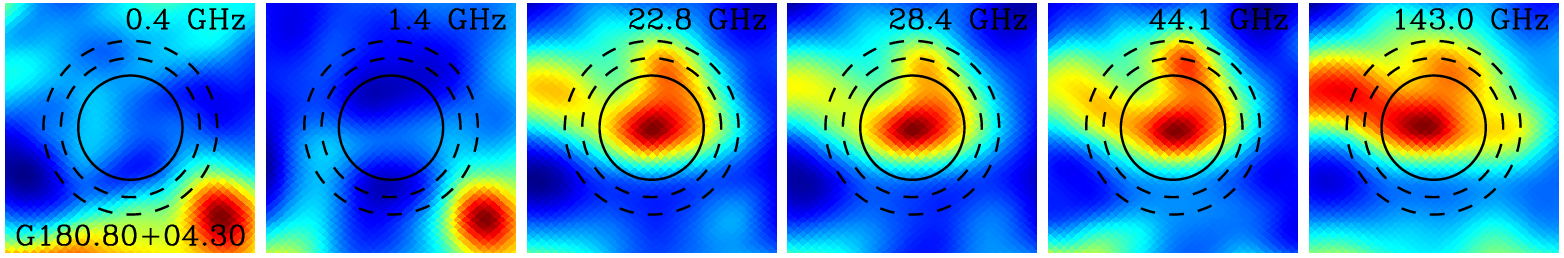}
\includegraphics[scale=0.65]{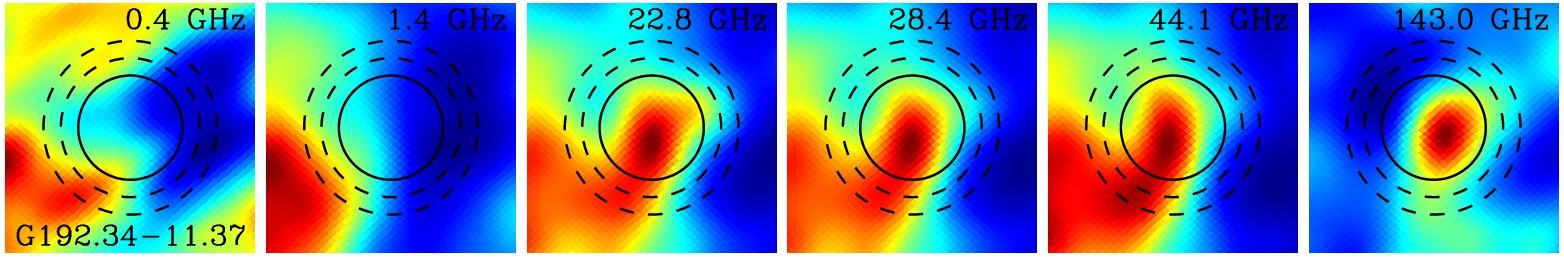}
\includegraphics[scale=0.65]{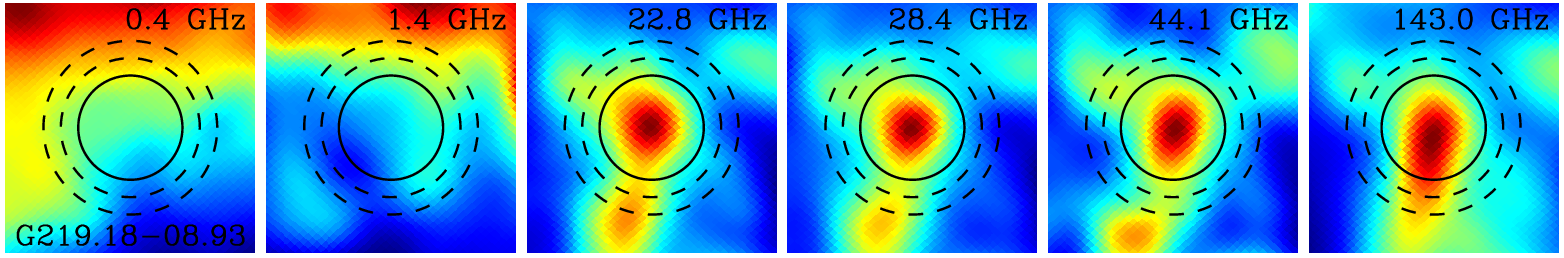}
\includegraphics[scale=0.65]{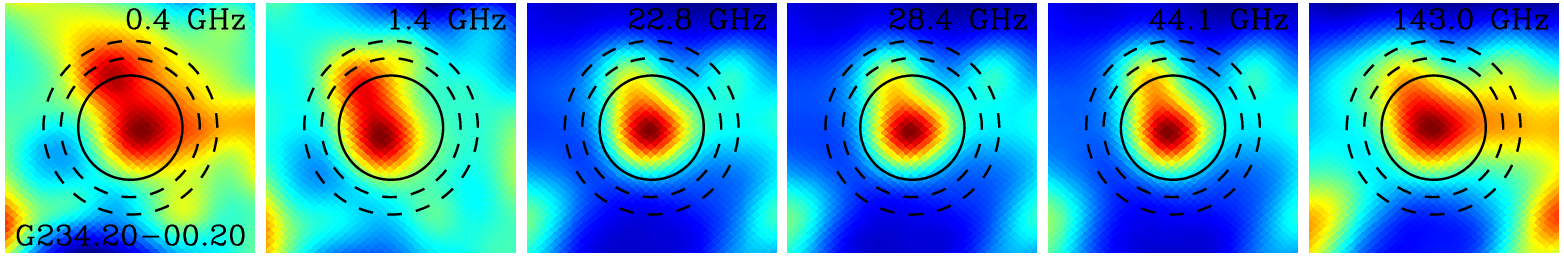}
\includegraphics[scale=0.65]{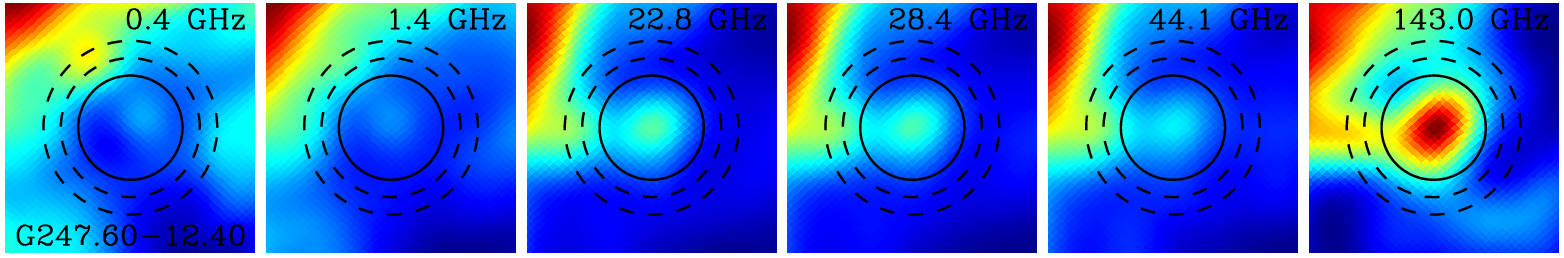}
\includegraphics[scale=0.65]{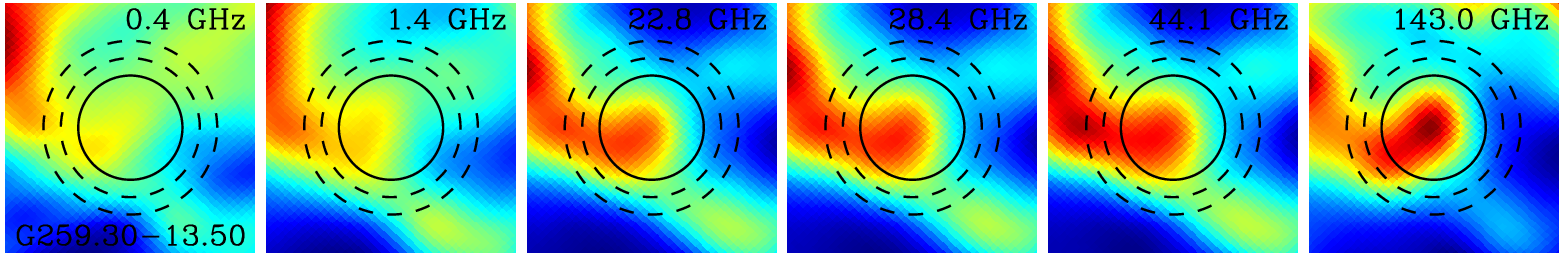}
\includegraphics[scale=0.65]{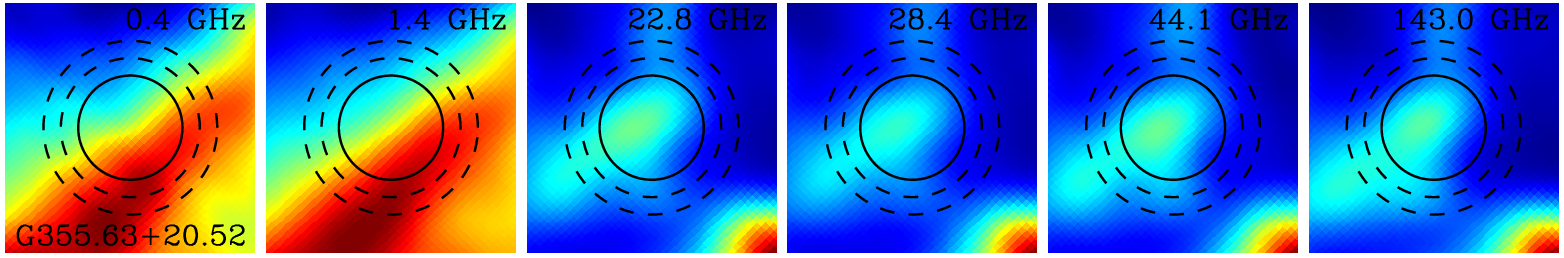}
\caption{Maps of example sources with significant ($\sigma_{\rm AME}>5$) excess emission. Each row is a gnomic map in Galactic coordinates, $5\deg$ on a side, and centred on each source, labelled in the 0.408\,GHz map. The maps from left to right are 0.408, 1.42, 22.7, 28.4, 44.1, and 143\,GHz. The {\it WMAP/Planck} maps have been CMB-subtracted. The colour-scale is linear, and ranges from the minimum to maximum value within each map. The aperture is shown as a solid line; the background annulus as a dashed line. The strong AME at frequencies around 30\,GHz is evident.}
\label{fig:maps_ameregions}
\end{center}
\end{figure*}
\begin{figure*}[tb]
\begin{center}
\includegraphics[scale=0.33]{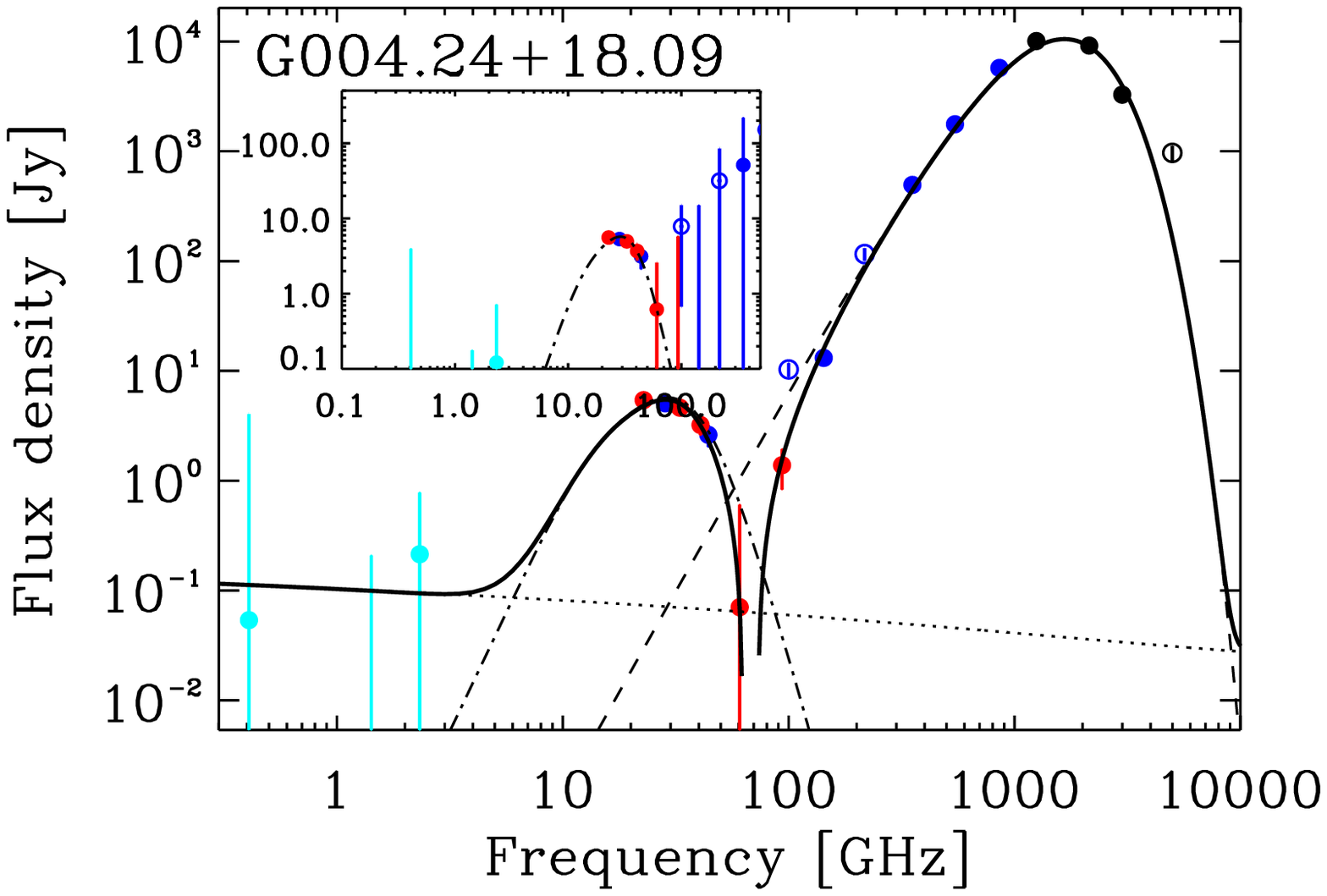}
\includegraphics[scale=0.33]{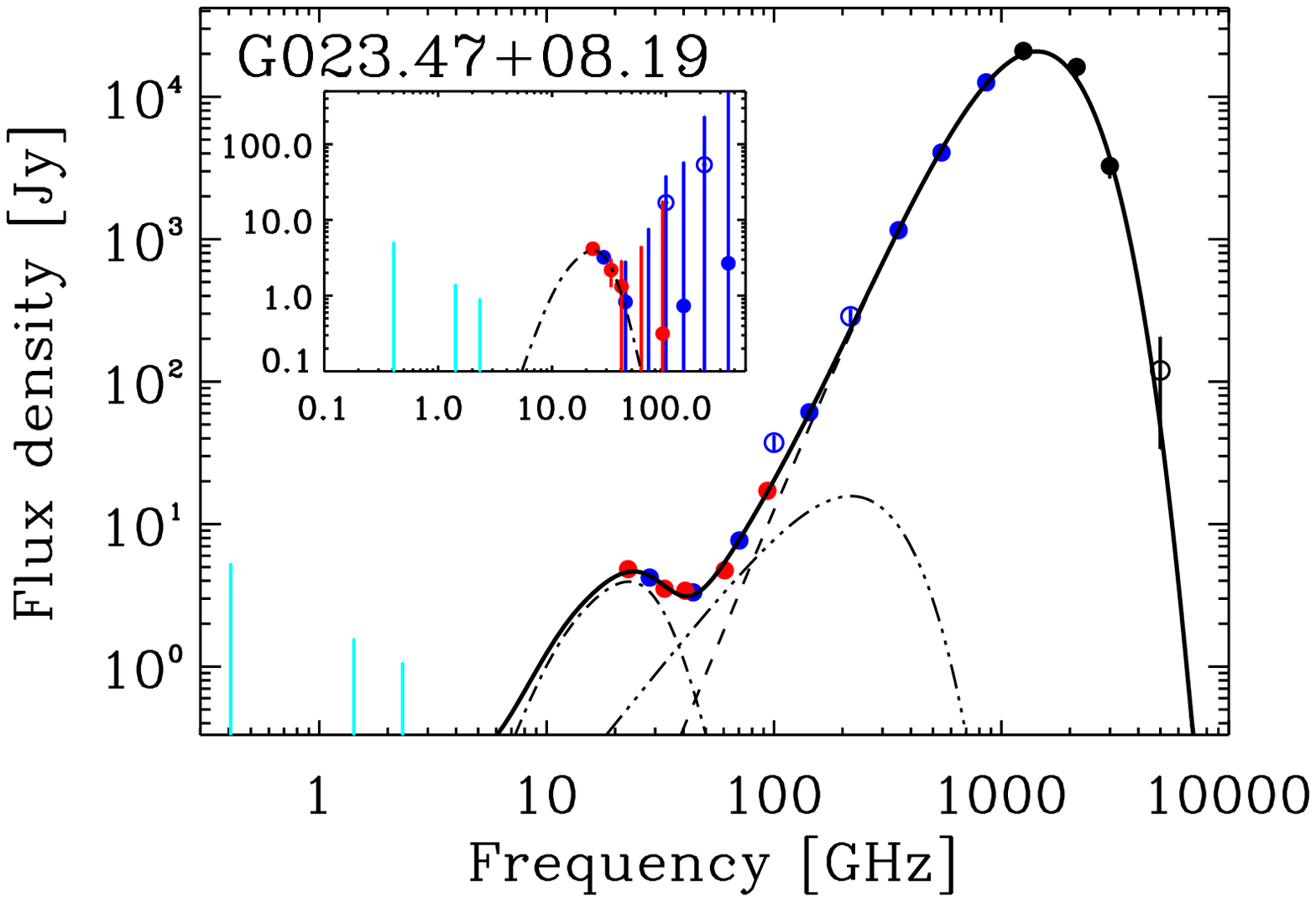}
\includegraphics[scale=0.33]{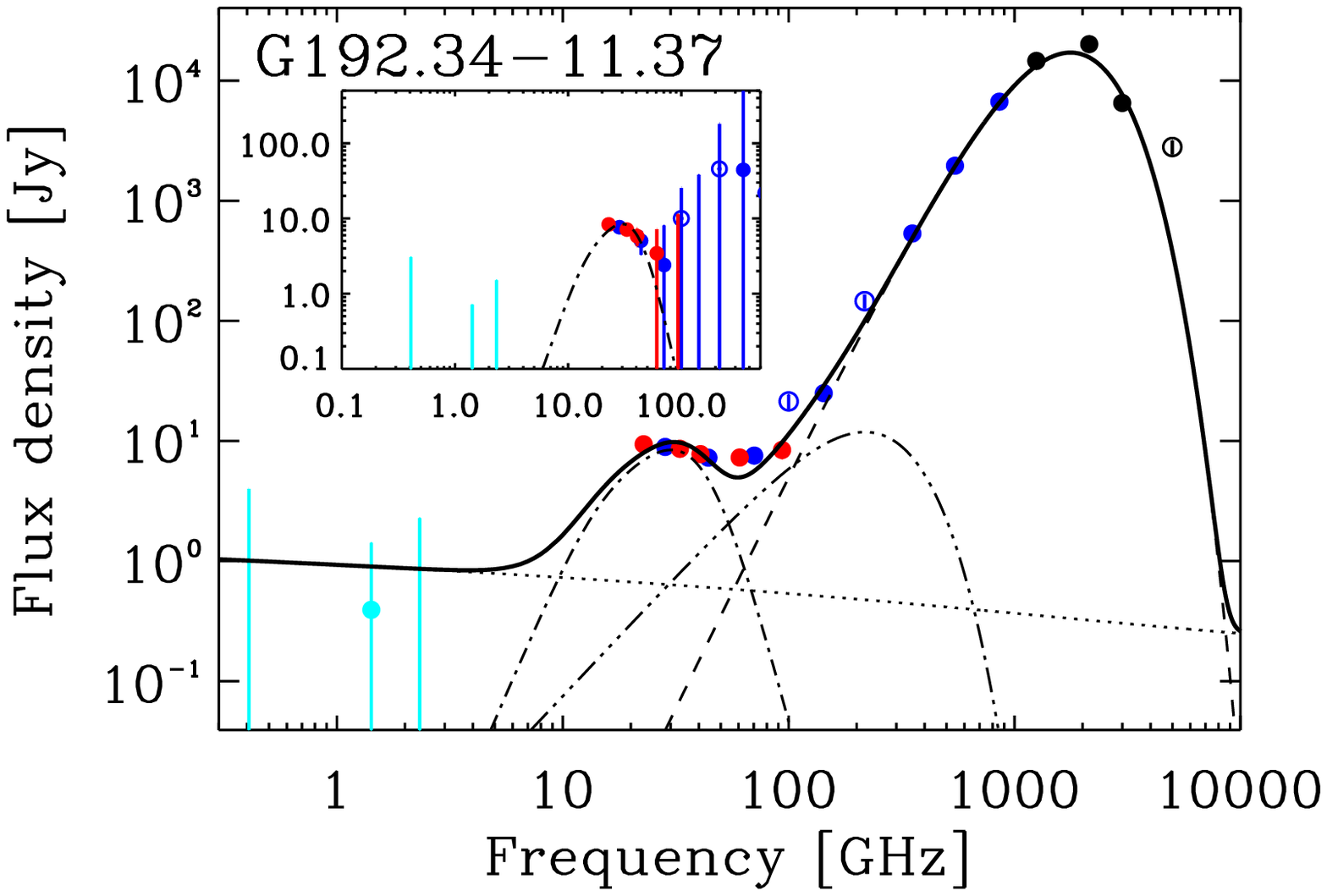}
\includegraphics[scale=0.33]{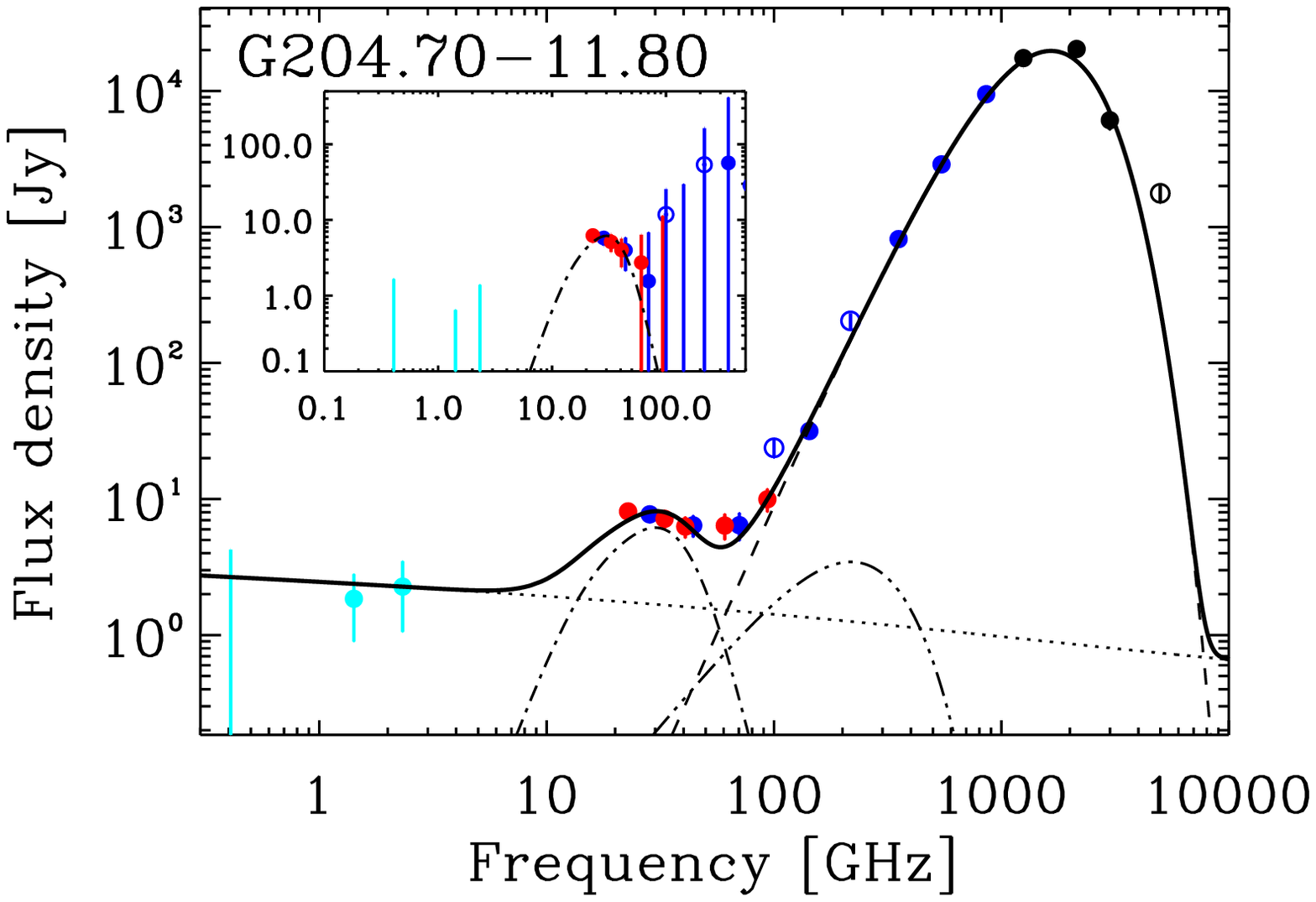}
\includegraphics[scale=0.33]{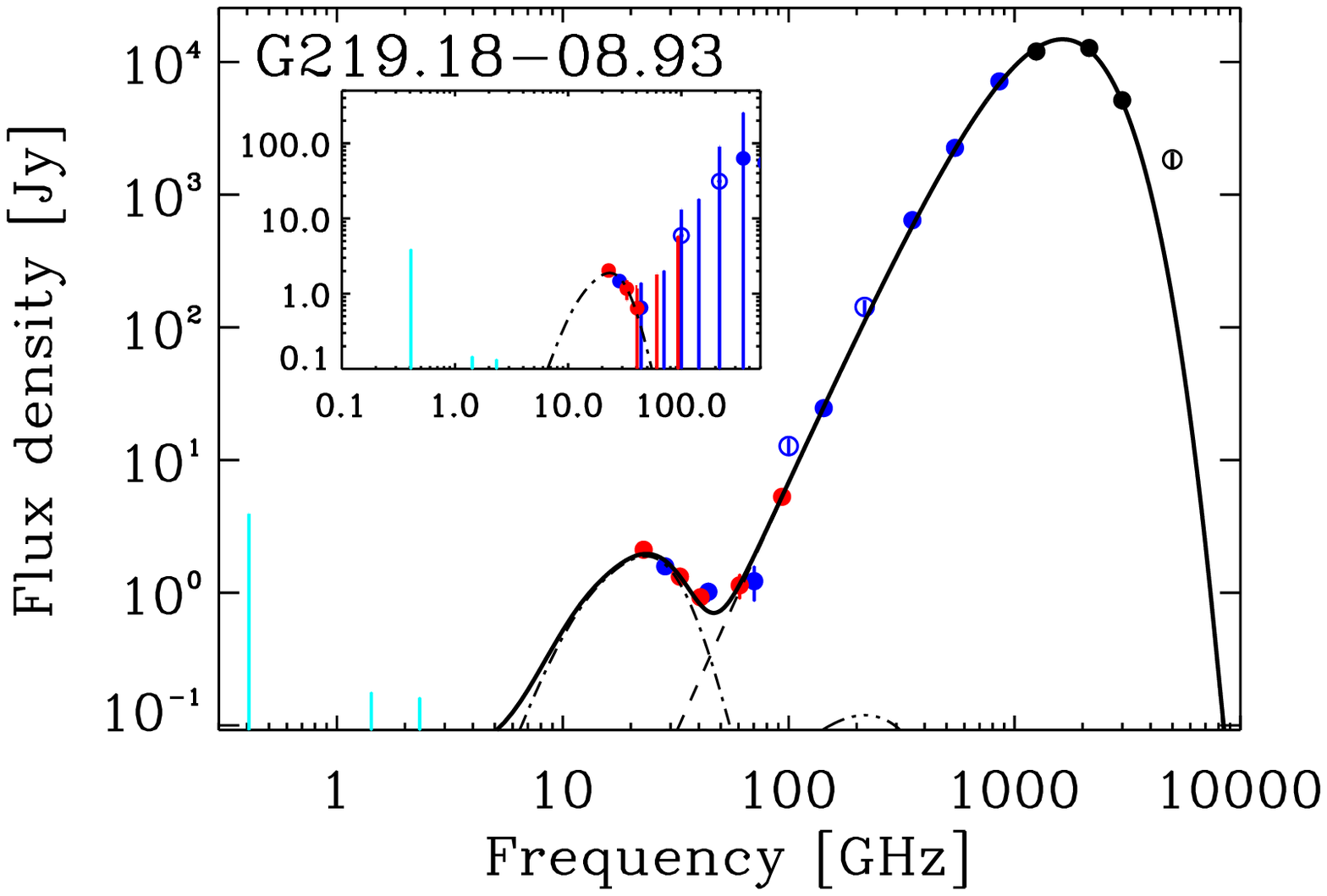}
\includegraphics[scale=0.33]{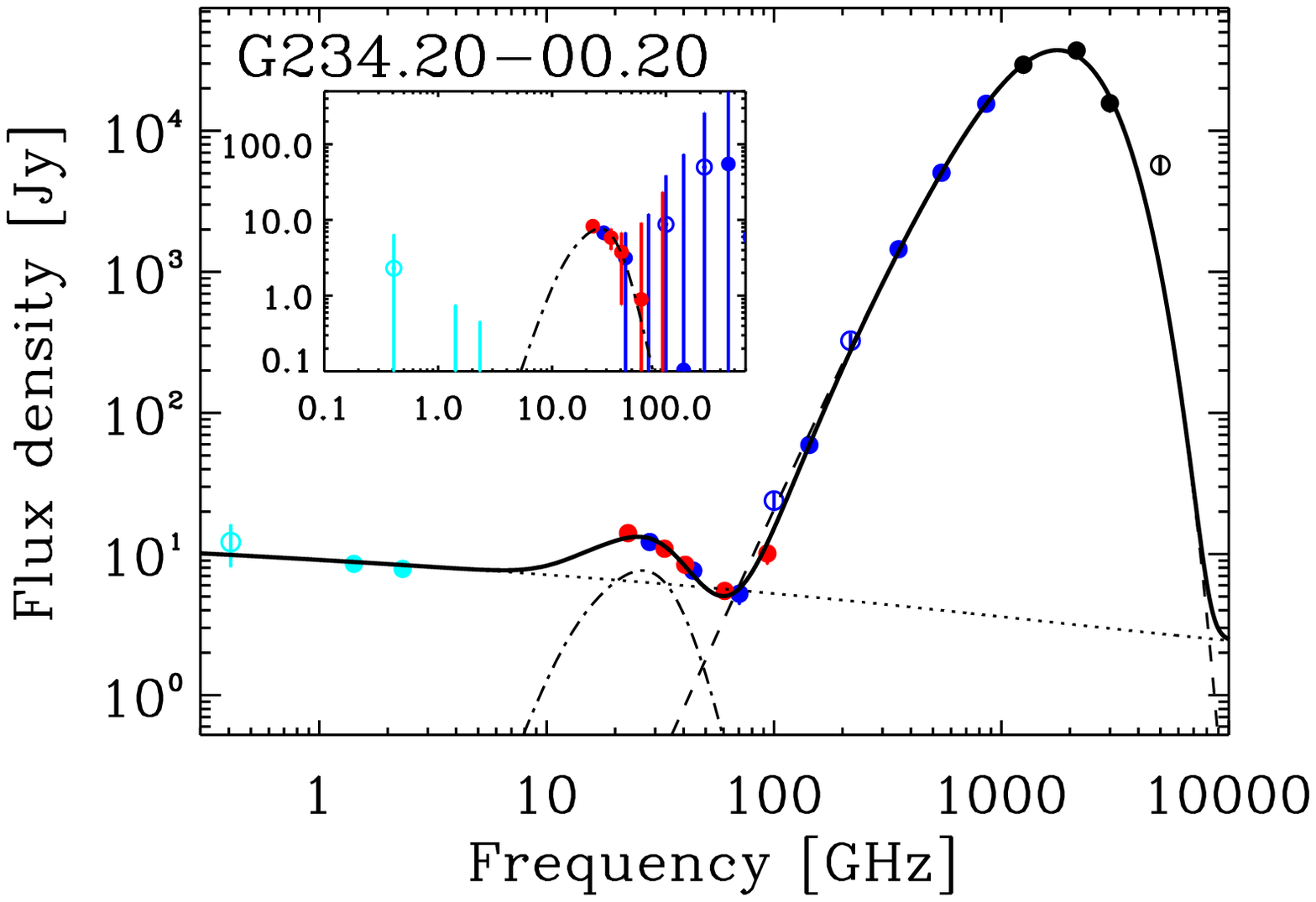}
\includegraphics[scale=0.33]{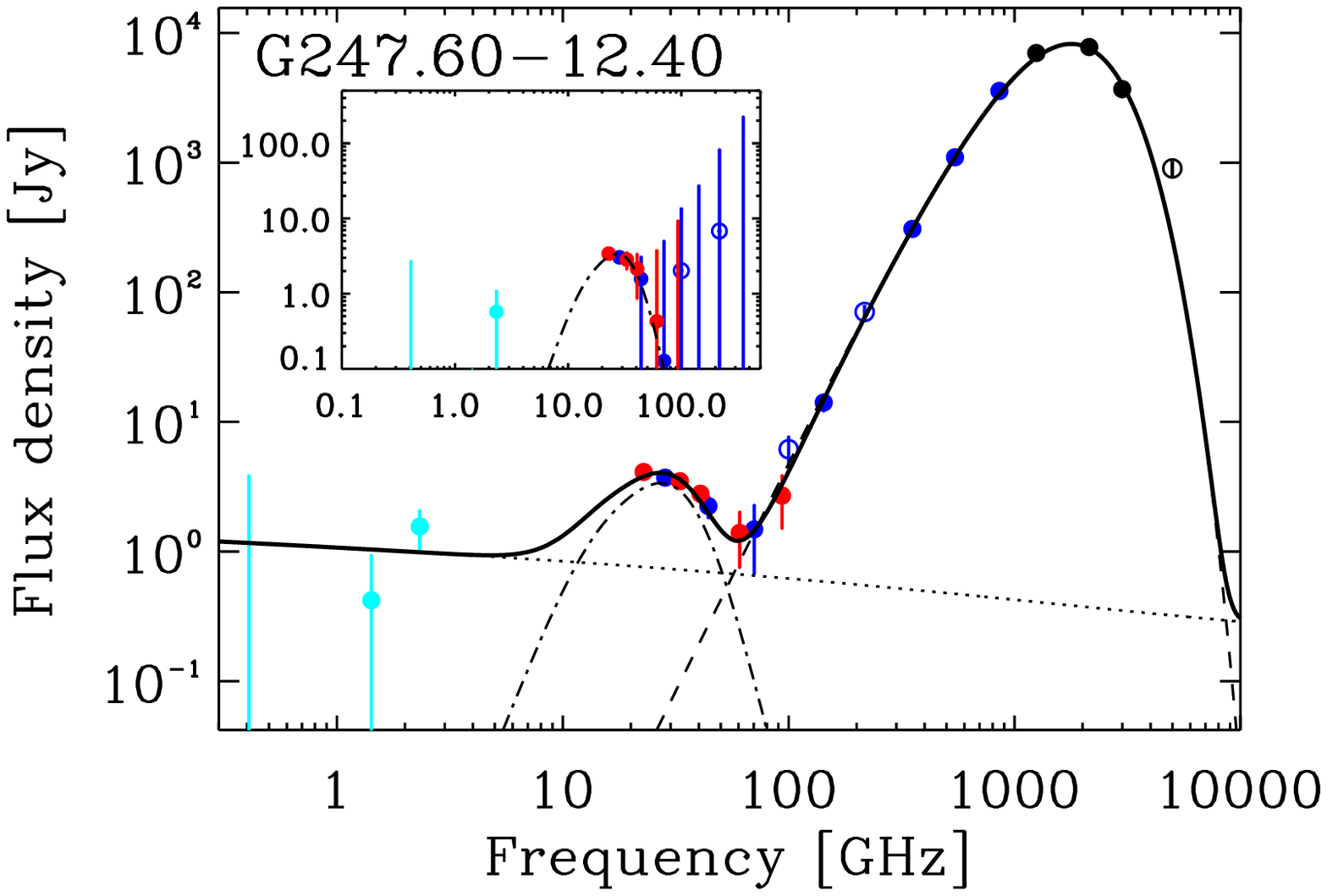}
\includegraphics[scale=0.33]{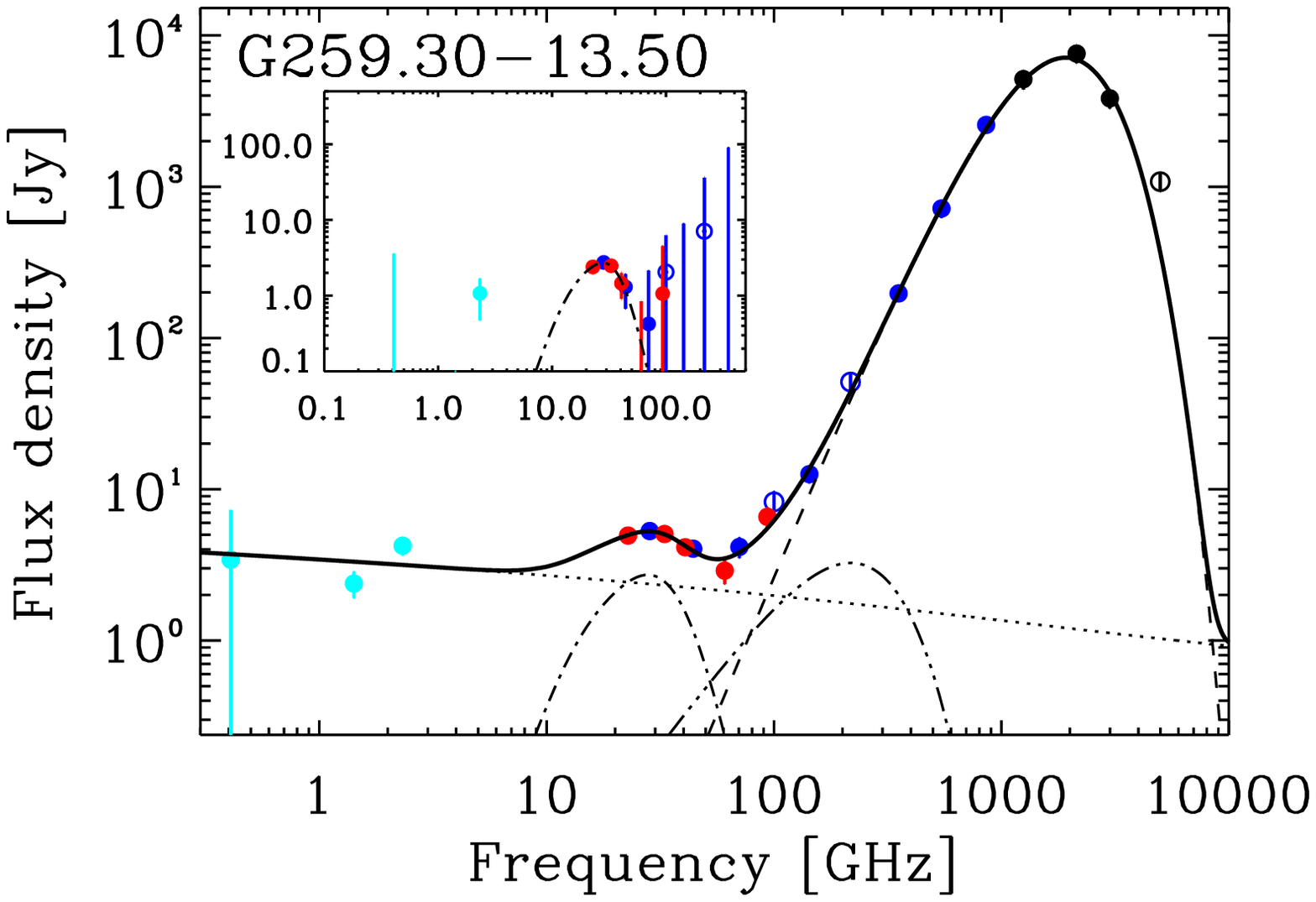}
\includegraphics[scale=0.33]{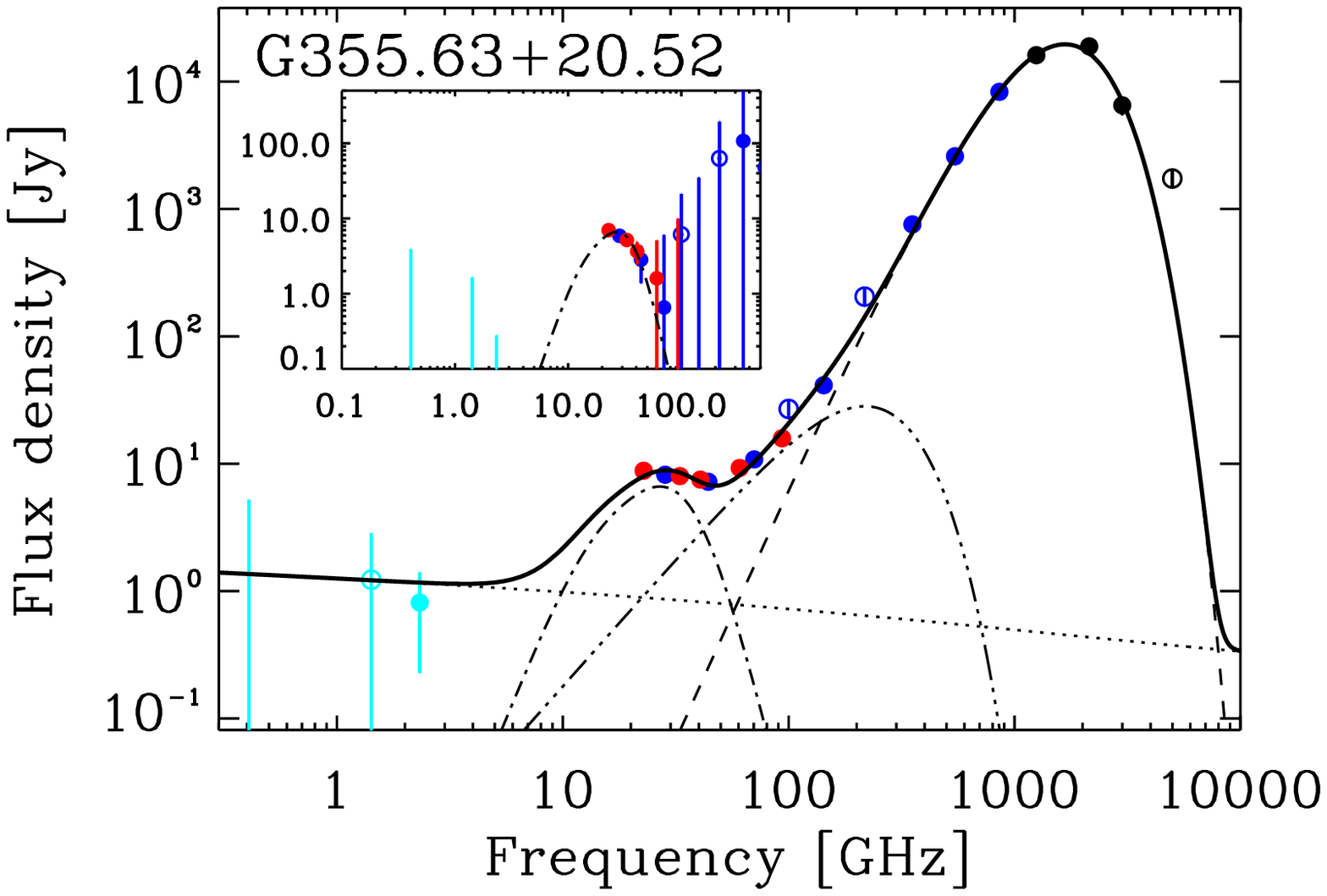}
\caption{SEDs for the sources with very significant AME and $f_{\rm max}^{\rm UC\hii}<0.25$. Data points are shown as circles with errors and are colour-coded for radio data (light blue), {\it WMAP} (red), \Planck\ (blue), and DIRBE/ \IRAS\ (black). The best-fitting model of free-free (dotted line), thermal dust (short dashed line), CMB (triple-dot dashed line), and spinning dust (dot-dashed line) is shown. Data included in the fit are shown as filled circles, while the other data are unfilled. The residual spectrum, after subtraction of free-free, synchrotron, CMB, and thermal dust components, is shown as the insert. The best-fitting spinning dust model is also shown.}
\label{fig:ame_seds}
\end{center}
\end{figure*}

\subsection{Maps and SEDs of most significant AME detections}
\label{sec:maps_seds}

Figure~\ref{fig:maps_ameregions} presents maps of some of the most statistically significant AME regions, excluding the previously known sources in Perseus and $\rho$~Ophiuchus. The {\it Planck} maps have been CMB-subtracted. The brighter emission at $20$--$60$\,GHz can sometimes be seen visually. In the most significant AME sources there is little or no discernible emission at low radio frequencies ($<2.3$\,GHz), but there is almost always strong thermal dust emission at high frequencies ($\gtrsim 143$\,GHz). Good examples of this include G005.40+36.50, G023.47+08.19, G192.34$-$11.37, and G219.18$-$8.93. In other sources the picture is not so clear, due to strong backgrounds (e.g., in G355.63+20.52) or because the source is not so well-defined.  For example, G004.24+18.09 appears as a spur emanating from a diffuse region. The images show that many of the AME sources are not compact, but rather diffuse and extended. This will be quantified in Sect.~\ref{sec:sizes}.

Figure~\ref{fig:ame_seds} shows SEDs for some of the most significant AME regions found in this study, excluding the previously known sources. The inset is the residual spectrum after subtraction of free-free, thermal dust, and CMB components. The error bars in the residual spectrum include the additional uncertainty in the model by simple propagation of errors. For these regions the residual spectrum shows significant flux density, although the uncertainties are larger than in the raw spectrum due to  subtraction of the model. Nevertheless, the best-fitting spinning dust spectrum is seen to be a remarkably good fit to these residuals. 

Similar to the example SEDs of the whole sample (Fig.~\ref{fig:example_seds}), the SEDs of the AME regions exhibit a wide range of phenomena. In some regions (e.g., G004.24+18.09, G023.47+08.19, G192.34$-$11.37, and G219.18$-$08.93), there is little radio emission at frequencies near 1\,GHz, resulting in the majority of the flux density at $30$\,GHz being assigned to spinning dust, and possibly a contribution from the CMB above about 70\,GHz. Other regions have a reasonably well-defined level of free-free emission, but the $20$--$100$\,GHz flux density is much higher than the extrapolation of the free-free and thermal dust models (e.g., G204.70$-$11.80 and G234.20$-$00.20).

\subsection{Robustness and validation}
\label{sec:robustness}

Throughout the analysis we have taken a conservative approach to the estimation of uncertainties. A critical part of the analysis is the extrapolation of synchrotron and free-free emission from low radio frequencies ($\approx\! 1$\,GHz) to {\it WMAP}/\Planck\ frequencies. We believe that our simple models of the synchrotron and free-free emission spectral laws are appropriate based on many other studies, both experimental and theoretical; see \cite{Bennett2003b} for an overview. The contribution from UC\hii\ regions at {\it WMAP}/\Planck\ frequencies could be important for a fraction of the sample, but is not thought to be a major contribution in general (see Sect.\,\ref{sec:uchii}).

We find that for the majority of the sources in this study, $\chi^2$/dof$<1$ (see Table~\ref{tab:main_list}), with a mean value $\bar{\chi^2}$/dof$=0.59$. For a number of sources it is clear that the uncertainties are overestimated, as the scatter in the flux densities relative to our model is much smaller than the error bars would suggest. In many cases this can be attributed to bright backgrounds near the region of interest. Also, the calculation of $\chi^2$ assumes that the errors are uncorrelated, while the errors will certainly be correlated to some degree due to similar backgrounds and also similar absolute calibration values from frequency to frequency. The contribution to the $\chi^2$ from various frequency ranges is found to be approximately equal (e.g., by comparing the $\chi^2$ values for data between 0.4 and 2.3\,GHz and between $353$ to $857$\,GHz). Therefore, we do not appear to be systematically over-estimating (or under-estimating) the errors in any particular range of frequencies.

We made a number of tests to check that our main results are not grossly affected by our assumptions and fitting methods. These include changing our assumed calibration uncertainties, aperture radii, background annulus radii, spinning dust model, and starting values for the fitting algorithm. In all cases, we find that the general trends presented here are unchanged and the SEDs do not change appreciably; however, we note that in a few individual cases, the spectral model does vary depending on some details of the analysis. These cases are mostly related to the low frequency components, specifically the free-free level, which is not always well-constrained by the data. The uncertainties reflect this, and we are confident in most of the AME detections presented here. Nevertheless, we recommend caution when examining individual sources in detail. Follow-up observations should be made for all our high significance regions.

\begin{figure}[tb]
\begin{center}
\includegraphics[scale=0.5]{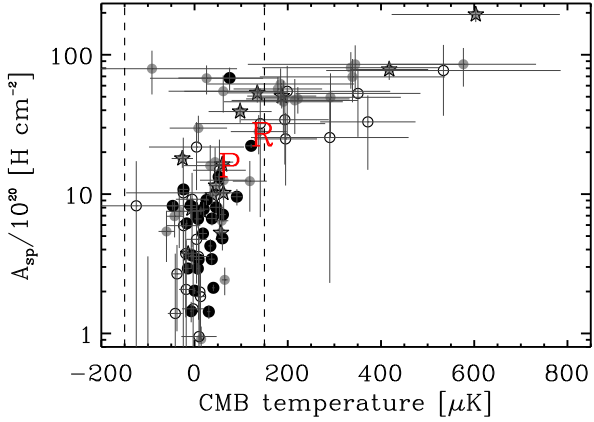}
\caption{Correlation plot between the AME amplitude, $A_{\rm sp}$, and the fitted CMB fluctuation temperature. Symbols are as in Fig.~\ref{fig:uchii_max2cm}. For most sources, the fitted CMB temperatures are within the expected range shown as dashed lines ($|\Delta T_{\rm CMB}|< 150\,\mu$K) and are not strongly correlated with the AME flux density. However, for some sources, there is a strong positive correlation with CMB temperatures that are higher than expected, but associated with larger uncertainties. }
\label{fig:cmb}
\end{center}
\end{figure}

For most sources, we find the CMB fluctuation temperatures are within the expected range $-150 < \Delta T_{\rm CMB} < 150\,\mu$K.  From Monte Carlo simulations, one would expect a fluctuation outside this range only 0.7\,\% of the time; however, in some cases the fitted CMB temperatures (see Table~\ref{tab:main_list}) are found to be larger than expected. Furthermore, a correlation between the AME amplitude and CMB is observed. Figure~\ref{fig:cmb} shows the correlation between the AME amplitude and CMB fluctuation temperature for the entire sample. The AME regions (solid filled circles) are mostly within the expected range, with relatively small uncertainties; however, the highest AME amplitudes ($A_{\rm sp}>20 \times 10^{20}\,$cm$^{-2}$) also have the highest CMB temperatures, which are well above what can be reasonably expected from the CMB alone ($\Delta T_{\rm CMB}>150\,\mu$K). As discussed in Sect.~\ref{sec:example_seds}, some regions exhibit a flattening of the thermal dust spectral index at frequencies in the range $100$--$353$\,GHz \citep{PIP96} that can be artificially accounted for by a stronger positive CMB fluctuation. This then results in a positive bias at frequencies $< 100$\,GHz, which increases the AME amplitude. Some of the sources with high CMB temperatures also have high $f_{\rm max}^{\rm UC\hii}$; these are shown as star symbols in Fig.~\ref{fig:cmb}. We do not believe this is a major effect on our most significant AME sample (i.e., $\sigma_{\rm AME}>5$ and $f_{\rm max}^{\rm UC\hii}<0.25$), since none of the AME sources has an anomalously high CMB temperature. Although the CMB has a stronger effect on the semi-significant AME regions, the uncertainties associated with it are larger than for the rest of the sample, as can be seen in Fig.~\ref{fig:cmb}. 

To test robustness, the entire analysis was repeated without fitting for a CMB component ($\Delta T_{\rm CMB}=0\,\mu$K); the results do not change substantially. We do find an additional source in G102.88$-$00.69 with a significance of $\sigma_{\rm AME}=6.9$, but this is clearly due to a negative CMB fluctuation ($\Delta T_{\rm CMB}=-60\pm17\,\mu$K). For the largest fitted CMB temperature of $\Delta T_{\rm CMB}=603\pm180\,\mu$K in G030.77--00.03, we find an AME amplitude of $(194\pm22) \times 10^{20}$\,cm$^{-2}$ ($8.5\,\sigma$). Without a CMB component ($\Delta T_{\rm CMB}=0$), the AME amplitude is $(154\pm20) \times 10^{20}$\,cm$^{-2}$ ($7.6\,\sigma$). In this case the CMB does not appear to be causing a large bias, although there is a $\approx\! 1$--$2\,\sigma$ bias in a few sources. This justifies our high cut-off threshold of $5\,\sigma$. We also verify that the general trends presented in our analysis still hold when not taking into account the CMB.

In summary, we are confident in the robustness of our AME detections, particularly those at $\sigma_{\rm AME}>5$. We have been conservative with the uncertainties in the photometry, and in most cases our SEDs do not change appreciably when changing the details of the analysis. In fact, we believe that many of the regions with $\sigma_{\rm AME}= 2$--$5$ (``semi-significant'') are likely to be ``real'' detections of AME, which can be seen in many of the subsequent plots to be consistent with the higher-significance AME sources.

\section{Statistical study of AME regions}  \label{sec:statistical_study}
\label{sec:stats}

In this section we study statistically a number of the observational-based parameters and correlations in AME and non-AME regions, and try to investigate the nature of the AME sources and the role that the environment plays.

\subsection{Nature of the sources}

\subsubsection{Angular sizes}
\label{sec:sizes}

First, we would like to know whether the sources are extended or compact relative to the analysis resolution of 1\deg. We have seen already that there is visual evidence that the strongest AME regions appear to be extended. This tendency for AME/spinning dust to originate mostly from the diffuse/extended emission has been seen in several previous studies. The majority of AME sources are diffuse, including the Perseus and $\rho$~Ophiuchi clouds; \cite{Tibbs2010} found that in the Perseus molecular cloud at least 90\,\% of the AME comes from diffuse extended emission. Surveys at higher resolution do not appear to detect strong AME; \cite{Scaife2008} found little or no detectable AME from a sample of relatively compact ($\lesssim 1\arcmin$) \hii\ regions.

To estimate the size of each source, an elliptical Gaussian is fitted to the pixels within the photometric aperture of radius $1\deg$, taking into account an offset and gradient in the background. We take the average size, defined as the mean of the major and minor FWHM axes derived from the Gaussian fit to the \Planck~28.4\,GHz map, and deconvolve this from the map resolution, as $\theta = \sqrt{{\rm FWHM}^2 - 1^2}$; these values are listed in Table~\ref{tab:main_list}. The uncertainties are estimated from the average noise level at 28.4\,GHz. We repeat the analysis at other frequencies and obtain similar results, with most sources agreeing in size to within $0\pdeg1$--$0\pdeg2$.  Derived values greater than $2\deg$ are not found to be very robust, but they  nevertheless indicate that the source is very extended.

\begin{figure}[tb]
\begin{center}
\includegraphics[scale=0.5]{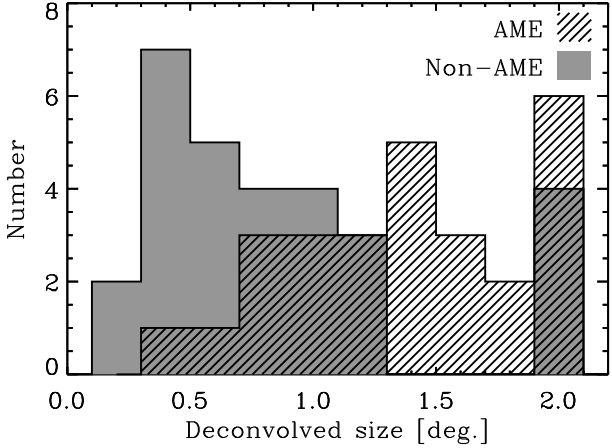}
\caption{Histogram of average deconvolved FWHM sizes for AME and non-AME sources, as derived from Gaussian fitting to the \Planck~28.4\,GHz map at $1\deg$ resolution. Sources with ${\rm FWHM} > 2$\deg are plotted as having a FWHM$=$2\deg. The most significant AME sources are shown as the hatched histogram, while the non-significant (non-AME) sources are shown as the grey histogram. The strongest AME sources are, in general, extended, while non-AME sources tend to be relatively compact.}
\label{fig:hist_sizes}
\end{center}
\end{figure}

Figure~\ref{fig:hist_sizes} shows the distribution of deconvolved sizes for the entire sample. There is a clear tendency for the AME sources to be more extended, while non-AME sources tend to be compact, relative to the beam. The AME sources have a mean size of $1\pdeg39\pm0\pdeg09$, while the non-AME sources are at $0\pdeg86\pm0\pdeg11$, with many of them essentially unresolved in maps at 1\deg\ resolution. Semi-significant sources have a mean size of $1\pdeg21\pm0\pdeg08$. As mentioned earlier, this is a trend that is becoming increasingly apparent in studies of AME. It could of course be a selection effect, since the dust grains around young and therefore compact \hii\ regions are known to be different to the dust grains in the ISM environment away from hot OB stars.  For example, the PAH population is depleted in close proximity to hot OB stars \citep{Boulanger1988,Povich2007}. This could be important for understanding the dust grain population and environments responsible for exciting the grain rotation. Indeed, \cite{planck2011-7.3} found that there was little or no AME associated with the warm neutral medium (WNM) or WIM phases.

\subsubsection{AME fraction and IRE -- \hii\ or molecular dust clouds?}
\label{sec:AME_fraction}

We would like to establish how much of the total emission, at a given frequency, is due to AME. At frequencies near 30\,GHz, the dominant emission will be due to either AME or free-free emission. \hii\ regions will exhibit strong free-free emission from warm ($T_e \approx\! 10^4$\,K) ionized gas, while dust clouds without massive star formation will have little or no associated free-free emission. 

\begin{figure}[tb]
\begin{center}
\includegraphics[scale=0.5]{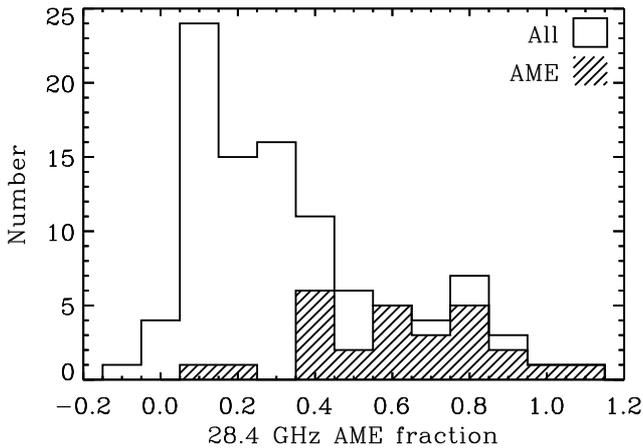}
\caption{Histogram of the AME fraction, $S_{\rm resid}^{28}/S_{28}$, at 28.4\,GHz. The total sample is shown as the unfilled histogram, while significant ($\sigma_{\rm AME}>5$) AME sources is the hatched region.}
\label{fig:ame_fraction}
\end{center}
\end{figure}

To calculate the AME fraction we subtract the non-AME components at 28.4\,GHz and propagate the uncertainties to leave the 28.4\,GHz AME residual, $S_{\rm resid}^{28}$, and its uncertainty, $\sigma_{\rm resid}^{28}$. We then estimate the AME fraction from the ratio of the AME residual to the total flux density at 28.4\,GHz.  The histogram of the 28.4\,GHz AME fraction is shown in Fig.~\ref{fig:ame_fraction}. As expected, the AME sources exhibit a much higher fraction of AME at 28.4\,GHz; in many, at least half of the 28.4\,GHz flux density could be due to AME.

The weighted mean AME fractions are $0.50\pm0.02$ (AME) and $0.06\pm0.02$ (non-AME).This can be compared with the study of \hii\ regions by \cite{Todorovic2010}, who found an average AME fraction at 30\,GHz of $0.41\pm0.10$.  \cite{Dickinson2009a} found that approximately half the 30\,GHz flux density from the \hii\ region RCW175 was due to AME. The average value in the Galactic plane appears to be lower; \cite{planck2011-7.3} found that $0.25\pm0.05$ of the 30\,GHz flux density was due to AME within the inner Galactic plane, while Planck Collaboration (in prep.) found a value of $0.42 \pm 0.02$. Although this is a considerable fraction, it is less than what has been observed in the diffuse Galactic foreground at high latitudes, where the dust-correlated AME is the dominant component and accounts for approximately $75\,\%$ of the total 30\,GHz emission \citep{Davies2006,Ghosh2012}. On the other hand, other measurements \citep{Dickinson2007,Scaife2008} have found much less or no AME as a fraction of the total flux density. Sources with a smaller AME fraction ($\lesssim 20\,\%$) may be prevalent, but are much more difficult to detect since the other non-AME components must be accurately removed first. The weighted average AME fraction for the semi-significant AME sources is $0.25\pm0.02$ in this study.

Most of our sources appear to be associated with \hii\ regions as well as molecular clouds within giant molecular complexes (GMCs). The fact that most bright FIR sources are coincident with \hii\ regions is well-known (e.g., \citealt{Myers1986}); however, a few AME sources in our sample appear to have very little free-free emission, leaving the bulk of 23\,GHz emission as AME. These regions tend to be associated with dark nebula \citep{Lynds1962} with no known \hii\ regions present. The highest significance examples of these are LDN137/141 (G004.24+18.09), LDN134 (G005.40+36.50), LDN1557 (G180.18+04.30), and LDN1582/1584 (G192.34$-$11.37). Many of the other AME regions do have strong free-free emission and can be associated with \hii\ regions. The most notable associations of dark nebulae and \hii\ regions are given in the Notes in Table~\ref{tab:main_list}. 

To investigate the nature of the heating of the dust, we estimate the infrared excess, defined by \cite{Mezger1978} as ${\rm IRE} \equiv L_{\rm FIR}/L_{\rm Ly\alpha}$. The IRE is a measure of the heating of dust associated with \hii\ regions. If ${\rm IRE} =1$, the FIR luminosity can be explained by stellar Ly$\alpha$ photons, which, after absorption by the gas and degradation to Ly$\alpha$, are absorbed by dust grains. If ${\rm IRE} >1$, either direct dust absorption of stellar photons or additional stars producing little ionization, or both, are needed \citep{Myers1986}.

\begin{figure}[tb]
\begin{center}
\includegraphics[scale=0.5]{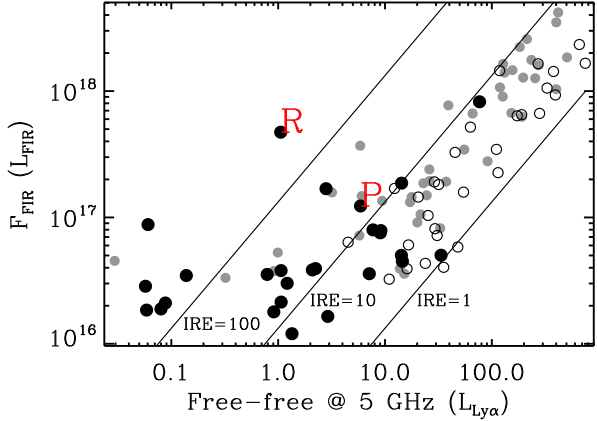}
\caption{Integrated FIR flux (proxy for FIR luminosity) plotted against the free-free flux density at 5\,GHz (proxy for Ly$\alpha$ luminosity). Symbols are as in Fig.~\ref{fig:uchii_max2cm}. Perseus (P) and $\rho$~Ophiuchi (R) clouds are marked. IRE values of 1, 10, and 100 are shown as straight lines. Most AME sources have ${\rm IRE}>4$.}
\label{fig:ire}
\end{center}
\end{figure}

As a proxy for $L_{\rm FIR}$, we estimate the total flux in the FIR by integrating the best-fitting thermal dust model across all frequencies. As a proxy for $L_{\rm Ly\alpha}$, we use the best-fitting free-free model to calculate the 5\,GHz flux density. Both these quantities depend on the square of the distance, which cancels out in the IRE. Figure~\ref{fig:ire} shows these two quantities plotted against each other.  There is a wide range of IRE values covering 1 to greater than $100$. Sources with ${\rm IRE} >100$ are unlikely to harbour strong \hii\ regions, and are mostly molecular dust clouds with no high-mass star formation; $\rho$~Ophiuchi lies in this region. AME sources tend to have higher IRE values compared to the rest of the sample with most AME sources at ${\rm IRE} \gtrsim 4$, with a median value of 14.8; most of the non-AME sources are at IRE$<10$, with a median value of 2.6. This suggests that AME comes from the molecular cloud dust or PDR, but not from within \hii\ regions themselves. This fits with the general idea that PAHs are destroyed in \hii\ gas, and that AME emissivity is lower in the ionized phase of the ISM.

\subsubsection{Dust properties}
\label{sec:dust_properties}

Next we look at the basic dust properties from the fitted model of thermal and spinning dust components and \IRAS\ colour ratios. We emphasize that the dust properties are strictly related to the big dust grains and are not thought to be directly responsible for the AME, which will be from the smallest dust grains and PAHs if due to spinning dust. Furthermore, for some sources it is possible that the FIR dust emission is not spatially coincident with the AME. Nevertheless, we expect that the global dust properties will be indicative of the environment close to where the AME is originating.

\begin{figure}[tb]
\begin{center}
\includegraphics[scale=0.5]{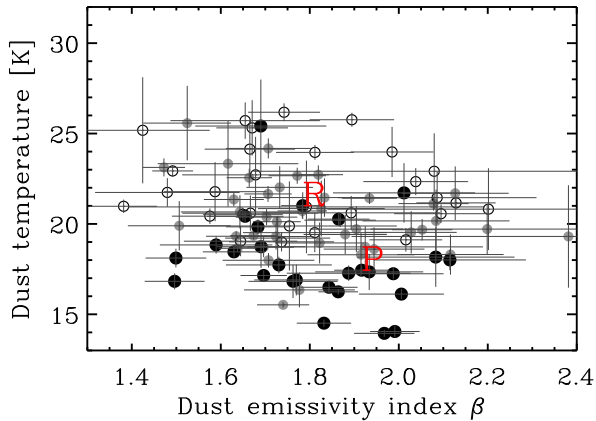}
\caption{Big grain dust temperature ($T_{\rm d}$) vs thermal dust emissivity index ($\beta_{\rm d}$). The symbols are as in Fig.~\ref{fig:uchii_max2cm}. Perseus (P) and $\rho$~Ophiuchi (R) clouds are marked. There is no strong trend with $\beta_{\rm d}$, while the most significant AME regions (filled black circles) tend to be associated with colder dust.}
\label{fig:td_beta}
\end{center}
\end{figure}

Figure~\ref{fig:td_beta} shows the dust temperature ($T_{\rm d}$) against the fitted thermal dust emissivity index ($\beta_{\rm d}$). The dust temperatures are as expected for evolved diffuse \hii\ regions and molecular clouds, in the range $\approx 15$--$30$\,K, with a weighted mean of 18.5\,K.  This is slightly cooler than the average expected for typical \hii\ regions, and is close to the average value found in the diffuse ISM \citep{Planck2013_allskydust}. This is probably due to the fact that we are measuring the average temperature over a relatively large area, and are therefore sensitive to the diffuse dust in the large-scale environment, as opposed to the dust in the direct vicinity of hot stars. The dust emissivity index is also in the expected range (for a single component fit) of $\beta_{\rm d}=1.4$--$2.4$, with a weighted mean of 1.76. This is similar to the value of 1.78 found in nearby molecular clouds in \cite{planck2011-7.13}. We note that more recent studies have found a flattening of the thermal dust emissivity index at frequencies around 100\,GHz to values in the range 1.5--1.7 \citep{PIP96}. This may make a small difference to the overall fits presented here, but will have a very small or negligible impact on the AME flux density at frequencies of 20--60\,GHz.

Figure~\ref{fig:td_beta} shows two trends. First, the most significant AME sources tend to be associated with cooler dust temperatures. Most of the AME sources are in the range $14 \le T_{\rm d} \le 20$\,K, with a weighted mean of 16.6\,K, compared to 22.0\,K for the non-AME regions. This is in line with the results of \cite{planck2011-7.3}, in which it was found that the AME sources are dominated by emission from the molecular phase and that hotter dust has little or no AME.  However, it is the opposite of what is found at smaller angular scales in the Perseus molecular cloud \citep{Tibbs2011}, where the AME-dominated regions are slightly warmer on average, ($20.2\pm0.5$)\,K, than the non-AME regions ($17.7\pm0.7$)\,K. The Perseus AME clouds are within the range of temperatures for AME sources in our sample. Second, there is a trend of the AME amplitude with increasing dust temperature, which is related to the interstellar radiation field (ISRF) $G_0$. This will be discussed in Sect.~\ref{sec:isrf_role}.

In terms of the dust emissivity index, there is no strong correlation, except that the AME regions tend to have a higher value of $\beta_{\rm d}$ that the non-AME regions; AME regions have $\bar{\beta_{\rm d}}=1.80\pm0.02$, while the non-AME regions have $\bar{\beta_{\rm d}}=1.72\pm0.02$. The non-AME regions are dominated by bright \hii\ regions, and the emissivity index range is similar to those found by \cite{Dupac2003} for clouds with relatively cool temperature ($T=11$--$20$\,K). For the ionized gas associated with pure \hii\ regions, there should be less AME \citep{planck2011-7.3}. 

If AME were due to PAHs and small grains, we would naively expect a tighter correlation of AME with the emission from the shorter wavelengths of \IRAS; the $12\micron$  and $25\micron$ bands trace smaller dust grains; however, \hii\ regions are known to be depleted of PAHs and the smallest grains, as traced by their $12\micron/25\micron$ ratios \citep{Boulanger1988,Chan1995,Povich2007}. Given that spinning dust requires small dust grains, one might expect a separation of AME and non-AME regions via their \IRAS\ ratios. 

\begin{figure}[tb]
\begin{center}
\includegraphics[scale=0.5]{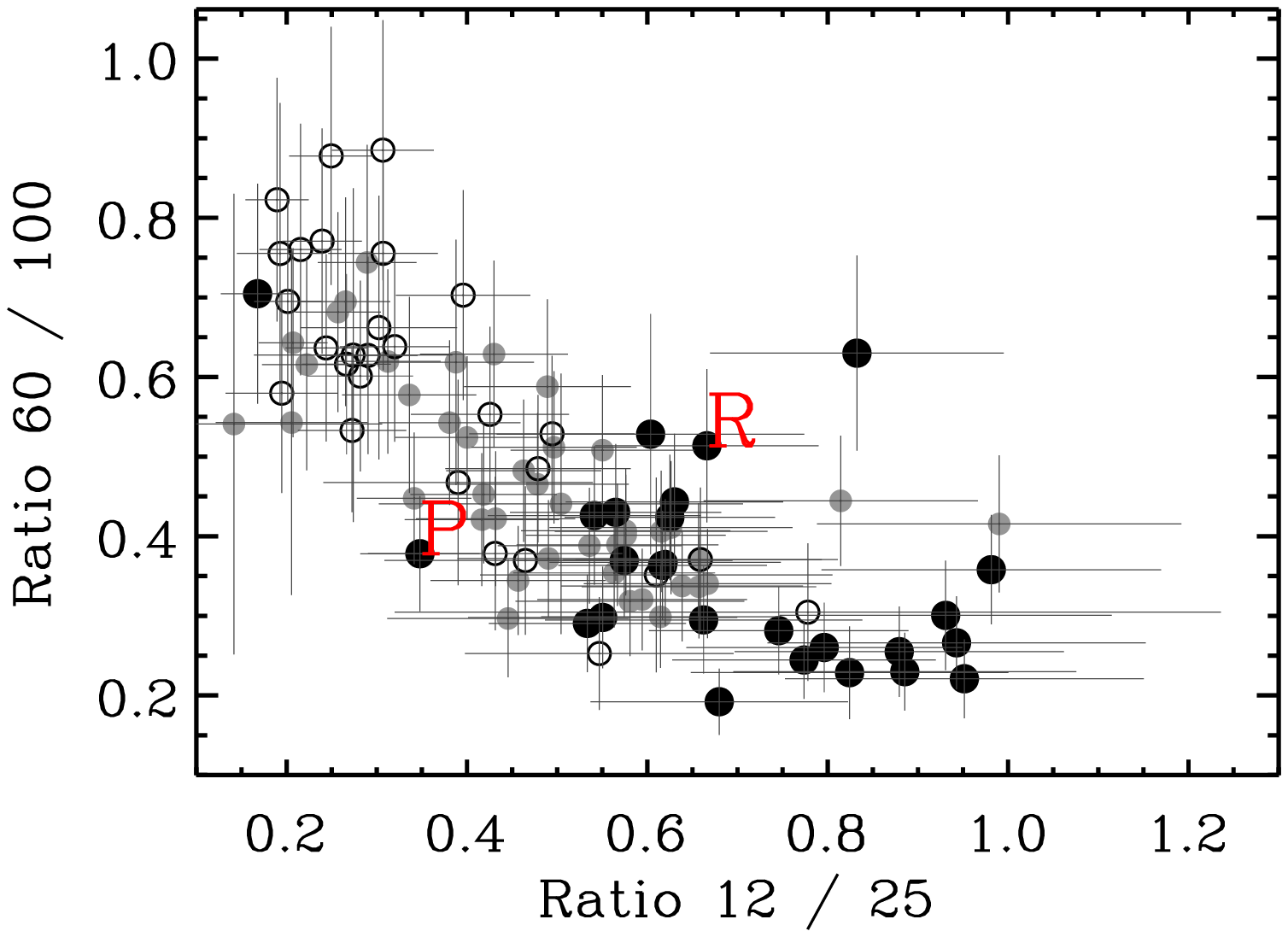}
\caption{Colour-colour plot of \IRAS\ $60\micron/100\micron$ vs.~\IRAS\ $12\micron/25\micron$. Symbols are as in Fig.~\ref{fig:uchii_max2cm}. The Perseus (P) and $\rho$~Ophiuchi (R) clouds are marked.}
\label{fig:iras_colours}
\end{center}
\end{figure}

Previous studies have found that there is an apparent sequence in the \IRAS\ colours given by the $12\micron/25\micron$ and $60\micron/100\micron$ ratios. This has been seen in \hii\ regions, ISM clouds \citep{Chan1995,Boulanger1988}, and external galaxies \citep{Helou1986}, where large $60\micron/100\micron$ ratio means smaller $12\micron/25\micron$ value and vice versa, i.e., an anti-correlation. The explanation is closely connected to the ISRF and the exciting star(s), since hotter grains have higher $60\micron/100\micron$ ratios, while close to the star the smallest grains (and PAHs) are destroyed, leading to smaller $12\micron/25\micron$ ratios. We see the same trend in our sample as a whole, as shown in Fig.~\ref{fig:iras_colours}. This is expected, since a large fraction of our sample contains a component from \hii\ regions.

It can also be seen that the AME sources generally have a higher $12\micron/25\micron$ ratio ($\approx\! 0.6$--1.0) than the rest of the sample ($\approx\! 0.2$--0.6). We interpret this as confirmation of the spinning dust model, where the very smallest grains (and PAHs) must be responsible for the bulk of the AME at frequencies $\gtrsim 20$\,GHz. At this resolution, we cannot rule out ionic/stellar contamination of the $12\micron$ \IRAS\ band, although the stellar contribution is expected to be minor. Previous studies \citep{Sellgren1985,Boulanger1988b} have shown that in bright nebulae, the bulk of the $12\micron$ emission comes from the PAH bands at 7.7, 8.8, and 11.3\micron; however, inspection of the sources with the highest $12\micron/25\micron$ ratios ($>0.8$) reveals that these are not the brightest free-free emitters of our sample but are in fact some of the most AME-dominated sources. These therefore do not appear to be dominated by ionized gas from \hii\ regions, which suggests that ionic/stellar contamination is not a prominent effect here. One exception to this is the source G239.40$-$04.70, which contains the luminous supergiant V$^{*}$~VY Canis Majoris, explaining the a $12\micron/25\micron$ ratio of 1.62; this source should be excluded from any mid-infrared analysis (see Sect.~\ref{sec:dust_correlations}).

\subsubsection{The interstellar radiation field: $G_0$}
\label{sec:G0}

An important parameter in studies of ISM clouds is the relative strength of the ISRF, $G_0$\footnote{$G_0$ is a common scaling used for measuring the ISRF, and is integrated between 5 and 13.6\,eV \citep{Mathis1983}. A standard value of $G_0=1$ is representative of the diffuse ISM (away from hot stars), and corresponds to $1.2\times10^{-3}$\, erg\,s$^{-1}$\,cm$^{-2}$.}. This parameter is important for spinning dust, since the ISRF plays a key role in rotationally exciting small grains, charging PAHs, and in the destruction of the smallest grains. While we assume the same form for the shape of the ISRF spectrum here, it should be noted that the intensity {\it and} hardness of the ISRF are important. 

An estimate of $G_0$ can be obtained from the equilibrium dust temperature of the big grains ($T_{\rm BG}$) compared to the average value of 17.5\,K, i.e., $G_0= (T_{\rm BG}/17.5\,{\rm K})^{(4+\beta_{\rm d})}$. We used the fitted dust temperature from our SEDs as a proxy for the big grain temperature, i.e., $T_{\rm BG} \approx\! T_{\rm d}$. We kept the dust emissivity index constant at $\beta_{\rm d}=2$ to protect against mixing of multiple dust components with a range of dust temperatures that could flatten the index. This allows an estimate of the average $G_0$ for each source (Table~\ref{tab:main_list}). As with other FIR properties, we note that in some regions these properties may not be related to the exact environment where the AME is arising. 

\begin{figure}[tb]
\begin{center}
\includegraphics[scale=0.5]{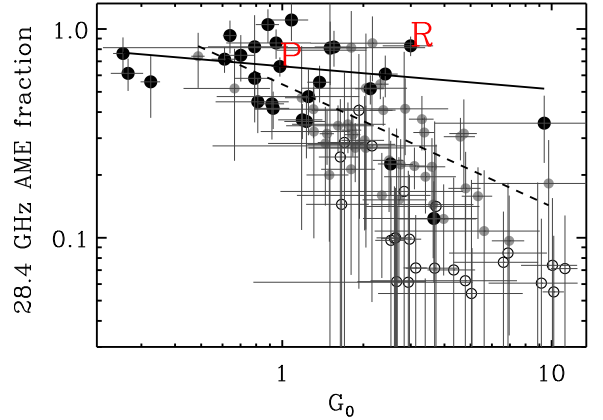}
\caption{AME fraction at 28.4\,GHz as a function of estimated $G_0$. The symbols are as in Fig.~\ref{fig:uchii_max2cm}. The Perseus (P) and $\rho$~Ophiuchi (R) clouds are indicated. The best-fitting power laws to the AME regions (solid line) and semi-significant AME regions (dashed line) are overplotted.}
\label{fig:amefraction_G0}
\end{center}
\end{figure}

Figure~\ref{fig:amefraction_G0} shows the AME fraction at 28.4\,GHz, defined as the AME residual flux density ($S_{\rm resid}^{28.4}$) divided by the total flux density at 28.4\,GHz, against the estimate of $G_0$. The most striking feature is that sources with low $G_0$ are those with the highest AME fractions; the majority of strong AME regions are at $G_0<4$, while non-AME regions are almost always at $G_0>1$. There is a general trend of decreasing AME fraction with $G_0$. When including semi-significant AME sources, there is also a definite trend of decreasing AME fraction with increasing $G_0$.  The best-fitting power law has a slope of $\gamma=-0.59\pm0.11$; however, taking only the strongest AME sources yields a slope of $\gamma=-0.11\pm0.04$. This suggests that there is a much flatter slope for AME strong sources, and that the AME sources are different in nature to the rest of the sources in the sample.

We caution that this trend could be entirely due to selection effects, because AME is more difficult to detect when there is strong free-free emission. Also, the thermal dust could be coming from a different region of space compared to the AME, and thus the relation between $G_0$ and AME might be entirely coincidental. Despite that, the effective free-free EM from our photometry and $G_0$ are positively correlated as one would expect. The role of the ISRF will be discussed further in Sect.~\ref{sec:isrf_role}.

\subsubsection{Peak frequency of spinning dust}
\label{sec:newspdustpeak}

As discussed in Sect.~\ref{sec:model_fitting}, as well as fitting for a spinning dust amplitude, we also fit for a possible shift in the spinning dust peak frequency. This allows us to determine the peak frequency $\nu_{\rm sp}$ for each SED, which is listed in Table~\ref{tab:main_list} for sources with $\sigma_{\rm AME}>2$. Visual inspection of the SEDs indicates that allowing this freedom provides a better fit to the data, where a few sources are clearly not peaking at the starting value of the fitted curve, at 28.1\,GHz. 

\begin{figure}[tb]
\begin{center}
\includegraphics[scale=0.5]{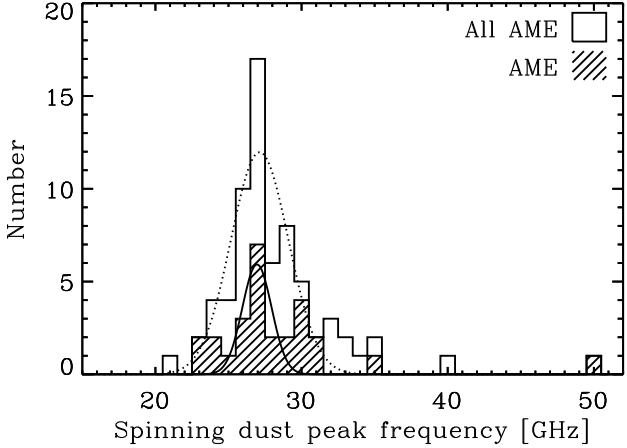}
\caption{Histogram of spinning dust peak frequencies, $\nu_{\rm sp}$. The unfilled histogram is for both AME and semi-significant AME regions, while the hatched region is for the AME sources only. The best-fitting Gaussians for these two histograms are overplotted. The source at $\nu_{\rm sp}= 50$\,GHz is G160.60$-$12.05 (the California Nebula).}
\label{fig:newspdustpeak}
\end{center}
\end{figure}

\begin{figure}[tb]
\begin{center}
\includegraphics[scale=0.5]{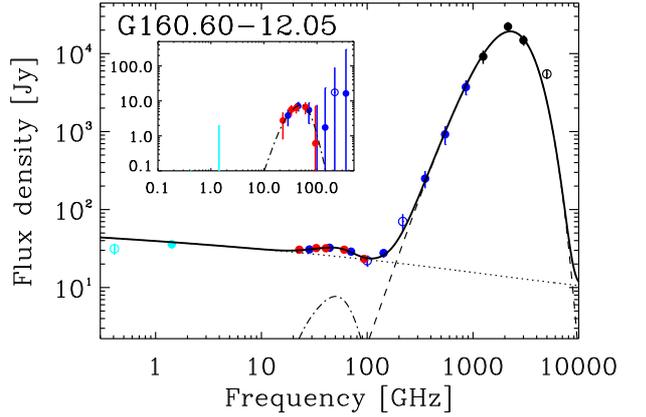}
\caption{SED for G160.60$-$12.05 (the California Nebula). The curves and symbols are as in Fig.~\ref{fig:ame_seds}. AME is detected at a level of $\sigma_{\rm AME}=5.0$, and the best-fitting spinning dust peak is shifted to $\nu_{\rm sp}=(50\pm17)$\,GHz.}
\label{fig:california}
\end{center}
\end{figure}

Figure~\ref{fig:newspdustpeak} shows the histogram of spinning dust peak frequencies. The AME sources peak in the range 20--35\,GHz, with a weighted mean of 27.9\,GHz. Although the formal uncertainties do not suggest that these are highly statistically significant, it is clear that the data sometimes prefer spinning dust that is at a different frequency from the starting value 28.1\,GHz. Examples of small shifts can be seen by careful examination of the SEDs in Fig.~\ref{fig:ame_seds}. A particularly clear example, previously commented on by \cite{planck2011-7.2}, is G160.60$-$12.05 (the California Nebula/NGC1499). The SED is plotted in Fig.~\ref{fig:california}. The spinning dust detection is at a level of $\sigma_{\rm AME}=5.1$, and the SED shows a peak at a higher frequency compared to the rest of the sample. The best-fitting peak frequency is $\nu_{\rm sp}= (50 \pm 17)$\,GHz. Notice that the reduced value is $\chi^2_r=0.33$, and thus the significance is likely underestimated. As discussed earlier, COSMOSOMAS data at 12--18\,GHz are not included here because the filtering required causes problems at frequencies above 100\,GHz, resulting in negative flux densities. This is due to a lack of dust emission in the centre of the nebula where most of the free-free emission comes from, and nearby dust emission appearing near a negative lobe caused by the COSMOSOMAS harmonic filtering; however, the shape of the spectrum at 12--18\,GHz is consistent with a free-free spectrum \citep{planck2011-7.2}.

The reality of this shift to higher frequency, and its possible explanation, are not clear and require a more detailed study. Despite that, the conditions in this diffuse \hii\ region may well be conducive to the production of higher frequency emission. The ISRF comes mostly from the ionizing star $\xi$~Per, where the density is $\approx\! 1$\,cm$^{-3}$, and goes up to $50$\,cm$^{-3}$ on the brightest filaments in the PDR \citep{Boulanger1988}, where the AME may be originating. Similarly, there is a range of temperatures, covering 20--50\,K. Our fitted parameters of $T_{\rm d}=(21.7\pm1.6)$\,K, $G_0=3.7\pm1.7$ are an average of these, and may well not be appropriate for conditions in the AME-emitting region, especially as much of the cooler dust emitting at $\approx\! 1000$\,GHz is coming from an adjacent dust cloud. Higher resolution observations of this cloud are needed to investigate this further.

\subsection{Dust correlations}
\label{sec:dust_correlations}

\begin{figure*}[tb]
\begin{center}
\includegraphics[width=0.33\textwidth]{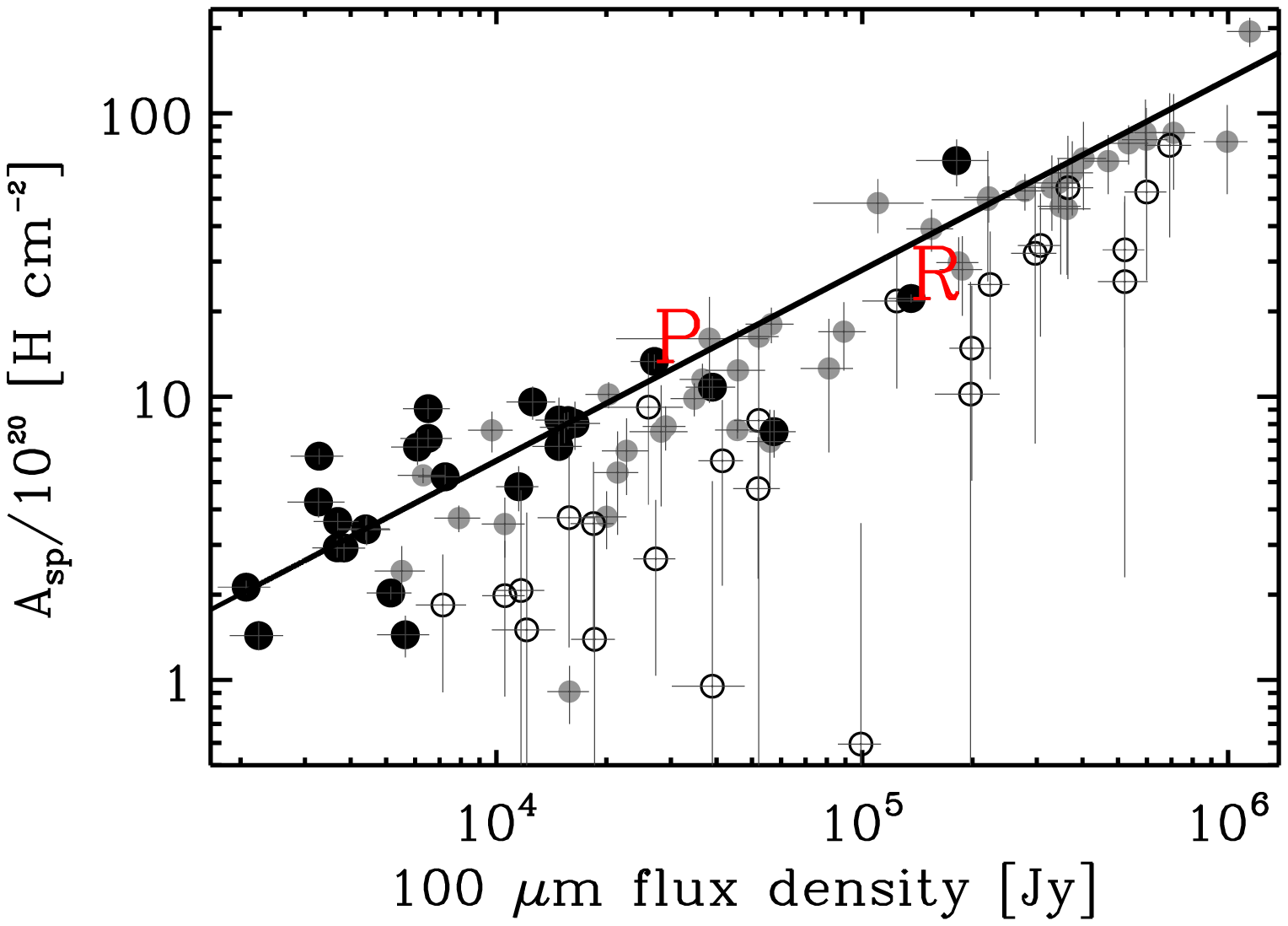}
\includegraphics[width=0.33\textwidth]{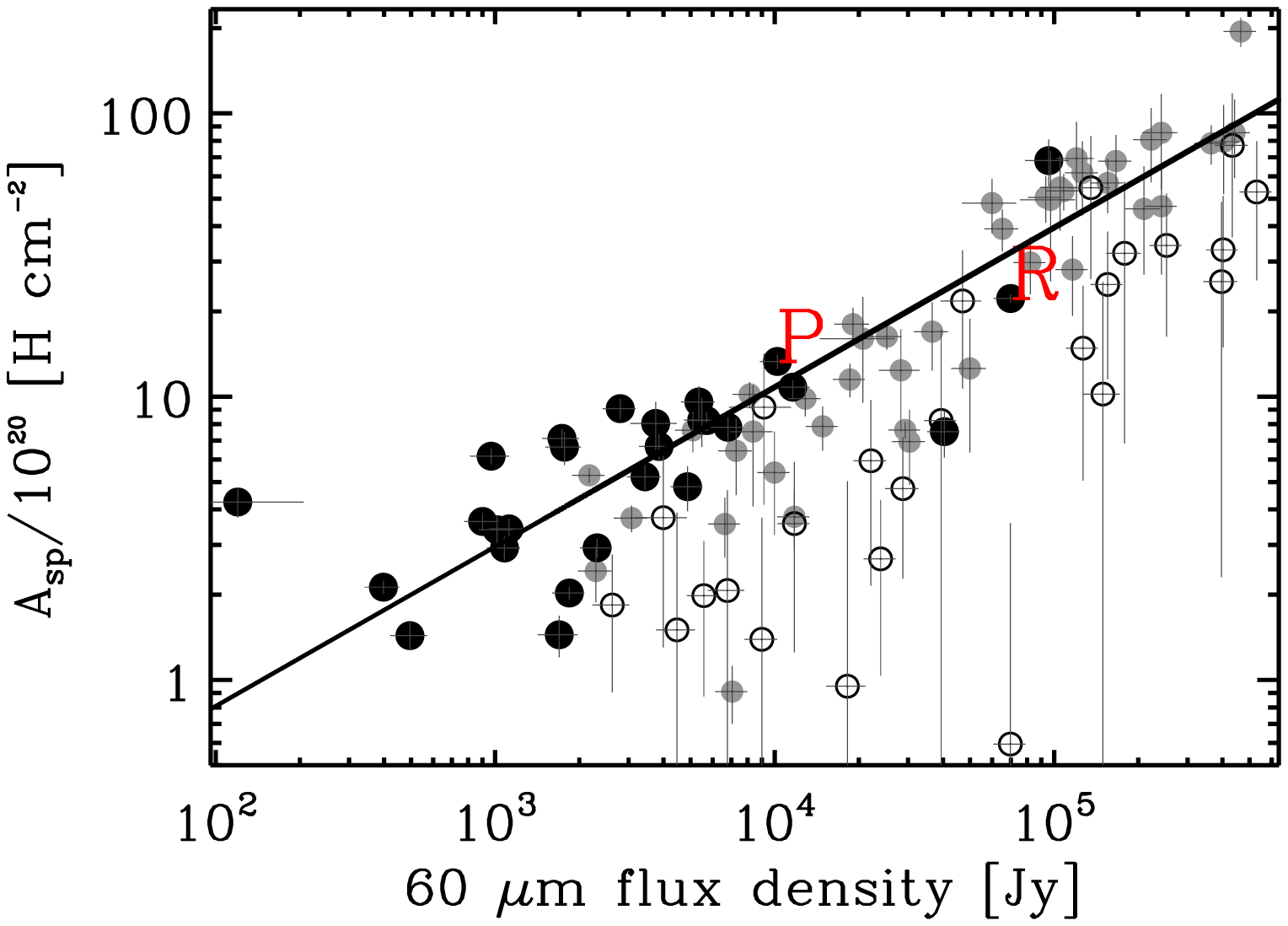}
\includegraphics[width=0.33\textwidth]{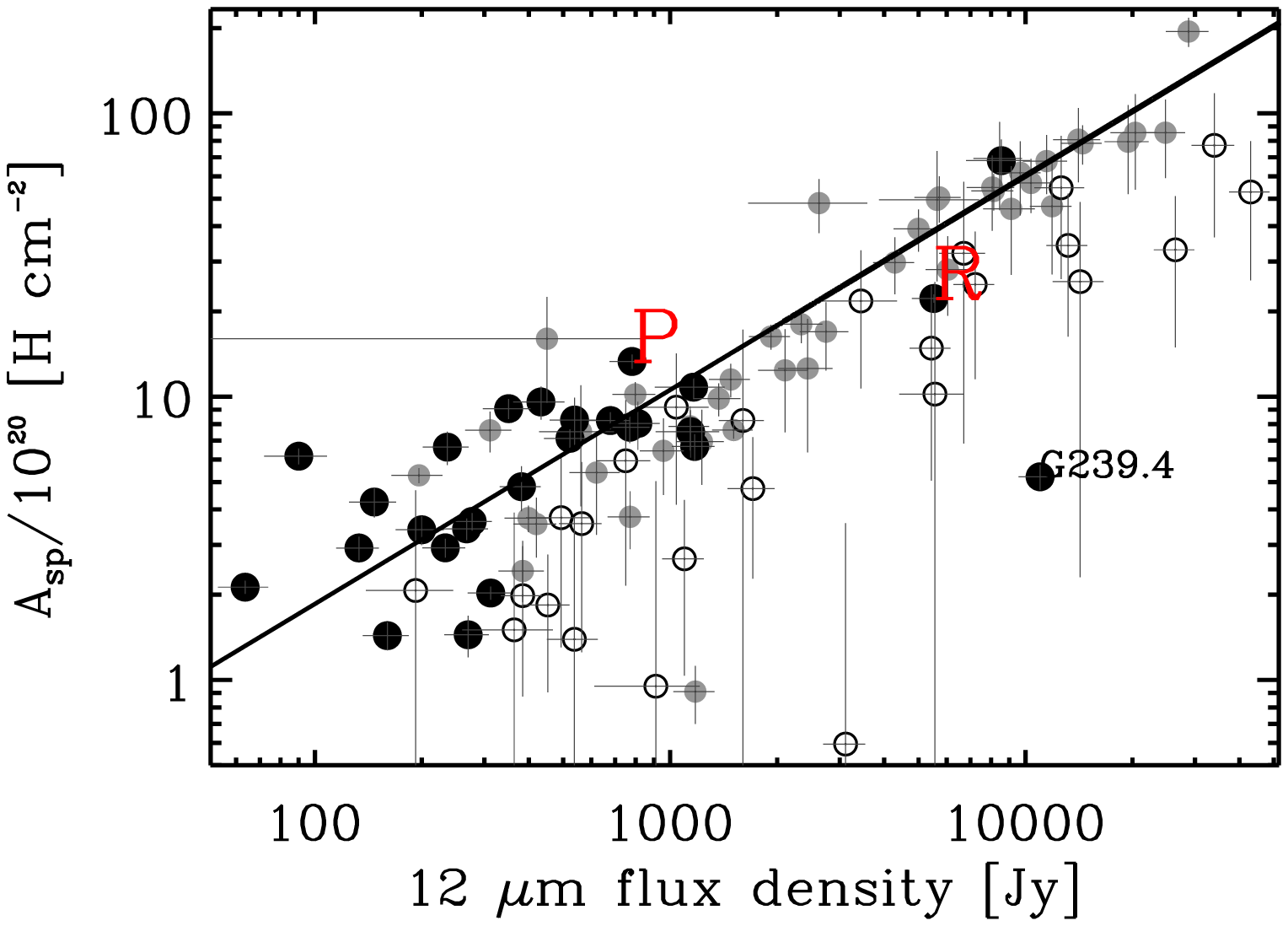}
\includegraphics[width=0.33\textwidth]{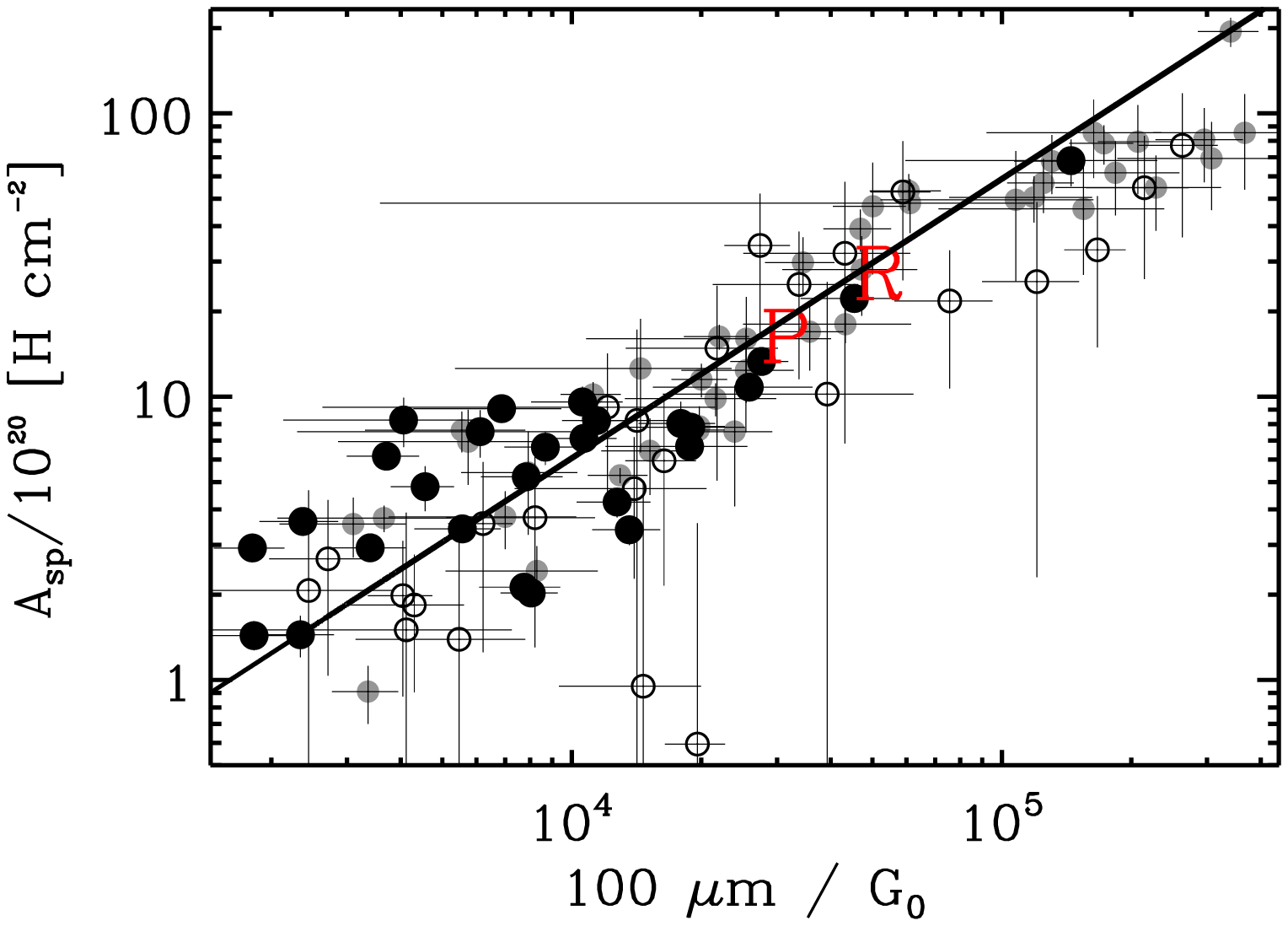}
\includegraphics[width=0.33\textwidth]{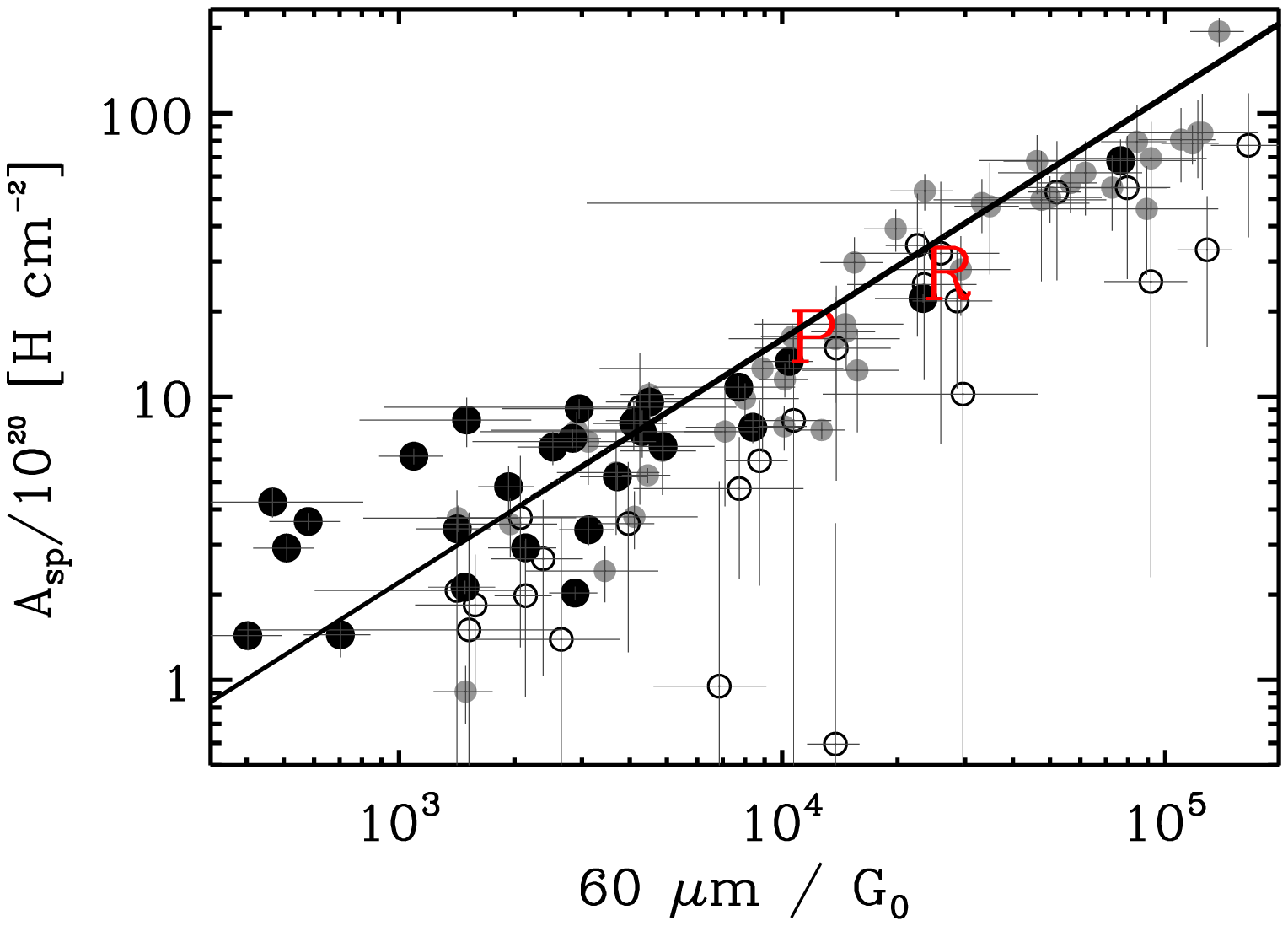}
\includegraphics[width=0.33\textwidth]{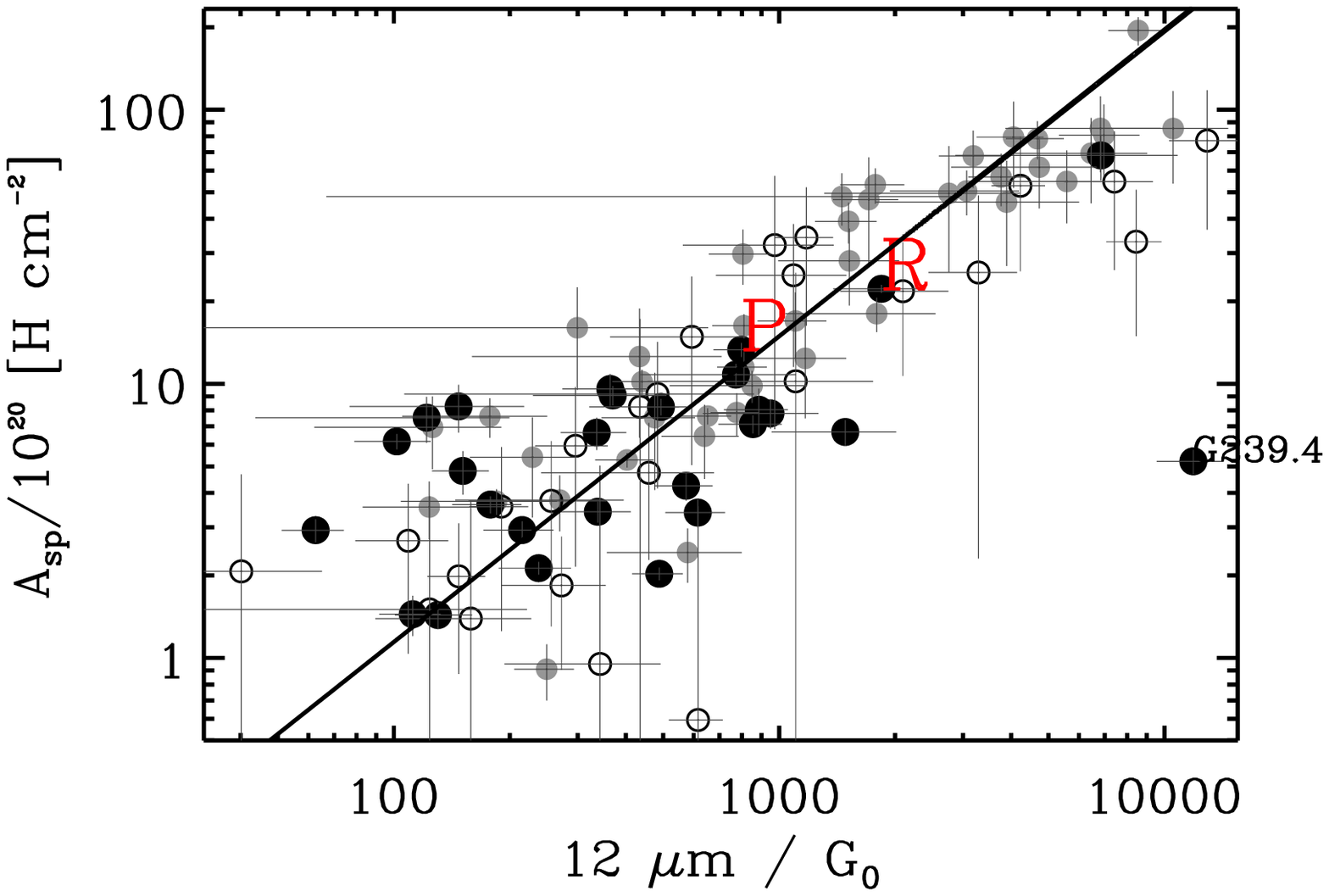}
\caption{{\it Top row}: AME amplitude ($A_{\rm sp}$) as a function of the $100\micron$ ({\it left}), $60\micron$ ({\it middle}), and $12\micron$ ({\it right}) flux density. {\it Bottom row}: The same as the top row, except that the FIR flux densities have been divided by $G_0$. Symbols are as in Fig.~\ref{fig:uchii_max2cm}. The best-fitting slope fitted to the significant AME regions only is plotted. The Perseus (P) and $\rho$~Ophiuchi (R) clouds are indicated. There is a strong correlation between the AME intensity and infrared tracers of dust, with an improved correlation after dividing by $G_0$. At $12\micron$ the fit is made excluding the outlier source G239.40$-$4.70 (marked).}
\label{fig:dust_correlations}
\end{center}
\end{figure*}

There should be a very strong correlation between the AME and the thermal dust emission, since the AME is thought to be due to spinning dust grains. Figure~\ref{fig:dust_correlations} shows the correlation of the AME amplitude ($A_{\rm sp}$) with the flux density at $100\micron$, $60\micron$, and $12\micron$ from the aperture photometry. A strong correlation is seen between the AME and IR/submillimetre brightness for the AME bright regions (black filled circles). The tight correlation also holds for the semi-significant AME regions (grey filled circles), while non-AME regions (unfilled circles) are weaker and more scattered, partly due to the intrinsic noise.

To quantify the strength of the correlations, we compute the Spearman rank correlation coefficient (in log space), $r_{\rm s}$, for the significant AME regions, which are listed in Table~\ref{tab:pearson}. This quantity, unlike the Pearson correlation coefficient, does not rely on a linear dependence between two variables (we obtained similar results using either statistic). Average uncertainties on $r_{\rm s}$ are calculated using 1000 Monte Carlo simulations based on their formal uncertainties. There is a relatively strong correlation for sources with significant AME ($r_{\rm s} \approx\! 0.8$), although non-AME sources are not less correlated. For the AME-bright regions, there is no strong preference for them to correlate better with any particular FIR band.

 \begin{table*}[tmb]
\begingroup
\newdimen\tblskip \tblskip=5pt
\caption{Spearman rank correlation coefficients between the AME amplitude ($A_{\rm sp}$) and IR/submillimetre flux densities. The correlation for AME regions is much stronger. In parentheses is the correlation coefficient after dividing the IR/submillimetre flux density by $G_0$ to account for the variation in the interstellar radiation field. Generally, the correlation is tighter after dividing by $G_0$. Note that the we have omitted the outlier source G239.40$-$04.70 at $12\micron$ and $25\micron$ (see text).}
\label{tab:pearson}
\nointerlineskip
\vskip -3mm
\footnotesize
\setbox\tablebox=\vbox{
 \newdimen\digitwidth 
 \setbox0=\hbox{\rm 0} 
 \digitwidth=\wd0 
 \catcode`*=\active 
 \def*{\kern\digitwidth}
 \newdimen\signwidth 
 \setbox0=\hbox{+} 
 \signwidth=\wd0 
 \catcode`!=\active 
 \def!{\kern\signwidth}
 \halign{\hbox to 0.9in{#\leaderfil}\tabskip 2.2em&
    \hfil#\hfil&
    \hfil#\hfil&
    \hfil#\hfil\tabskip 0pt\cr
\noalign{\doubleline\vskip 2pt}
\omit&\multispan3\hfil Spearman rank correlation coefficient, $r_{\rm s}$\hfil\cr
\noalign{\vskip -3pt}
\omit&\multispan3\hrulefill\cr
\omit\hfil Wavelength\hfil&All&AME&Non-AME   \cr
\noalign{\vskip 4pt\hrule\vskip 6pt}
100\micron&$0.85\pm0.02$ ~~($0.90\pm 0.02$)&$0.83\pm 0.03$ ~~($0.68\pm 0.06$)&$0.84\pm 0.07$ ~~($0.81\pm 0.08$)\cr
*60\micron&$0.81\pm 0.03$ ~~($0.89\pm 0.02$)&$0.80\pm 0.03$ ~~($0.78\pm 0.05$)&$0.81\pm 0.07$ ~~($0.82\pm 0.08$)\cr
*25\micron&$0.82\pm 0.03$ ~~($0.89\pm 0.03$)&$0.84\pm 0.03$ ~~($0.74\pm 0.05$)&$0.82\pm 0.07$ ~~($0.85\pm 0.08$)\cr
*12\micron&$0.84\pm 0.03$ ~~($0.86\pm 0.03$)&$0.78\pm 0.03$ ~~($0.58\pm 0.06$)&$0.85\pm 0.07$ ~~($0.83\pm 0.08$)\cr
\noalign{\vskip 3pt\hrule\vskip 4pt}
}}
\endPlancktablewide
\endgroup
\end{table*}

If the AME is due to the smallest dust grains emitting electric dipole radiation, the best correlation should be with the shorter \IRAS\ wavelengths at $\lambda < 60\micron$. This has been observed in previous studies at higher angular resolution (e.g., \citealt{Casassus2006,Ysard2010b}). We do not find such a clear trend, with the worst correlation occurring for the \IRAS\ $12\micron$ band. Figure~\ref{fig:dust_correlations} shows the correlation of the AME amplitude against the $12\micron$ flux density. There is a reasonably strong correlation but there is also considerable scatter.

It has been discussed in several works (e.g., \citealt{Ali-Hamoud2009,Ysard2010b}) that spinning dust emissivity (per column density) at frequencies $10$--$30$\,GHz is not particularly sensitive to the ISRF intensity, $G_0$. On the other hand, the thermal dust emission is strongly dependent on $G_{0}$, since it is the UV radiation that governs the dust grain temperature.  The nanoscale dust particles (PAHs and VSGs) are proportional to $G_0$ \citep{Sellgren1985} when $G_0 \lesssim 10$. Previous studies (e.g., \citealt{Ysard2010b,Vidal2011}) have found that a better correlation can be obtained with the infrared by dividing far-infrared flux density by $G_0$.

The Spearman rank correlation coefficients for all infrared bands divided by $G_0$ are given in Table~\ref{tab:pearson} in parentheses. In general, we do not see a significant increase in the correlation after dividing by $G_0$. There is a general increase within the sample as a whole ($r_{\rm s} \approx 0.9$), but no improvement for the AME sample.
 
We suspect that some of the correlation scatter with the \IRAS\ bands, particularly at $12\micron$ and $25\micron$, is due to contamination from stellar emission or from fine-structure lines. \cite{Shipman1996} predicted that for bright \hii\ regions, line contamination could account for most of the flux densities at $12\micron$ and might even affect the longer \IRAS\ wavelengths. \cite{Sellgren1985} found that the $12\micron$ band flux densities in bright reflection nebulae could be explained by approximately equal amounts of continuum and line emission from PAHs. Infrared spectra of the Omega nebula (M17) showed that PAHs dominate the mid-infrared in the neutral PDR beyond the ionized gas of \hii\ regions \citep{Povich2007}. Also, for many regions in our sample, we are looking at a complex integration of environments including molecular clouds, CNM, and WIM. On the other hand, spinning dust predictions are local (i.e., for a particular environment). 

Evidence of a particular example of contamination can be seen in Fig.~\ref{fig:dust_correlations}, where there is a source (G239.40$-$04.70) that has a much higher $12\micron$ flux density than the rest of the sample. This is due to one of the most luminous supergiants known, V$^{*}$ VY Canis Majoris (CMa), which has a $12\micron$ flux density of $10\,000$\,Jy \citep{Helou1988} and accounts for the bulk of the flux density in this band. We therefore omit this source for the correlation values (Table~\ref{tab:pearson}) and power-law fits (Fig.~\ref{fig:dust_correlations}) at $12\micron$ and $25\micron$. Although this is an extreme object, this clear case of contamination does question the robustness of the \IRAS\ $12\micron$ flux density, and partly the $25\micron$ band, for low latitude sight-lines.

We also investigated the ratio of $8\micron/24\micron$ using the {\it Spitzer} data. This ratio is diagnostic of the PAH fraction compared to that of the small grains. Only 24 sources had coverage and of these only three are AME regions. This makes it difficult to distinguish any trends between AME or non-AME regions. We find no distinction in the $8\micron/24\micron$ ratio for AME vs non-AME regions; however, we note that $\rho$~Ophiuchi has one of the highest ratios, $1.09 \pm 0.15$, compared to the sample weighted average of $0.28\pm0.01$. One particular outlier is G023.47+08.19, which has a much higher ratio of $10.6\pm1.5$ due to the source being very faint in the {\it Spitzer} maps.

\subsection{AME emissivity}
\label{sec:ame_emissivity}

Next, we would like to compare the AME amplitude to values in the literature. A common way of normalizing the emissivity is to take the ratio of AME flux density to $100\micron$ (3000\,GHz) flux density. Since the far-infrared emission is optically thin for a given dust temperature and composition, this will be proportional to the column density of dust along the line of sight. If AME is due to spinning dust emission, we would therefore expect a strong correlation with the $100\micron$ brightness (Fig.~\ref{fig:dust_correlations}). 

\begin{figure}[tb]
\begin{center}
\includegraphics[width=88mm]{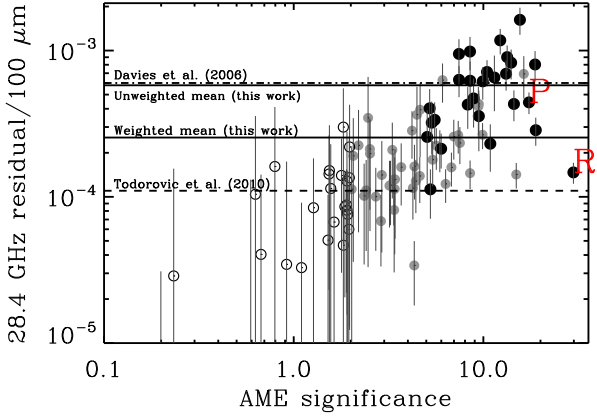}
\caption{Ratio of 28.4\,GHz residual to the $100\micron$ flux density as a function of the AME significance. The symbols are as in Fig.~\ref{fig:uchii_max2cm}. The Perseus (P) and $\rho$~Ophiuchi (R) clouds are indicated. The weighted and unweighted average ratios for the significant AME regions are shown as the solid horizontal lines. The average ratio from \cite{Todorovic2010} (dashed line) and \cite{Davies2006} (dot-dashed line) are also shown.}
\label{fig:s30residual_100micron}
\end{center}
\end{figure}

We begin by choosing the 28.4\,GHz AME flux density, defined as the residual at 28.4\,GHz after subtracting the non-AME components for each source. This allows a comparison with previous works, where frequencies near $30\,$GHz have been used extensively. Note that using the 28.4\,GHz AME residual is almost identical to using the AME amplitude ($A_{\rm sp}$) directly from the fit, since they are highly correlated. Figure~\ref{fig:s30residual_100micron} shows the ratio of 28.4\,GHz residual flux density to the $100\micron$ (3000\,GHz) flux density as a function of the AME significance, $\sigma_{\rm AME}$. The ratio $S_{\rm resid}^{\rm 28.4\,GHz}/S_{100}$ for the AME regions has a large range of values covering the range (1--15)$\times 10^{-4}$, with a weighted average of $(2.5\pm 0.2) \times 10^{-4}$. This is higher than in the \hii\ regions of \cite{Todorovic2010}, but less than the average from high latitude AME \citep{Davies2006}, as shown in Fig.~\ref{fig:s30residual_100micron}. It can be seen that the weighted average is actually lower than many of the AME regions, due to a few sources having a relatively small uncertainty and lower emissivity values of $(1$--$2) \times 10^{-4}$, including $\rho$~Ophiuchi (G353.05+16.90) at $(1.5 \pm 0.2) \times 10^{4}$. The unweighted average is $(5.8 \pm 0.7) \times 10^{-4}$, which is consistent with the $6.2 \times 10^{-4}$ value of \cite{Davies2006} for the diffuse high latitude AME.

The best-fitting power law to the $S_{100}$ vs.~AME amplitude ($A_{\rm sp}$) yields a power-law index of $+0.67 \pm 0.03$ for AME regions, suggesting that there is not a simple, one-to-one ratio between the AME brightness and $100\micron$ brightness. This is not surprising, given that different dust temperatures will affect a given frequency, particularly near the peak of the thermal dust spectrum where it is very sensitive on the dust temperature \citep{Tibbs2012b}. Instead, as suggested by \cite{Finkbeiner2004}, the thermal dust optical depth is expected to be a better diagnostic of the AME emissivity, since it is proportional to the column density of dust via $\tau_\nu = N_{\rm H} \kappa_\nu \mu m_{\rm H}$, where $N_{\rm H}$ is the column density, $\kappa$ is the dust mass absorption coefficient, $\mu$ is the mean molecular weight, and $m_{\rm H}$ is the mass of hydrogen \citep{Boulanger1996,Martin2012}. The fit against the dust optical depth at a wavelength of $250\micron$ ($\tau_{250}$), which is proportional to $N_{\rm H}$, yields a slope much closer to unity ($+1.03\pm 0.03$). This suggests that the AME is emitting at approximately the same level per unit column density in all the bright AME regions chosen in our sample. \cite{planck2011-7.3} also found that AME emits at the same level per $N_{\rm H}$ throughout the Galactic plane. We therefore define our default AME emissivity as the AME amplitude divided by the thermal dust optical depth, $A_{\rm sp}/\tau_{250}$. In this case, the units are formally $10^{20}$\,cm$^{-2}$, but we note that the interpretation of this value depends on the spinning dust model that is fitted. The actual dust column density calculated from the optical depth depends on the range of dust temperatures along the line-of-sight \citep{Ysard2012}, as well as details of the dust grains including the dust opacity (which in turn depends on many factors such as dust composition).

\begin{figure}[tb]
\begin{center}
\includegraphics[scale=0.5]{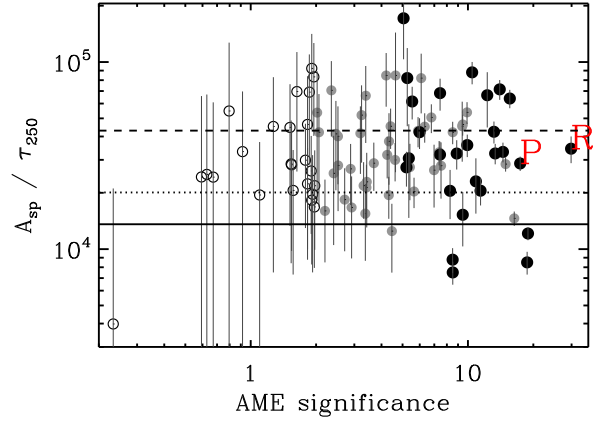}
\caption{Ratio of AME amplitude ($A_{\rm sp}$) to dust optical depth ($\tau_{250}$), as a function of the AME significance $\sigma_{\rm AME}$. The symbols are as in Fig.~\ref{fig:uchii_max2cm}. The Perseus (P) and $\rho$~Ophiuchi (R) cloudss are indicated. The horizontal lines represent averages: weighted average of significant AME regions (solid line), unweighted average of significant AME regions (dashed line) and weighted average of non-significant AME regions (dotted line).}
\label{fig:sdnh_tau}
\end{center}
\end{figure}

Figure~\ref{fig:sdnh_tau} plots the AME emissivity $A_{\rm sp}/\tau_{250}$ against the significance, $\sigma_{\rm AME}$. There is considerable scatter in the emissivity, some of which is due to the intrinsic uncertainty in the measurement; however, for AME sources there is more variation than can be accounted for by the uncertainty alone. This will be investigated via correlations with other parameters. Remarkably, unlike the equivalent plot for the $100\micron$-based emissivity, there is no clear trend with AME significance.  This is an important result.  It shows that AME emissivities based on $100\micron$ data are likely to be biased due to the effect of varying  dust temperature. This plot also suggests that AME is emitting at approximately the same level per unit dust column density, not only for the strong AME sources, but for most of the sources within our sample.  However, we caution that the uncertainties are large and including an AME component will inevitably result in a positive bias. Furthermore, some sources have a possible contribution from UC\hii\ regions as well (Table~\ref{tab:main_list}). The weighted average emissivity for significant AME sources is $(1.36\pm0.05) \times 10^4$, while the non-AME sources are at $(2.0\pm0.5)\times 10^4$. 

It is worth noting that for the AME regions the weighted average is smaller than the unweighted average of $(4.3 \pm 0.6) \times 10^{4}$, due to four high significance sources that appear to have a lower emissivity ($A_{\rm sp}/\tau_{250} <10^4$) than the rest of the sample. These are G005.40+36.50, G023.47+08.19, G133.27+09.05 and G219.18$-$08.93.  All are associated with dark nebulae (see Table~\ref{tab:main_list} and Sect.~\ref{sec:discussion_nature}). These sources are high latitude sources with very weak free-free emission and a low intensity radiation field ($G_0=0.3$--$0.6$), leading to cold big grain dust temperatures ($T_{\rm d}= 14$--16\,K).

\begin{figure}[tb]
\begin{center}
\includegraphics[scale=0.5]{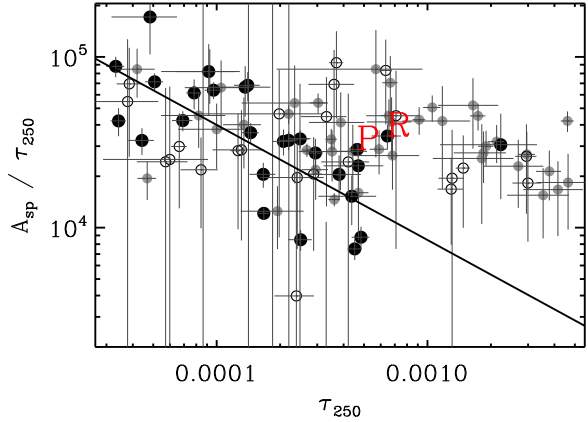}
\caption{Ratio of AME amplitude ($A_{\rm sp}$) to dust optical depth ($\tau_{250}$), as a function of $\tau_{250}$. The symbols are as in Fig.~\ref{fig:uchii_max2cm}. The Perseus (P) and $\rho$~Ophiuchi (R) clouds are indicated. The solid line shows the best-fitting power law for the significant AME regions.}
\label{fig:sdnh_tau_tau}
\end{center}
\end{figure}

We have investigated the correlation of AME emissivity with a wide range of parameters and properties in this study. In general, we find little or no correlation with most parameters. Previous works \citep{Lagache2003,Vidal2011} found evidence of an anti-correlation of the AME emissivity with the column density. The interpretation is that this is what would be expected with AME arising from small spinning dust grains; as the density ($n_{\rm H}$) increases, dust grain growth becomes more efficient and thus the small grain population decreases. Figure~\ref{fig:sdnh_tau_tau} shows the AME emissivity against the optical depth, $\tau_{250}$. For the significant AME regions, we also find a similar trend. The anti-correlation ($r_{\rm s}=-0.72 \pm 0.07$) has a best-fitting power-law slope of $\gamma=-0.67\pm0.04$, similar to the $-0.54 \pm 0.10$ value found by \cite{Vidal2011}.

The effect we see is somewhat surprising given the difficulty of estimating density in the potential complex mix of environments in a $1\deg$ beam, especially at low latitudes where there may be multiple objects along the line-of-sight. Nevertheless, this is one of the few observational trends associated with AME that has come to light in several independent studies.  It is consistent with small grains as AME carriers, because they are expected to be depleted in denser environments by coagulation onto larger grains.

\subsection{The role of the ISRF}
\label{sec:isrf_role}

The ISRF plays an important role in the ISM. It is a major source of local heating, with UV photons being absorbed by dust grains and re-radiated in the IR. It also has important effects on the composition of the ISM, since it can charge PAHs and also destroy PAHs and VSGs when the intensity is high enough. For spinning dust, the population of PAHs and VSGs is fundamental in determining the AME, and the formation of ions (e.g., C$^{+}$, H$^{+}$) can be important for rotationally exciting small dust grains. The ISRF also plays a role in rotationally exciting small dust particles through the absorption of infrared photons; these particles can be dominant in some environments (e.g., when $G_0 =1$ and $n_{\rm H} < 10$\,cm$^{-3}$). 

\begin{figure}[tb]
\begin{center}
\includegraphics[scale=0.5]{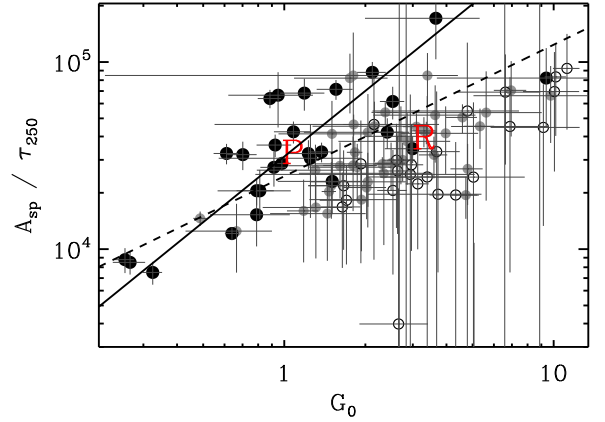}
\caption{AME emissivity as a function of $G_0$. The symbols are as in Fig.~\ref{fig:uchii_max2cm}. The straight lines are the best-fitting power laws to the AME (solid line) and including semi-significant AME (dashed line) regions.}
\label{fig:G0_AME}
\end{center}
\end{figure}

In the recent studies by \cite{Tibbs2011,Tibbs2012}, the AME brightness was found to correlate with $G_0$ more than any other parameter, including the PAH fraction. Figure~\ref{fig:G0_AME} shows the AME emissivity as a function of $G_0$ for our sample. We also find that the AME emissivity is related to $G_0$. For the AME regions there is a positive correlation ($r_{\rm s}=0.63 \pm 0.07$) with a fitted power-law index $\gamma=+0.74\pm0.10$. We verified that the correlation is due primarily to variations in $T_{\rm d}$, and not $\beta_{\rm d}$, by setting $\beta_{\rm d}$ to the fitted values.  We found a similar positive trend but with larger scatter ($r_{\rm s}=0.49 \pm 0.12$ and $\gamma=+1.32 \pm 0.10$). Surprisingly, for semi-significant regions (grey filled circles), the same trend is also visible, but with a flatter slope of $\gamma=+0.70\pm0.04$, and the non-AME sources are also correlated. This may suggest that there is AME in many of our sources or that there is a systematic effect that is driving this result. As before, we only consider the AME sources with $\sigma_{\rm AME}>5$ and $f^{\rm UC\hii}_{\rm max}<0.25$ to be reliable. Given the larger scatter, at this point we can only say that there is a positive correlation with a slope $\gamma \approx +1$; we will adopt this value in Sect.~\ref{sec:modelling}.

We may not be seeing AME at high $G_0$ (and high $T_{\rm d}$) because the PAHs/VSGs require radiation to spin, but too much UV destroys them, as happens inside very bright \hii\ regions (with higher EM, $T_{\rm d}$ and $G_0$). The grains need to be shielded, hence to be close to the PDR of the \hii\ regions (which is the coldest part of an \hii\ region). It could also be a selection effect due to brighter free-free emission associated with high values of $G_0$ that would make AME more difficult to detect.

\begin{figure}[tb]
\begin{center}
\includegraphics[scale=0.5]{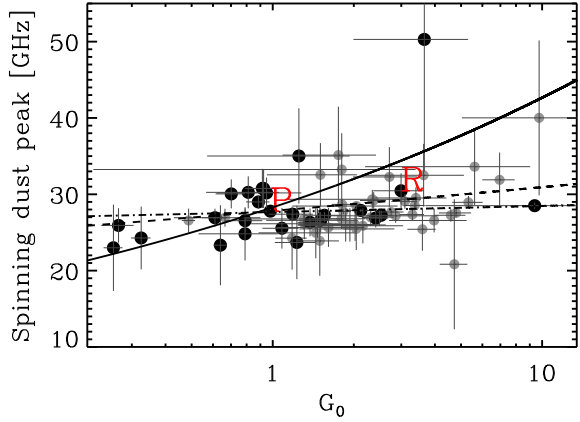}
\caption{Spinning dust peak frequency as a function of $G_0$. The symbols are as in Fig.~\ref{fig:uchii_max2cm}. The solid line is the best-fitting power law for the AME regions with all data points equally weighted. The dashed line is the weighted fit to the AME regions, while the dot-dashed line is the weighted fit to AME and semi-significant AME regions.}
\label{fig:newspdustpeak_G0}
\end{center}
\end{figure}

In some environments (e.g., low density), it is expected that as the ISRF increases (up to a certain point), the smaller grains will receive more rotational excitation, resulting in a spinning dust spectrum shifting to higher frequencies (e.g., \citealt{Ysard2011}). Figure~\ref{fig:newspdustpeak_G0} plots the spinning dust peak frequency against $G_0$. Although not strong, there is tentative evidence for a correlation between the two quantities ($r_{\rm s}=0.41\pm0.15$). The majority of the high signal-to-noise ratio detections are clustered near the centre of this plot ($\nu_{\rm sp}=25$--30\,GHz and $G_{0}=1$--3), making a slope difficult to constrain; a weighted fit is flat within the uncertainties (dashed line).  However, if we make an unweighted fit to the AME sources, assigning median errors for $G_0$ and $\nu_{\rm sp}$, we find a positive slope of $+0.18\pm0.02$. Omitting the California Nebula, which is driving this fit, gives a slope of $+0.05\pm 0.02$. The precise value of the AME emissivity does depend on the assumptions on $\beta_{\rm d}$ that affect the estimation of $G_0$; however, in this case, using the fitted values for $\beta_{\rm d}$ actually results in a tighter positive correlation, with $r_{\rm s}=0.65 \pm 0.15$ and $\gamma=+0.08 \pm 0.02$ (omitting the California Nebula).

As discussed in the next section, the increase of AME emissivity and peak frequency with the increase of $G_0$ is compatible with model predictions. For $0.1 \lesssim G_0 \lesssim 10$, gas-grain interactions dominate the process for rotational excitation over photon emission. This might explain why we see a decrease of the AME intensity when $T_{\rm d}$ increases. Alternatively, there could be an observational bias related to the fact that cooler regions, with a weaker radiation field due to less massive stars, will typically have less free-free emission. Thus we may be selecting cool regions preferentially; indeed sources with larger values of $T_{\rm d}$ appear to have larger uncertainties in the AME amplitude. A complete flux-density-limited sample would be required to investigate this possibility. Nevertheless, this result agrees with previous studies that found AME arising preferentially from denser and colder molecular/atomic gas rather than from the ionized gas phase \citep{planck2011-7.2, planck2011-7.3}. 

\subsection{Spinning dust modelling}
\label{sec:modelling}

We investigate here whether the behaviour of the AME presently observed can be explained by spinning dust emission, i.e., rotational emission of PAHs. We focus on the relationship between the AME and the intensity of the radiation field $G_0$. The present results indicate a trend between the AME emissivity at 28.4\,GHz  (noted $j_{28}$) and $G_0$ (Fig.~\ref{fig:G0_AME}), where $j_{28}=S^{28}_{\rm resid}/\tau_{250}=j_0\, G_0^{+1.0}$, with $j_0\simeq 3 \times 10^{-24}$ MJy\,sr\mo\,cm$^{2}$ per H atom. This trend is found for the significant AME regions.

Models show that the spinning dust emissivity increases with $G_0$ and $n_{\rm H}$, the gas density \citep{Ali-Hamoud2009,Ysard2010a,Hoang2011}. In general, the spinning dust emissivity and therefore $j_{28}$ are functions of $n_{\rm H}$ and $G_0$. In the CNM, excitation of spinning dust by photons and collisions have similar rates. For the present sample, we expect both processes to contribute to the spinning dust emissivity. To model the spinning dust emission in our sample, we describe the emission region in the following way. Observations suggest that PAHs are depleted in the ionized gas (e.g., \citealt{Salgado2012,Dobler2009}) and in the case of \hii\ regions, PAHs are observed to be located in a dense, neutral shell surrounding the ionized gas. We assume that this shell dominates the thermal dust emission and is therefore heated by a mean radiation field of intensity $G_0$. We note that the dust optical depth $\tau_{250}$ does not show any trend with $G_0$ ($r=-0.01 \pm 0.13$), suggesting that the column density of dust  is rather constant, probably corresponding to the shell material directly heated by the star.  The AME thus comes from a shell around the \hii\ region with density $n_{\rm H}$ and radiation field $G_0$. From the observed trend of $j_{28}$ with $G_0$, and assuming that the AME is due to spinning dust, models can constrain the gas density $n_{\rm H}$. 

\begin{figure}[tb]
\begin{center}
\includegraphics[scale=0.5]{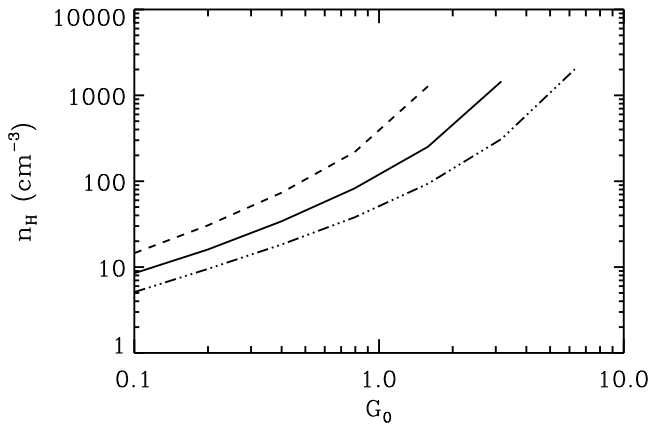}
\caption{Density estimates (solid black line) from spinning dust models that
 follow the observed trend between the AME emissivity $j_{28}$ and
 $G_0$. We also show the influence of AME emissivities half ($j_0=1.5 \times 10^{-24}$\,MJy\,sr\mo\,cm$^2$per H atom ; triple-dot-dashed line) and twice as large ($j_0=6 \times 10^{-24}$\,MJy\,sr\mo\,cm$^2$ per H atom; dashed line).}
\label{fig:density_AME}
\end{center}
\end{figure}

Using the {\tt SPDUST} code \citep{Silsbee2011} and the gas ion scheme of \cite{Ysard2011}, we ran model grids for the spinning dust emissivity in the 28.4\,GHz LFI band with the following hypotheses. We assume the PAH properties to be the same for all values of $G_0$ and $n_{\rm H}$, namely that they enclose 65\,ppm of carbon and have a log-normal size distribution with centroid 0.6\,nm and width 0.4\,nm. In addition, we assume the gas temperature to be constant for all $n_{\rm H}$ and $G_0$, namely $T_{\rm gas}=60$\,K\footnote{From {\tt CLOUDY} simulations (as described in \citealt{Ysard2010a}) we find $T_{\rm gas}$ to be between 50 and 70\,K along the $n_H$--$G_0$ trend of Fig.~\ref{fig:density_AME}.}. In Fig.~\ref{fig:density_AME}, we show the relationship between $G_0$ and $n_{\rm H}$ that results from the $j_{28}$--$G_0$ trend. We see that $n_{\rm H}$ increases with $G_0$, suggesting that density plays an important role in the AME pumping. As expected, higher AME emissivities require higher densities. This trend possibly reflects the presence of more massive stars in denser clouds. The emissivity rises with $n_{\rm H}$ but levels off for $n_{\rm H}>10^3$\,cm$^{-3}$ (see figure 8 of \citealt{Ysard2011}) due to recombination of H$^+$ and C$^+$ ions. When $j_0=6 \times 10^{-24}$ MJy\,sr\mo\,cm$^2$ per H atom, the model emissivity cannot match the observed level even for very high density  ($n_{\rm H}=10^6$\,cm$^{-3}$); this is why the dashed curve is interrupted in Fig.~\ref{fig:density_AME}.

We emphasize that this $n_{\rm H}$--$G_0$ trend is qualitative, because of several important uncertainties. First, while estimating $G_0$ from the emergent dust SED, we may be biased towards low values because of the contribution of large column densities of cold dust. Also, we assume a constant PAH abundance, whereas it is known that PAHs are depleted in dense gas \citep{Arab2012,Compiegne2008}. This analysis is thus preliminary, and microwave data at higher angular resolution, as well as radiative transfer modelling of such regions, are required to further our understanding of the AME and its possible use in characterizing the physical conditions in the emitting region.


\section{Discussion and conclusions} 
\label{sec:conclusions}

\subsection{Detection of AME}
\label{sec:discussion_detection}

In this paper we have increased the number of known candidate AME targets by a factor of several. We have also presented evidence for AME being exhibited in a much larger number of Galactic clouds, albeit at a lower significance. The spectral coverage of radio, submillimetre, and far-infrared data have allowed us to estimate the SED to high precision to determine the continuum components that are contributing at each frequency. These sources are a mix of \hii\ regions, molecular clouds, and dark clouds, often in cloud complexes. The SEDs can be fitted by a combination of synchrotron (in a few cases), optically thin free-free emission, CMB, thermal dust, and spinning dust. 

Of the 98 regions, 42 are highly significant $(>5\,\sigma$) in terms of AME amplitude. Of these 42 sources, 15 contain candidate UC\hii\ regions (based on \IRAS\ colour ratios) that could be contributing a portion ($>25\,\%$) of the excess emission and are dropped from the significant AME list. This leaves 27 regions that appear to emit AME. Among the clearest AME detections are the well-known $\rho$~Ophiuchi and Perseus molecular clouds. A further 27 sources are AME significant at the level of 2--5\,$\sigma$. Only 29 of the sample are non-significant ($<2\,\sigma$) although the sample is far from complete. 

The AME detections are shown to be robust. In general, we have been conservative in the assignment of uncertainties. The average reduced $\chi^2$ value of 0.59 indicates that we may be over-estimating our uncertainties by about 30\,\%; however, given the complexity of assigning flux densities to diffuse emission in the presence of complex and bright backgrounds, we choose to be conservative in this low resolution analysis. We consider the effect of fitting for a CMB component and its relation to a possible flattening of the thermal dust index at frequencies $<300$\,GHz. Although there is some evidence for this effect, it is shown not to be a major effect on the AME properties. The excess emission is readily fitted by spinning dust models that peak at frequencies around $25$--30\,GHz and have column densities of about $10^{21}$\,cm$^{-2}$, reasonable for diffuse ISM clouds. We have not attempted to fit for other components, such as magneto-dipole radiation \citep{Draine1999,Draine2012,Draine2013}, which could also account for the excess emission.

A number of the sources in the sample have been discussed in the literature. \cite{Genova-Santos2011} used {\it WMAP} data to study the reflection nebula associated with the Pleiades (G166.44$-$24.08) and found ($2.15\pm0.12$)\,Jy at 22.8\,GHz in a 1\deg\ radius aperture. Of this, the bulk was found to be from AME. In our analysis, we find a similar total flux density of ($1.89\pm0.13$)\,Jy at 22.8\,GHz, but the AME level is much lower at ($0.83\pm0.19$)\,Jy. The results are different mostly due to the different level of free-free emission assigned to the cloud due to the large background at low frequencies. Although it does not enter our significant AME list, with $\sigma_{\rm AME}=4.3$, this region is likely to be emitting appreciable amounts of AME at frequencies $\approx 20$--40\,GHz.

The source G201.62+01.63 was first claimed to emit AME by \cite{Finkbeiner2002}, but was later shown by \cite{Dickinson2006} to be consistent with free-free emission on angular scales of $6\arcmin$ with an upper limit of 24\,\% AME at 31\,GHz. More recently, \cite{Rubino-Martin2012a} detected AME at the $2.8\,\sigma$ level. In the present analysis, this source is highly significant at $\sigma_{\rm AME}=7.4$. The analysis of \cite{Dickinson2006} was of the bright, more compact ($6\arcmin$) source in the centre of the nebula, which contains approximately half of the flux density of the more extended region covering $30\arcmin$. It is possible that most of the AME is originating within this extended region that the 1\deg\ analysis is more sensitive to. G204.70$-$11.80 is associated with LDN1622 and LDN1621, which are both known to be dominated by dust-correlated emission at 31\,GHz \citep{Casassus2006,Dickinson2010}.

\subsection{Nature of the sources in our sample}
\label{sec:discussion_nature}

Given the low angular resolution ($1\deg$) of the present analysis, it is somewhat difficult to disentangle the various sources and environments for the majority of the AME regions, since in many cases there are multiple sources within the beam.  Furthermore, the sample is not complete and there likely remains selection effects and biases. However, in some cases the nature of the source is clear. For example, some of the non-AME regions are clearly dominated by bright \hii\ regions, such as G209.01$-$19.38 (M42) and G267.95$-$1.06 (RCW38). 

Of the AME regions, many have a contribution of free-free emission associated with \hii\ regions.  However, a small fraction of these do not have a bright associated \hii\ region, and a large fraction of the 20--60\,GHz flux density is due to AME. These sight-lines do contain known dark nebula and may be responsible for the AME. The best examples of these are G004.24+18.09 (LDN137/141), G005.40+36.50 (LDN134), G023.47+08.19 (LDN462), G133.27+09.05 (LDN1358/1355/1357), G142.35+01.35 (DNe TGU H942), G180.18+04.30 (LDN1557), G192.34$-$11.37 (LDN1582/1584), G219.18$-$08.93 (DNe H1544/1546), and G231.83$-$02.00 (DNe TGU H1593). These sources are dominated by AME at 23\,GHz, but the AME emissivity (per unit dust column density) is typically lower than the average for the sample.

The AME regions are, in general, spatially extended even at an angular resolution of $1\deg$, while non-AME regions tend to be relatively compact. This is a trend that is becoming increasingly familiar; AME sources are typically diffuse (e.g., \citealt{Davies2006,Casassus2006,Casassus2008,Dickinson2010,planck2011-7.2,planck2011-7.3}), while compact objects show little evidence for AME (e.g., \citealt{Scaife2009}), or a much smaller fraction of the total flux density (e.g., \citealt{Dickinson2007,Tibbs2010}).

A tight correlation is observed between the AME amplitude and dust continuum tracers in the submillimetre (353\,GHz) and far-infrared (12, 25, 60, 100\micron). In agreement with other studies, dividing the infrared fluxes by the intensity of the interstellar radiation field $G_0$ improves the correlation. We find that the majority of the AME regions have a high ratio of $S_{12}/S_{25}=0.06$--$1.0$, compared to the typical range $0.2$--$0.6$. From inspection of these regions, they do not appear to contain a strong ionized component, and therefore probably do not have strong ionic/stellar contamination of the $12\micron$ band. Instead, we believe that this is indicative of PAHs as the carriers of spinning dust emission. Comparison with {\it Spitzer} data is inconclusive because of the limited coverage of the data.

We find that the AME sources favour slightly cooler regions with $T_{\rm d}=14$--20\,K and a dust emissivity index of $\beta_{\rm d} \approx\! +1.8$, while the non-AME sources appear to be warmer ($T_{\rm d}=20$--$27$\,K) and prefer a slightly flatter emissivity index $\beta_{\rm d} \approx\! +1.7$. This is consistent with the picture that AME is arising primarily from the colder neutral ISM phases, rather than the warmer ionized gas associated with bona fide \hii\ regions. In \hii\ regions the PAH population is expected to be depleted, thus reducing the primary carriers of spinning dust emission. The lack of AME associated with the WNM/WIM phase is in agreement with the findings of \cite{planck2011-7.3}. The ISRF intensity, parameterized by $G_0$, appears to play an important role for AME. We find that most AME sources prefer lower levels of the ISRF in the range $0.3 \lesssim G_0 \lesssim 3$, while non-AME regions are generally higher ($G_0 \approx 1$--10). 

For the AME emissivity relative to $100\micron$, we find that the levels are similar to those found in the diffuse ISM at high Galactic latitudes, as well as from previous studies of \hii\ regions, with typical values of a few $10^{-4}$. A more robust emissivity is found by forming the ratio between the AME flux density and the thermal dust optical depth $\tau_{250}$, due to its non-dependancy on dust temperature. Remarkably, we find that most of the sources in our sample (AME and non-AME regions) have a similar AME emissivity relative to $\tau_{250}$, although there is a large scatter at all significance levels. This suggests that AME may be inherent to most of the sources in our sample, but for some sources the AME is more difficult to detect, such as when there is a dominant free-free component. 

We investigated the AME emissivity with various observational/physical parameters from our sample. In general, we do not find strong trends. We believe this is partly due to the aforementioned complexity and mix of environments and sources within each region, in addition to the relatively small sample size; however, a few weak trends are observed. 

In agreement with previous studies \citep{Lagache2003,Vidal2011}, we find that the AME emissivity decreases with the column density, as traced by the thermal dust optical depth. This might be explained by dust coagulation in denser regions, reducing the population of smallest grains and therefore reducing the number of spinning dust carriers.

We observe a correlation of $G_0$ with the AME emissivity. A weak correlation of $G_0$ with the peak frequency of the spinning dust frequency is also observed. Both these trends may be understood in terms of radiative and collisional excitation of the rotation of small grains; however, we caution that there may also be an observational bias, due to the fact that an increasing $G_0$ generally means an increasing free-free component that makes it more difficult to detect AME. On the other hand, spinning dust modelling of the AME emissivity assuming a fixed small grain abundance suggests that the density increases with $G_0$, a trend expected in massive star forming regions.

Finally, we note that the sources within our sample cover a wide range of distances (from a few kpc to $15$\,kpc); however, the majority of the significant AME regions (that do not have a potential UC\hii\ component) tend to be located several degrees away from the Galactic plane, resulting in them being relatively nearby ($d<1$\,kpc). The semi-significant AME candidates in our sample cover a wide range of Galactocentric distances, showing that AME is not just a local phenomenon, as was previously shown by \cite{planck2011-7.3}.

\subsection{Final remarks}
\label{sec:discussion_final}

To make further progress in this field it is now vital that we complement these studies in two ways. First, more accurately calibrated data at frequencies not covered in this analysis will help determine more precisely each of the emission components. In particular, more data at frequencies 5--20\,GHz (e.g., the 5\,GHz C-Band All-Sky Survey, C-BASS; \citealt{King2010}, or the multi-frequency 11--19\,GHz Q-U-I JOint TEnerife CMB experiment, QUIJOTE; \citealt{Rubino-Martin2012b}) would improve the constraints on free-free emission, which is a large fraction of the 20--60\,GHz brightness in many of the sources under study. Also, the inclusion of the \Planck\ 100 and 217\,GHz channels will dramatically improve the constraints where we see evidence of flattening of the low frequency tail of thermal dust, as well as a (minor) contribution from CMB fluctuations on scales of $1\deg$. Second and perhaps more importantly, higher resolution (few arcmin and better) data are required to study different environments in more detail. As discussed earlier, we typically see a mix of environments and multiple sources when working at an angular resolution of $1\deg$. The relationships between the gas and dust can then be inferred and the contribution of AME from different sources can be assessed by comparison with high resolution radio and infrared data. Furthermore, on scales $\ll1\deg$ the CMB becomes negligible and background emission can, at least in some areas of the sky, be less problematic. High resolution follow-up observations of the AME candidates, such as with AMI \citep{Zwart2008} at a frequency of $15$\,GHz, will determine whether the AME is in fact due to diffuse dust and whether UC\hii\ regions are a major component. With more accurate SEDs and spatial information, we will be able to better constrain important quantities such as the intensity of the radiation field and the abundance of small grains. The polarization properties of AME are also of great importance in identifying the emission mechanisms responsible, such as magneto-dipole radiation \citep{Draine2013}. Any significant level of polarized emission, in the frequency range $\approx 30$--300\,GHz, will likely be a major foreground for future sensitive CMB B-mode experiments \citep{Rubino-Martin2012a,Armitage-Caplan2012}.


\begin{acknowledgements}
We thank the anonymous referee for providing useful comments. We thank Justin Jonas for providing the 2326\,MHz Hart\-RAO map. We acknowledge the use of the MPIfR Survey Sampler website at \url{http://www.mpifr-bonn.mpg.de/survey.html} and the Legacy Archive for Microwave Background Data Analysis (LAMBDA); support for LAMBDA is provided by the NASA Office of Space Science. This research has made use of the NASA/IPAC Extragalactic Database (NED), which is operated by the Jet Propulsion Laboratory, California Institute of Technology, under contract with the National Aeronautics and Space Administration. This research also makes use of the SIMBAD database, operated at CDS, Strasbourg, France. We acknowledge the use of NASA's SkyView facility (\url{http://skyview.gsfc.nasa.gov}) located at NASA Goddard Space Flight Center. CD acknowledges an STFC Advanced Fellowship, an EU Marie-Curie IRG grant under the FP7, and an ERC Starting Grant (no. 307209).

The development of \Planck\ has been supported by: ESA; CNES and CNRS/INSU-IN2P3-INP (France); ASI, CNR, and INAF (Italy); NASA and DoE (USA); STFC and UKSA (UK); CSIC, MICINN, JA and RES (Spain); Tekes, AoF and CSC (Finland); DLR and MPG (Germany); CSA (Canada); DTU Space (Denmark); SER/SSO (Switzerland); RCN (Norway); SFI (Ireland); FCT/MCTES (Portugal); and PRACE (EU). A description of the Planck Collaboration and a list of its members, including the technical or scientific activities in which they have been involved, can be found at \url{http://www.sciops.esa.int/index.php?project=planck&page=Planck_Collaboration}.

\end{acknowledgements}


\bibliographystyle{aa}
\bibliography{Planck_bib,clive_refs}
\raggedright 


\include{main_table_hii_ame}


\end{document}

%% file: Planck.tex
\def\setsymbol#1#2{\expandafter\def\csname #1\endcsname{#2}}
\def\getsymbol#1{\csname #1\endcsname}

\def\Planck{\textit{Planck}}

\def\allearlypapers{\nocite{planck2011-1.1, planck2011-1.3, planck2011-1.4, planck2011-1.5, planck2011-1.6, planck2011-1.7, planck2011-1.10, planck2011-1.10sup, planck2011-5.1a, planck2011-5.1b, planck2011-5.2a, planck2011-5.2b, planck2011-5.2c, planck2011-6.1, planck2011-6.2, planck2011-6.3a, planck2011-6.4a, planck2011-6.4b, planck2011-6.6, planck2011-7.0, planck2011-7.2, planck2011-7.3, planck2011-7.7a, planck2011-7.7b, planck2011-7.12, planck2011-7.13}}

\newbox\tablebox    \newdimen\tablewidth
\def\leaderfil{\leaders\hbox to 5pt{\hss.\hss}\hfil}
\def\endPlancktablewide{\tablewidth=\textwidth 
    $$\hss\copy\tablebox\hss$$
    \vskip-\lastskip\vskip -2pt}
\def\tablenote#1 #2\par{\begingroup \parindent=0.8em
    \abovedisplayshortskip=0pt\belowdisplayshortskip=0pt
    \noindent
    $$\hss\vbox{\hsize\tablewidth \hangindent=\parindent \hangafter=1 \noindent
    \hbox to \parindent{$^#1$\hss}\strut#2\strut\par}\hss$$
    \endgroup}
\def\doubleline{\vskip 3pt\hrule \vskip 1.5pt \hrule \vskip 5pt}

\def\L2{\ifmmode L_2\else $L_2$\fi}

\def\DeltaT{\ifmmode \Delta T\else $\Delta T$\fi}
\def\deltat{\ifmmode \Delta t\else $\Delta t$\fi}
\def\fknee{\ifmmode f_{\rm knee}\else $f_{\rm knee}$\fi}
\def\Fmax{\ifmmode F_{\rm max}\else $F_{\rm max}$\fi}
\def\solar{\ifmmode{\rm M}_{\mathord\odot}\else${\rm M}_{\mathord\odot}$\fi}
\def\Msolar{\ifmmode{\rm M}_{\mathord\odot}\else${\rm M}_{\mathord\odot}$\fi}
\def\Lsolar{\ifmmode{\rm L}_{\mathord\odot}\else${\rm L}_{\mathord\odot}$\fi}

\def\inv{\ifmmode^{-1}\else$^{-1}$\fi}
\def\mo{\ifmmode^{-1}\else$^{-1}$\fi}
\def\sup#1{\ifmmode ^{\rm #1}\else $^{\rm #1}$\fi}
\def\expo#1{\ifmmode \times 10^{#1}\else $\times 10^{#1}$\fi}
\def\,{\thinspace}
\def\lsim{\mathrel{\raise .4ex\hbox{\rlap{$<$}\lower 1.2ex\hbox{$\sim$}}}}
\def\gsim{\mathrel{\raise .4ex\hbox{\rlap{$>$}\lower 1.2ex\hbox{$\sim$}}}}

\def\simprop{\mathrel{\raise .4ex\hbox{\rlap{$\propto$}\lower 1.2ex\hbox{$\sim$}}}}
\def\deg{\ifmmode^\circ\else$^\circ$\fi}
\def\pdeg{\ifmmode $\setbox0=\hbox{$^{\circ}$}\rlap{\hskip.11\wd0 .}$^{\circ}
          \else \setbox0=\hbox{$^{\circ}$}\rlap{\hskip.11\wd0 .}$^{\circ}$\fi}
\def\arcs{\ifmmode {^{\scriptstyle\prime\prime}}
          \else $^{\scriptstyle\prime\prime}$\fi}
\def\arcm{\ifmmode {^{\scriptstyle\prime}}
          \else $^{\scriptstyle\prime}$\fi}
\newdimen\sa  \newdimen\sb
\def\parcs{\sa=.07em \sb=.03em
     \ifmmode \hbox{\rlap{.}}^{\scriptstyle\prime\kern -\sb\prime}\hbox{\kern -\sa}
     \else \rlap{.}$^{\scriptstyle\prime\kern -\sb\prime}$\kern -\sa\fi}
\def\parcm{\sa=.08em \sb=.03em
     \ifmmode \hbox{\rlap{.}\kern\sa}^{\scriptstyle\prime}\hbox{\kern-\sb}
     \else \rlap{.}\kern\sa$^{\scriptstyle\prime}$\kern-\sb\fi}
\def\ra[#1 #2 #3.#4]{#1\sup{h}#2\sup{m}#3\sup{s}\llap.#4}
\def\dec[#1 #2 #3.#4]{#1\deg#2\arcm#3\arcs\llap.#4}
\def\deco[#1 #2 #3]{#1\deg#2\arcm#3\arcs}
\def\rra[#1 #2]{#1\sup{h}#2\sup{m}}

\def\dots{\relax\ifmmode \ldots\else $\ldots$\fi}
\def\WHzsr{\ifmmode $W\,Hz\mo\,sr\mo$\else W\,Hz\mo\,sr\mo\fi}
\def\mHz{\ifmmode $\,mHz$\else \,mHz\fi}
\def\GHz{\ifmmode $\,GHz$\else \,GHz\fi}
\def\mKs{\ifmmode $\,mK\,s$^{1/2}\else \,mK\,s$^{1/2}$\fi}
\def\muKs{\ifmmode \,\mu$K\,s$^{1/2}\else \,$\mu$K\,s$^{1/2}$\fi}
\def\muKRJs{\ifmmode \,\mu$K$_{\rm RJ}$\,s$^{1/2}\else \,$\mu$K$_{\rm RJ}$\,s$^{1/2}$\fi}
\def\muKHz{\ifmmode \,\mu$K\,Hz$^{-1/2}\else \,$\mu$K\,Hz$^{-1/2}$\fi}
\def\MJysr{\ifmmode \,$MJy\,sr\mo$\else \,MJy\,sr\mo\fi}
\def\MJysrmK{\ifmmode \,$MJy\,sr\mo$\,mK$_{\rm CMB}\mo\else \,MJy\,sr\mo\,mK$_{\rm CMB}\mo$\fi}
\def\microns{\ifmmode \,\mu$m$\else \,$\mu$m\fi}
\def\micron{\microns}
\def\muK{\ifmmode \,\mu$K$\else \,$\mu$\hbox{K}\fi}
\def\microK{\ifmmode \,\mu$K$\else \,$\mu$\hbox{K}\fi}
\def\muW{\ifmmode \,\mu$W$\else \,$\mu$\hbox{W}\fi}
\def\kms{\ifmmode $\,km\,s$^{-1}\else \,km\,s$^{-1}$\fi}
\def\kmsMpc{\ifmmode $\,\kms\,Mpc\mo$\else \,\kms\,Mpc\mo\fi}
\setsymbol{LFI:center:frequency:70GHz:units}{70.3\,GHz}
\setsymbol{LFI:center:frequency:44GHz:units}{44.1\,GHz}
\setsymbol{LFI:center:frequency:30GHz:units}{28.5\,GHz}

\setsymbol{LFI:center:frequency:70GHz}{70.3}
\setsymbol{LFI:center:frequency:44GHz}{44.1}
\setsymbol{LFI:center:frequency:30GHz}{28.5}

\setsymbol{LFI:center:frequency:LFI18:Rad:M:units}{71.7\GHz}
\setsymbol{LFI:center:frequency:LFI19:Rad:M:units}{67.5\GHz}
\setsymbol{LFI:center:frequency:LFI20:Rad:M:units}{69.2\GHz}
\setsymbol{LFI:center:frequency:LFI21:Rad:M:units}{70.4\GHz}
\setsymbol{LFI:center:frequency:LFI22:Rad:M:units}{71.5\GHz}
\setsymbol{LFI:center:frequency:LFI23:Rad:M:units}{70.8\GHz}
\setsymbol{LFI:center:frequency:LFI24:Rad:M:units}{44.4\GHz}
\setsymbol{LFI:center:frequency:LFI25:Rad:M:units}{44.0\GHz}
\setsymbol{LFI:center:frequency:LFI26:Rad:M:units}{43.9\GHz}
\setsymbol{LFI:center:frequency:LFI27:Rad:M:units}{28.3\GHz}
\setsymbol{LFI:center:frequency:LFI28:Rad:M:units}{28.8\GHz}
\setsymbol{LFI:center:frequency:LFI18:Rad:S:units}{70.1\GHz}
\setsymbol{LFI:center:frequency:LFI19:Rad:S:units}{69.6\GHz}
\setsymbol{LFI:center:frequency:LFI20:Rad:S:units}{69.5\GHz}
\setsymbol{LFI:center:frequency:LFI21:Rad:S:units}{69.5\GHz}
\setsymbol{LFI:center:frequency:LFI22:Rad:S:units}{72.8\GHz}
\setsymbol{LFI:center:frequency:LFI23:Rad:S:units}{71.3\GHz}
\setsymbol{LFI:center:frequency:LFI24:Rad:S:units}{44.1\GHz}
\setsymbol{LFI:center:frequency:LFI25:Rad:S:units}{44.1\GHz}
\setsymbol{LFI:center:frequency:LFI26:Rad:S:units}{44.1\GHz}
\setsymbol{LFI:center:frequency:LFI27:Rad:S:units}{28.5\GHz}
\setsymbol{LFI:center:frequency:LFI28:Rad:S:units}{28.2\GHz}

\setsymbol{LFI:center:frequency:LFI18:Rad:M}{71.7}
\setsymbol{LFI:center:frequency:LFI19:Rad:M}{67.5}
\setsymbol{LFI:center:frequency:LFI20:Rad:M}{69.2}
\setsymbol{LFI:center:frequency:LFI21:Rad:M}{70.4}
\setsymbol{LFI:center:frequency:LFI22:Rad:M}{71.5}
\setsymbol{LFI:center:frequency:LFI23:Rad:M}{70.8}
\setsymbol{LFI:center:frequency:LFI24:Rad:M}{44.4}
\setsymbol{LFI:center:frequency:LFI25:Rad:M}{44.0}
\setsymbol{LFI:center:frequency:LFI26:Rad:M}{43.9}
\setsymbol{LFI:center:frequency:LFI27:Rad:M}{28.3}
\setsymbol{LFI:center:frequency:LFI28:Rad:M}{28.8}
\setsymbol{LFI:center:frequency:LFI18:Rad:S}{70.1}
\setsymbol{LFI:center:frequency:LFI19:Rad:S}{69.6}
\setsymbol{LFI:center:frequency:LFI20:Rad:S}{69.5}
\setsymbol{LFI:center:frequency:LFI21:Rad:S}{69.5}
\setsymbol{LFI:center:frequency:LFI22:Rad:S}{72.8}
\setsymbol{LFI:center:frequency:LFI23:Rad:S}{71.3}
\setsymbol{LFI:center:frequency:LFI24:Rad:S}{44.1}
\setsymbol{LFI:center:frequency:LFI25:Rad:S}{44.1}
\setsymbol{LFI:center:frequency:LFI26:Rad:S}{44.1}
\setsymbol{LFI:center:frequency:LFI27:Rad:S}{28.5}
\setsymbol{LFI:center:frequency:LFI28:Rad:S}{28.2}

\setsymbol{LFI:white:noise:sensitivity:70GHz:units}{134.7\muKs}
\setsymbol{LFI:white:noise:sensitivity:44GHz:units}{164.7\muKs}
\setsymbol{LFI:white:noise:sensitivity:30GHz:units}{143.4\muKs}

\setsymbol{LFI:white:noise:sensitivity:70GHz}{134.7}
\setsymbol{LFI:white:noise:sensitivity:44GHz}{164.7}
\setsymbol{LFI:white:noise:sensitivity:30GHz}{143.4}

\setsymbol{LFI:white:noise:sensitivity:LFI18:Rad:M:units}{512.0\muKs}
\setsymbol{LFI:white:noise:sensitivity:LFI19:Rad:M:units}{581.4\muKs}
\setsymbol{LFI:white:noise:sensitivity:LFI20:Rad:M:units}{590.8\muKs}
\setsymbol{LFI:white:noise:sensitivity:LFI21:Rad:M:units}{455.2\muKs}
\setsymbol{LFI:white:noise:sensitivity:LFI22:Rad:M:units}{492.0\muKs}
\setsymbol{LFI:white:noise:sensitivity:LFI23:Rad:M:units}{507.7\muKs}
\setsymbol{LFI:white:noise:sensitivity:LFI24:Rad:M:units}{462.2\muKs}
\setsymbol{LFI:white:noise:sensitivity:LFI25:Rad:M:units}{413.6\muKs}
\setsymbol{LFI:white:noise:sensitivity:LFI26:Rad:M:units}{478.6\muKs}
\setsymbol{LFI:white:noise:sensitivity:LFI27:Rad:M:units}{277.7\muKs}
\setsymbol{LFI:white:noise:sensitivity:LFI28:Rad:M:units}{312.3\muKs}
\setsymbol{LFI:white:noise:sensitivity:LFI18:Rad:S:units}{465.7\muKs}
\setsymbol{LFI:white:noise:sensitivity:LFI19:Rad:S:units}{555.6\muKs}
\setsymbol{LFI:white:noise:sensitivity:LFI20:Rad:S:units}{623.2\muKs}
\setsymbol{LFI:white:noise:sensitivity:LFI21:Rad:S:units}{564.1\muKs}
\setsymbol{LFI:white:noise:sensitivity:LFI22:Rad:S:units}{534.4\muKs}
\setsymbol{LFI:white:noise:sensitivity:LFI23:Rad:S:units}{542.4\muKs}
\setsymbol{LFI:white:noise:sensitivity:LFI24:Rad:S:units}{399.2\muKs}
\setsymbol{LFI:white:noise:sensitivity:LFI25:Rad:S:units}{392.6\muKs}
\setsymbol{LFI:white:noise:sensitivity:LFI26:Rad:S:units}{418.6\muKs}
\setsymbol{LFI:white:noise:sensitivity:LFI27:Rad:S:units}{302.9\muKs}
\setsymbol{LFI:white:noise:sensitivity:LFI28:Rad:S:units}{285.3\muKs}

\setsymbol{LFI:white:noise:sensitivity:LFI18:Rad:M}{512.0}
\setsymbol{LFI:white:noise:sensitivity:LFI19:Rad:M}{581.4}
\setsymbol{LFI:white:noise:sensitivity:LFI20:Rad:M}{590.8}
\setsymbol{LFI:white:noise:sensitivity:LFI21:Rad:M}{455.2}
\setsymbol{LFI:white:noise:sensitivity:LFI22:Rad:M}{492.0}
\setsymbol{LFI:white:noise:sensitivity:LFI23:Rad:M}{507.7}
\setsymbol{LFI:white:noise:sensitivity:LFI24:Rad:M}{462.2}
\setsymbol{LFI:white:noise:sensitivity:LFI25:Rad:M}{413.6}
\setsymbol{LFI:white:noise:sensitivity:LFI26:Rad:M}{478.6}
\setsymbol{LFI:white:noise:sensitivity:LFI27:Rad:M}{277.7}
\setsymbol{LFI:white:noise:sensitivity:LFI28:Rad:M}{312.3}
\setsymbol{LFI:white:noise:sensitivity:LFI18:Rad:S}{465.7}
\setsymbol{LFI:white:noise:sensitivity:LFI19:Rad:S}{555.6}
\setsymbol{LFI:white:noise:sensitivity:LFI20:Rad:S}{623.2}
\setsymbol{LFI:white:noise:sensitivity:LFI21:Rad:S}{564.1}
\setsymbol{LFI:white:noise:sensitivity:LFI22:Rad:S}{534.4}
\setsymbol{LFI:white:noise:sensitivity:LFI23:Rad:S}{542.4}
\setsymbol{LFI:white:noise:sensitivity:LFI24:Rad:S}{399.2}
\setsymbol{LFI:white:noise:sensitivity:LFI25:Rad:S}{392.6}
\setsymbol{LFI:white:noise:sensitivity:LFI26:Rad:S}{418.6}
\setsymbol{LFI:white:noise:sensitivity:LFI27:Rad:S}{302.9}
\setsymbol{LFI:white:noise:sensitivity:LFI28:Rad:S}{285.3}

\setsymbol{LFI:knee:frequency:70GHz:units}{29.5\mHz}
\setsymbol{LFI:knee:frequency:44GHz:units}{56.2\mHz}
\setsymbol{LFI:knee:frequency:30GHz:units}{113.7\mHz}

\setsymbol{LFI:knee:frequency:70GHz}{29.5}
\setsymbol{LFI:knee:frequency:44GHz}{56.2}
\setsymbol{LFI:knee:frequency:30GHz}{113.7}

\setsymbol{LFI:knee:frequency:LFI18:Rad:M:units}{16.3\mHz}
\setsymbol{LFI:knee:frequency:LFI19:Rad:M:units}{15.1\mHz}
\setsymbol{LFI:knee:frequency:LFI20:Rad:M:units}{18.7\mHz}
\setsymbol{LFI:knee:frequency:LFI21:Rad:M:units}{37.2\mHz}
\setsymbol{LFI:knee:frequency:LFI22:Rad:M:units}{12.7\mHz}
\setsymbol{LFI:knee:frequency:LFI23:Rad:M:units}{34.6\mHz}
\setsymbol{LFI:knee:frequency:LFI24:Rad:M:units}{46.2\mHz}
\setsymbol{LFI:knee:frequency:LFI25:Rad:M:units}{24.9\mHz}
\setsymbol{LFI:knee:frequency:LFI26:Rad:M:units}{67.6\mHz}
\setsymbol{LFI:knee:frequency:LFI27:Rad:M:units}{187.4\mHz}
\setsymbol{LFI:knee:frequency:LFI28:Rad:M:units}{122.2\mHz}
\setsymbol{LFI:knee:frequency:LFI18:Rad:S:units}{17.7\mHz}
\setsymbol{LFI:knee:frequency:LFI19:Rad:S:units}{22.0\mHz}
\setsymbol{LFI:knee:frequency:LFI20:Rad:S:units}{8.7\mHz}
\setsymbol{LFI:knee:frequency:LFI21:Rad:S:units}{25.9\mHz}
\setsymbol{LFI:knee:frequency:LFI22:Rad:S:units}{15.8\mHz}
\setsymbol{LFI:knee:frequency:LFI23:Rad:S:units}{129.8\mHz}
\setsymbol{LFI:knee:frequency:LFI24:Rad:S:units}{100.9\mHz}
\setsymbol{LFI:knee:frequency:LFI25:Rad:S:units}{38.9\mHz}
\setsymbol{LFI:knee:frequency:LFI26:Rad:S:units}{58.9\mHz}
\setsymbol{LFI:knee:frequency:LFI27:Rad:S:units}{104.4\mHz}
\setsymbol{LFI:knee:frequency:LFI28:Rad:S:units}{40.7\mHz}

\setsymbol{LFI:knee:frequency:LFI18:Rad:M}{16.3}
\setsymbol{LFI:knee:frequency:LFI19:Rad:M}{15.1}
\setsymbol{LFI:knee:frequency:LFI20:Rad:M}{18.7}
\setsymbol{LFI:knee:frequency:LFI21:Rad:M}{37.2}
\setsymbol{LFI:knee:frequency:LFI22:Rad:M}{12.7}
\setsymbol{LFI:knee:frequency:LFI23:Rad:M}{34.6}
\setsymbol{LFI:knee:frequency:LFI24:Rad:M}{46.2}
\setsymbol{LFI:knee:frequency:LFI25:Rad:M}{24.9}
\setsymbol{LFI:knee:frequency:LFI26:Rad:M}{67.6}
\setsymbol{LFI:knee:frequency:LFI27:Rad:M}{187.4}
\setsymbol{LFI:knee:frequency:LFI28:Rad:M}{122.2}
\setsymbol{LFI:knee:frequency:LFI18:Rad:S}{17.7}
\setsymbol{LFI:knee:frequency:LFI19:Rad:S}{22.0}
\setsymbol{LFI:knee:frequency:LFI20:Rad:S}{8.7}
\setsymbol{LFI:knee:frequency:LFI21:Rad:S}{25.9}
\setsymbol{LFI:knee:frequency:LFI22:Rad:S}{15.8}
\setsymbol{LFI:knee:frequency:LFI23:Rad:S}{129.8}
\setsymbol{LFI:knee:frequency:LFI24:Rad:S}{100.9}
\setsymbol{LFI:knee:frequency:LFI25:Rad:S}{38.9}
\setsymbol{LFI:knee:frequency:LFI26:Rad:S}{58.9}
\setsymbol{LFI:knee:frequency:LFI27:Rad:S}{104.4}
\setsymbol{LFI:knee:frequency:LFI28:Rad:S}{40.7}

\setsymbol{LFI:slope:70GHz:units}{$-1.03$\mHz}
\setsymbol{LFI:slope:44GHz:units}{$-0.89$\mHz}
\setsymbol{LFI:slope:30GHz:units}{$-0.87$\mHz}

\setsymbol{LFI:slope:70GHz}{$-1.03$}
\setsymbol{LFI:slope:44GHz}{$-0.89$}
\setsymbol{LFI:slope:30GHz}{$-0.87$}

\setsymbol{LFI:slope:LFI18:Rad:M:units}{$-1.04$\mHz}
\setsymbol{LFI:slope:LFI19:Rad:M:units}{$-1.09$\mHz}
\setsymbol{LFI:slope:LFI20:Rad:M:units}{$-0.69$\mHz}
\setsymbol{LFI:slope:LFI21:Rad:M:units}{$-1.56$\mHz}
\setsymbol{LFI:slope:LFI22:Rad:M:units}{$-1.01$\mHz}
\setsymbol{LFI:slope:LFI23:Rad:M:units}{$-0.96$\mHz}
\setsymbol{LFI:slope:LFI24:Rad:M:units}{$-0.83$\mHz}
\setsymbol{LFI:slope:LFI25:Rad:M:units}{$-0.91$\mHz}
\setsymbol{LFI:slope:LFI26:Rad:M:units}{$-0.95$\mHz}
\setsymbol{LFI:slope:LFI27:Rad:M:units}{$-0.87$\mHz}
\setsymbol{LFI:slope:LFI28:Rad:M:units}{$-0.88$\mHz}
\setsymbol{LFI:slope:LFI18:Rad:S:units}{$-1.15$\mHz}
\setsymbol{LFI:slope:LFI19:Rad:S:units}{$-1.00$\mHz}
\setsymbol{LFI:slope:LFI20:Rad:S:units}{$-0.95$\mHz}
\setsymbol{LFI:slope:LFI21:Rad:S:units}{$-0.92$\mHz}
\setsymbol{LFI:slope:LFI22:Rad:S:units}{$-1.01$\mHz}
\setsymbol{LFI:slope:LFI23:Rad:S:units}{$-0.95$\mHz}
\setsymbol{LFI:slope:LFI24:Rad:S:units}{$-0.73$\mHz}
\setsymbol{LFI:slope:LFI25:Rad:S:units}{$-1.16$\mHz}
\setsymbol{LFI:slope:LFI26:Rad:S:units}{$-0.79$\mHz}
\setsymbol{LFI:slope:LFI27:Rad:S:units}{$-0.82$\mHz}
\setsymbol{LFI:slope:LFI28:Rad:S:units}{$-0.91$\mHz}

\setsymbol{LFI:slope:LFI18:Rad:M}{$-1.04$}
\setsymbol{LFI:slope:LFI19:Rad:M}{$-1.09$}
\setsymbol{LFI:slope:LFI20:Rad:M}{$-0.69$}
\setsymbol{LFI:slope:LFI21:Rad:M}{$-1.56$}
\setsymbol{LFI:slope:LFI22:Rad:M}{$-1.01$}
\setsymbol{LFI:slope:LFI23:Rad:M}{$-0.96$}
\setsymbol{LFI:slope:LFI24:Rad:M}{$-0.83$}
\setsymbol{LFI:slope:LFI25:Rad:M}{$-0.91$}
\setsymbol{LFI:slope:LFI26:Rad:M}{$-0.95$}
\setsymbol{LFI:slope:LFI27:Rad:M}{$-0.87$}
\setsymbol{LFI:slope:LFI28:Rad:M}{$-0.88$}
\setsymbol{LFI:slope:LFI18:Rad:S}{$-1.15$}
\setsymbol{LFI:slope:LFI19:Rad:S}{$-1.00$}
\setsymbol{LFI:slope:LFI20:Rad:S}{$-0.95$}
\setsymbol{LFI:slope:LFI21:Rad:S}{$-0.92$}
\setsymbol{LFI:slope:LFI22:Rad:S}{$-1.01$}
\setsymbol{LFI:slope:LFI23:Rad:S}{$-0.95$}
\setsymbol{LFI:slope:LFI24:Rad:S}{$-0.73$}
\setsymbol{LFI:slope:LFI25:Rad:S}{$-1.16$}
\setsymbol{LFI:slope:LFI26:Rad:S}{$-0.79$}
\setsymbol{LFI:slope:LFI27:Rad:S}{$-0.82$}
\setsymbol{LFI:slope:LFI28:Rad:S}{$-0.91$}

\setsymbol{LFI:FWHM:70GHz:units}{13\parcm01}
\setsymbol{LFI:FWHM:44GHz:units}{27\parcm92}
\setsymbol{LFI:FWHM:30GHz:units}{32\parcm65}

\setsymbol{LFI:FWHM:70GHz}{13.01}
\setsymbol{LFI:FWHM:44GHz}{27.92}
\setsymbol{LFI:FWHM:30GHz}{32.65}

\setsymbol{LFI:FWHM:LFI18:units}{13\parcm39}
\setsymbol{LFI:FWHM:LFI19:units}{13\parcm01}
\setsymbol{LFI:FWHM:LFI20:units}{12\parcm75}
\setsymbol{LFI:FWHM:LFI21:units}{12\parcm74}
\setsymbol{LFI:FWHM:LFI22:units}{12\parcm87}
\setsymbol{LFI:FWHM:LFI23:units}{13\parcm27}
\setsymbol{LFI:FWHM:LFI24:units}{22\parcm98}
\setsymbol{LFI:FWHM:LFI25:units}{30\parcm46}
\setsymbol{LFI:FWHM:LFI26:units}{30\parcm31}
\setsymbol{LFI:FWHM:LFI27:units}{32\parcm65}
\setsymbol{LFI:FWHM:LFI28:units}{32\parcm66}

\setsymbol{LFI:FWHM:LFI18}{13.39}
\setsymbol{LFI:FWHM:LFI19}{13.01}
\setsymbol{LFI:FWHM:LFI20}{12.75}
\setsymbol{LFI:FWHM:LFI21}{12.74}
\setsymbol{LFI:FWHM:LFI22}{12.87}
\setsymbol{LFI:FWHM:LFI23}{13.27}
\setsymbol{LFI:FWHM:LFI24}{22.98}
\setsymbol{LFI:FWHM:LFI25}{30.46}
\setsymbol{LFI:FWHM:LFI26}{30.31}
\setsymbol{LFI:FWHM:LFI27}{32.65}
\setsymbol{LFI:FWHM:LFI28}{32.66}

\setsymbol{LFI:FWHM:uncertainty:LFI18:units}{0.170\arcm}
\setsymbol{LFI:FWHM:uncertainty:LFI19:units}{0.174\arcm}
\setsymbol{LFI:FWHM:uncertainty:LFI20:units}{0.170\arcm}
\setsymbol{LFI:FWHM:uncertainty:LFI21:units}{0.156\arcm}
\setsymbol{LFI:FWHM:uncertainty:LFI22:units}{0.164\arcm}
\setsymbol{LFI:FWHM:uncertainty:LFI23:units}{0.171\arcm}
\setsymbol{LFI:FWHM:uncertainty:LFI24:units}{0.652\arcm}
\setsymbol{LFI:FWHM:uncertainty:LFI25:units}{1.075\arcm}
\setsymbol{LFI:FWHM:uncertainty:LFI26:units}{1.131\arcm}
\setsymbol{LFI:FWHM:uncertainty:LFI27:units}{1.266\arcm}
\setsymbol{LFI:FWHM:uncertainty:LFI28:units}{1.287\arcm}

\setsymbol{LFI:FWHM:uncertainty:LFI18}{0.170}
\setsymbol{LFI:FWHM:uncertainty:LFI19}{0.174}
\setsymbol{LFI:FWHM:uncertainty:LFI20}{0.170}
\setsymbol{LFI:FWHM:uncertainty:LFI21}{0.156}
\setsymbol{LFI:FWHM:uncertainty:LFI22}{0.164}
\setsymbol{LFI:FWHM:uncertainty:LFI23}{0.171}
\setsymbol{LFI:FWHM:uncertainty:LFI24}{0.652}
\setsymbol{LFI:FWHM:uncertainty:LFI25}{1.075}
\setsymbol{LFI:FWHM:uncertainty:LFI26}{1.131}
\setsymbol{LFI:FWHM:uncertainty:LFI27}{1.266}
\setsymbol{LFI:FWHM:uncertainty:LFI28}{1.287}

\setsymbol{HFI:center:frequency:100GHz:units}{100\,GHz}
\setsymbol{HFI:center:frequency:143GHz:units}{143\,GHz}
\setsymbol{HFI:center:frequency:217GHz:units}{217\,GHz}
\setsymbol{HFI:center:frequency:353GHz:units}{353\,GHz}
\setsymbol{HFI:center:frequency:545GHz:units}{545\,GHz}
\setsymbol{HFI:center:frequency:857GHz:units}{857\,GHz}

\setsymbol{HFI:center:frequency:100GHz}{100}
\setsymbol{HFI:center:frequency:143GHz}{143}
\setsymbol{HFI:center:frequency:217GHz}{217}
\setsymbol{HFI:center:frequency:353GHz}{353}
\setsymbol{HFI:center:frequency:545GHz}{545}
\setsymbol{HFI:center:frequency:857GHz}{857}

\setsymbol{HFI:Ndetectors:100GHz}{8}
\setsymbol{HFI:Ndetectors:143GHz}{11}
\setsymbol{HFI:Ndetectors:217GHz}{12}
\setsymbol{HFI:Ndetectors:353GHz}{12}
\setsymbol{HFI:Ndetectors:545GHz}{3}
\setsymbol{HFI:Ndetectors:857GHz}{4}

\setsymbol{HFI:FWHM:Maps:100GHz:units}{9\parcm88}
\setsymbol{HFI:FWHM:Maps:143GHz:units}{7\parcm18}
\setsymbol{HFI:FWHM:Maps:217GHz:units}{4\parcm87}
\setsymbol{HFI:FWHM:Maps:353GHz:units}{4\parcm65}
\setsymbol{HFI:FWHM:Maps:545GHz:units}{4\parcm72}
\setsymbol{HFI:FWHM:Maps:857GHz:units}{4\parcm39}
\setsymbol{HFI:FWHM:Maps:100GHz}{9.88}
\setsymbol{HFI:FWHM:Maps:143GHz}{7.18}
\setsymbol{HFI:FWHM:Maps:217GHz}{4.87}
\setsymbol{HFI:FWHM:Maps:353GHz}{4.65}
\setsymbol{HFI:FWHM:Maps:545GHz}{4.72}
\setsymbol{HFI:FWHM:Maps:857GHz}{4.39}

\setsymbol{HFI:beam:ellipticity:Maps:100GHz}{1.15}
\setsymbol{HFI:beam:ellipticity:Maps:143GHz}{1.01}
\setsymbol{HFI:beam:ellipticity:Maps:217GHz}{1.06}
\setsymbol{HFI:beam:ellipticity:Maps:353GHz}{1.05}
\setsymbol{HFI:beam:ellipticity:Maps:545GHz}{1.14}
\setsymbol{HFI:beam:ellipticity:Maps:857GHz}{1.19}

\setsymbol{HFI:FWHM:Mars:100GHz:units}{9\parcm37}
\setsymbol{HFI:FWHM:Mars:143GHz:units}{7\parcm04}
\setsymbol{HFI:FWHM:Mars:217GHz:units}{4\parcm68}
\setsymbol{HFI:FWHM:Mars:353GHz:units}{4\parcm43}
\setsymbol{HFI:FWHM:Mars:545GHz:units}{3\parcm80}
\setsymbol{HFI:FWHM:Mars:857GHz:units}{3\parcm67}

\setsymbol{HFI:FWHM:Mars:100GHz}{9.37}
\setsymbol{HFI:FWHM:Mars:143GHz}{7.04}
\setsymbol{HFI:FWHM:Mars:217GHz}{4.68}
\setsymbol{HFI:FWHM:Mars:353GHz}{4.43}
\setsymbol{HFI:FWHM:Mars:545GHz}{3.80}
\setsymbol{HFI:FWHM:Mars:857GHz}{3.67}

\setsymbol{HFI:beam:ellipticity:Mars:100GHz}{1.18}
\setsymbol{HFI:beam:ellipticity:Mars:143GHz}{1.03}
\setsymbol{HFI:beam:ellipticity:Mars:217GHz}{1.14}
\setsymbol{HFI:beam:ellipticity:Mars:353GHz}{1.09}
\setsymbol{HFI:beam:ellipticity:Mars:545GHz}{1.25}
\setsymbol{HFI:beam:ellipticity:Mars:857GHz}{1.03}

\setsymbol{HFI:CMB:relative:calibration:100GHz}{$\lsim 1\%$}
\setsymbol{HFI:CMB:relative:calibration:143GHz}{$\lsim 1\%$}
\setsymbol{HFI:CMB:relative:calibration:217GHz}{$\lsim 1\%$}
\setsymbol{HFI:CMB:relative:calibration:353GHz}{$\lsim 1\%$}
\setsymbol{HFI:CMB:relative:calibration:545GHz}{}
\setsymbol{HFI:CMB:relative:calibration:857GHz}{}

\setsymbol{HFI:CMB:absolute:calibration:100GHz}{$\lsim 2\%$}
\setsymbol{HFI:CMB:absolute:calibration:143GHz}{$\lsim 2\%$}
\setsymbol{HFI:CMB:absolute:calibration:217GHz}{$\lsim 2\%$}
\setsymbol{HFI:CMB:absolute:calibration:353GHz}{$\lsim 2\%$}
\setsymbol{HFI:CMB:absolute:calibration:545GHz}{}
\setsymbol{HFI:CMB:absolute:calibration:857GHz}{}

\setsymbol{HFI:FIRAS:gain:calibration:accuracy:statistical:100GHz}{}
\setsymbol{HFI:FIRAS:gain:calibration:accuracy:statistical:143GHz}{}
\setsymbol{HFI:FIRAS:gain:calibration:accuracy:statistical:217GHz}{}
\setsymbol{HFI:FIRAS:gain:calibration:accuracy:statistical:353GHz}{2.5\%}
\setsymbol{HFI:FIRAS:gain:calibration:accuracy:statistical:545GHz}{1\%}
\setsymbol{HFI:FIRAS:gain:calibration:accuracy:statistical:857GHz}{0.5\%}

\setsymbol{HFI:FIRAS:gain:calibration:accuracy:systematic:100GHz}{}
\setsymbol{HFI:FIRAS:gain:calibration:accuracy:systematic:143GHz}{}
\setsymbol{HFI:FIRAS:gain:calibration:accuracy:systematic:217GHz}{}
\setsymbol{HFI:FIRAS:gain:calibration:accuracy:systematic:353GHz}{}
\setsymbol{HFI:FIRAS:gain:calibration:accuracy:systematic:545GHz}{7\%}
\setsymbol{HFI:FIRAS:gain:calibration:accuracy:systematic:857GHz}{7\%}

\setsymbol{HFI:FIRAS:zero:point:accuracy:100GHz:units}{0.8\MJysr}
\setsymbol{HFI:FIRAS:zero:point:accuracy:143GHz:units}{}
\setsymbol{HFI:FIRAS:zero:point:accuracy:217GHz:units}{}
\setsymbol{HFI:FIRAS:zero:point:accuracy:353GHz:units}{1.4\MJysr}
\setsymbol{HFI:FIRAS:zero:point:accuracy:545GHz:units}{2.2\MJysr}
\setsymbol{HFI:FIRAS:zero:point:accuracy:857GHz:units}{1.7\MJysr}

\setsymbol{HFI:FIRAS:zero:point:accuracy:100GHz}{0.8}
\setsymbol{HFI:FIRAS:zero:point:accuracy:143GHz}{}
\setsymbol{HFI:FIRAS:zero:point:accuracy:217GHz}{}
\setsymbol{HFI:FIRAS:zero:point:accuracy:353GHz}{1.4}
\setsymbol{HFI:FIRAS:zero:point:accuracy:545GHz}{2.2}
\setsymbol{HFI:FIRAS:zero:point:accuracy:857GHz}{1.7}

\setsymbol{HFI:unit:conversion:100GHz:units}{0.2415\MJysrmK}
\setsymbol{HFI:unit:conversion:143GHz:units}{0.3694\MJysrmK}
\setsymbol{HFI:unit:conversion:217GHz:units}{0.4811\MJysrmK}
\setsymbol{HFI:unit:conversion:353GHz:units}{0.2883\MJysrmK}
\setsymbol{HFI:unit:conversion:545GHz:units}{0.05826\MJysrmK}
\setsymbol{HFI:unit:conversion:857GHz:units}{0.002238\MJysrmK}

\setsymbol{HFI:unit:conversion:100GHz}{0.2415}
\setsymbol{HFI:unit:conversion:143GHz}{0.3694}
\setsymbol{HFI:unit:conversion:217GHz}{0.4811}
\setsymbol{HFI:unit:conversion:353GHz}{0.2883}
\setsymbol{HFI:unit:conversion:545GHz}{0.05826}
\setsymbol{HFI:unit:conversion:857GHz}{0.002238}

\setsymbol{HFI:colour:correction:alpha=-2:V1.01:100GHz}{0.9893}
\setsymbol{HFI:colour:correction:alpha=-2:V1.01:143GHz}{0.9759}
\setsymbol{HFI:colour:correction:alpha=-2:V1.01:217GHz}{1.0007}
\setsymbol{HFI:colour:correction:alpha=-2:V1.01:353GHz}{1.0028}
\setsymbol{HFI:colour:correction:alpha=-2:V1.01:545GHz}{1.0019}
\setsymbol{HFI:colour:correction:alpha=-2:V1.01:857GHz}{0.9889}

\setsymbol{HFI:colour:correction:alpha=0:V1.01:100GHz}{1.0008}
\setsymbol{HFI:colour:correction:alpha=0:V1.01:143GHz}{1.0148}
\setsymbol{HFI:colour:correction:alpha=0:V1.01:217GHz}{0.9909}
\setsymbol{HFI:colour:correction:alpha=0:V1.01:353GHz}{0.9888}
\setsymbol{HFI:colour:correction:alpha=0:V1.01:545GHz}{0.9878}
\setsymbol{HFI:colour:correction:alpha=0:V1.01:857GHz}{1.0014}

%% file: AuthorList_PIP_77_Proj_7_10_Peel_authors_and_institutes.tex
\author{\small
Planck Collaboration:
P.~A.~R.~Ade\inst{77}
\and
N.~Aghanim\inst{53}
\and
M.~I.~R.~Alves\inst{53}
\and
M.~Arnaud\inst{66}
\and
F.~Atrio-Barandela\inst{18}
\and
J.~Aumont\inst{53}
\and
C.~Baccigalupi\inst{76}
\and
A.~J.~Banday\inst{82, 10}
\and
R.~B.~Barreiro\inst{60}
\and
E.~Battaner\inst{83}
% ...
}
\institute{\small
APC, AstroParticule et Cosmologie, Universit\'{e} Paris Diderot, CNRS/IN2P3, CEA/lrfu, Observatoire de Paris, Sorbonne Paris Cit\'{e}, 10, rue Alice Domon et L\'{e}onie Duquet, 75205 Paris Cedex 13, France\\
\and
Aalto University Mets\"{a}hovi Radio Observatory and Dept of Radio Science and Engineering, P.O. Box 13000, FI-00076 AALTO, Finland\\
\and
African Institute for Mathematical Sciences, 6-8 Melrose Road, Muizenberg, Cape Town, South Africa\\
\and
Agenzia Spaziale Italiana Science Data Center, Via del Politecnico snc, 00133, Roma, Italy\\
\and
Agenzia Spaziale Italiana, Viale Liegi 26, Roma, Italy\\
\and
Astrophysics Group, Cavendish Laboratory, University of Cambridge, J J Thomson Avenue, Cambridge CB3 0HE, U.K.\\
\and
Astrophysics \& Cosmology Research Unit, School of Mathematics, Statistics \& Computer Science, University of KwaZulu-Natal, Westville Campus, Private Bag X54001, Durban 4000, South Africa\\
\and
Atacama Large Millimeter/submillimeter Array, ALMA Santiago Central Offices, Alonso de Cordova 3107, Vitacura, Casilla 763 0355, Santiago, Chile\\
\and
CITA, University of Toronto, 60 St. George St., Toronto, ON M5S 3H8, Canada\\
\and
CNRS, IRAP, 9 Av. colonel Roche, BP 44346, F-31028 Toulouse cedex 4, France\\
\and
California Institute of Technology, Pasadena, California, U.S.A.\\
\and
Centro de Estudios de F\'{i}sica del Cosmos de Arag\'{o}n (CEFCA), Plaza San Juan, 1, planta 2, E-44001, Teruel, Spain\\
\and
Computational Cosmology Center, Lawrence Berkeley National Laboratory, Berkeley, California, U.S.A.\\
\and
Consejo Superior de Investigaciones Cient\'{\i}ficas (CSIC), Madrid, Spain\\
\and
DSM/Irfu/SPP, CEA-Saclay, F-91191 Gif-sur-Yvette Cedex, France\\
\and
DTU Space, National Space Institute, Technical University of Denmark, Elektrovej 327, DK-2800 Kgs. Lyngby, Denmark\\
\and
D\'{e}partement de Physique Th\'{e}orique, Universit\'{e} de Gen\`{e}ve, 24, Quai E. Ansermet,1211 Gen\`{e}ve 4, Switzerland\\
\and
Departamento de F\'{\i}sica Fundamental, Facultad de Ciencias, Universidad de Salamanca, 37008 Salamanca, Spain\\
\and
Departamento de F\'{\i}sica, Universidad de Oviedo, Avda. Calvo Sotelo s/n, Oviedo, Spain\\
\and
Department of Astrophysics/IMAPP, Radboud University Nijmegen, P.O. Box 9010, 6500 GL Nijmegen, The Netherlands\\
\and
Department of Electrical Engineering and Computer Sciences, University of California, Berkeley, California, U.S.A.\\
\and
Department of Physics \& Astronomy, University of British Columbia, 6224 Agricultural Road, Vancouver, British Columbia, Canada\\
\and
Department of Physics and Astronomy, Dana and David Dornsife College of Letter, Arts and Sciences, University of Southern California, Los Angeles, CA 90089, U.S.A.\\
\and
Department of Physics and Astronomy, University College London, London WC1E 6BT, U.K.\\
\and
Department of Physics, Gustaf H\"{a}llstr\"{o}min katu 2a, University of Helsinki, Helsinki, Finland\\
\and
Department of Physics, Princeton University, Princeton, New Jersey, U.S.A.\\
\and
Department of Physics, University of California, Santa Barbara, California, U.S.A.\\
\and
Department of Physics, University of Illinois at Urbana-Champaign, 1110 West Green Street, Urbana, Illinois, U.S.A.\\
\and
Dipartimento di Fisica e Astronomia G. Galilei, Universit\`{a} degli Studi di Padova, via Marzolo 8, 35131 Padova, Italy\\
\and
Dipartimento di Fisica e Scienze della Terra, Universit\`{a} di Ferrara, Via Saragat 1, 44122 Ferrara, Italy\\
\and
Dipartimento di Fisica, Universit\`{a} La Sapienza, P. le A. Moro 2, Roma, Italy\\
\and
Dipartimento di Fisica, Universit\`{a} degli Studi di Milano, Via Celoria, 16, Milano, Italy\\
\and
Dipartimento di Fisica, Universit\`{a} degli Studi di Trieste, via A. Valerio 2, Trieste, Italy\\
\and
Dipartimento di Fisica, Universit\`{a} di Roma Tor Vergata, Via della Ricerca Scientifica, 1, Roma, Italy\\
\and
Discovery Center, Niels Bohr Institute, Blegdamsvej 17, Copenhagen, Denmark\\
\and
Dpto. Astrof\'{i}sica, Universidad de La Laguna (ULL), E-38206 La Laguna, Tenerife, Spain\\
\and
European Southern Observatory, ESO Vitacura, Alonso de Cordova 3107, Vitacura, Casilla 19001, Santiago, Chile\\
\and
European Space Agency, ESAC, Planck Science Office, Camino bajo del Castillo, s/n, Urbanizaci\'{o}n Villafranca del Castillo, Villanueva de la Ca\~{n}ada, Madrid, Spain\\
\and
European Space Agency, ESTEC, Keplerlaan 1, 2201 AZ Noordwijk, The Netherlands\\
\and
Helsinki Institute of Physics, Gustaf H\"{a}llstr\"{o}min katu 2, University of Helsinki, Helsinki, Finland\\
\and
INAF - Osservatorio Astronomico di Padova, Vicolo dell'Osservatorio 5, Padova, Italy\\
\and
INAF - Osservatorio Astronomico di Roma, via di Frascati 33, Monte Porzio Catone, Italy\\
\and
INAF - Osservatorio Astronomico di Trieste, Via G.B. Tiepolo 11, Trieste, Italy\\
\and
INAF Istituto di Radioastronomia, Via P. Gobetti 101, 40129 Bologna, Italy\\
\and
INAF/IASF Bologna, Via Gobetti 101, Bologna, Italy\\
\and
INAF/IASF Milano, Via E. Bassini 15, Milano, Italy\\
\and
INFN, Sezione di Bologna, Via Irnerio 46, I-40126, Bologna, Italy\\
\and
INFN, Sezione di Roma 1, Universit\`{a} di Roma Sapienza, Piazzale Aldo Moro 2, 00185, Roma, Italy\\
\and
INFN/National Institute for Nuclear Physics, Via Valerio 2, I-34127 Trieste, Italy\\
\and
IPAG: Institut de Plan\'{e}tologie et d'Astrophysique de Grenoble, Universit\'{e} Joseph Fourier, Grenoble 1 / CNRS-INSU, UMR 5274, Grenoble, F-38041, France\\
\and
Imperial College London, Astrophysics group, Blackett Laboratory, Prince Consort Road, London, SW7 2AZ, U.K.\\
\and
Infrared Processing and Analysis Center, California Institute of Technology, Pasadena, CA 91125, U.S.A.\\
\and
Institut d'Astrophysique Spatiale, CNRS (UMR8617) Universit\'{e} Paris-Sud 11, B\^{a}timent 121, Orsay, France\\
\and
Institut d'Astrophysique de Paris, CNRS (UMR7095), 98 bis Boulevard Arago, F-75014, Paris, France\\
\and
Institute for Space Sciences, Bucharest-Magurale, Romania\\
\and
Institute of Astronomy and Astrophysics, Academia Sinica, Taipei, Taiwan\\
\and
Institute of Astronomy, University of Cambridge, Madingley Road, Cambridge CB3 0HA, U.K.\\
\and
Institute of Theoretical Astrophysics, University of Oslo, Blindern, Oslo, Norway\\
\and
Instituto de Astrof\'{\i}sica de Canarias, C/V\'{\i}a L\'{a}ctea s/n, La Laguna, Tenerife, Spain\\
\and
Instituto de F\'{\i}sica de Cantabria (CSIC-Universidad de Cantabria), Avda. de los Castros s/n, Santander, Spain\\
\and
Jet Propulsion Laboratory, California Institute of Technology, 4800 Oak Grove Drive, Pasadena, California, U.S.A.\\
\and
Jodrell Bank Centre for Astrophysics, Alan Turing Building, School of Physics and Astronomy, The University of Manchester, Oxford Road, Manchester, M13 9PL, U.K.\\
\and
Kavli Institute for Cosmology Cambridge, Madingley Road, Cambridge, CB3 0HA, U.K.\\
\and
LAL, Universit\'{e} Paris-Sud, CNRS/IN2P3, Orsay, France\\
\and
LERMA, CNRS, Observatoire de Paris, 61 Avenue de l'Observatoire, Paris, France\\
\and
Laboratoire AIM, IRFU/Service d'Astrophysique - CEA/DSM - CNRS - Universit\'{e} Paris Diderot, B\^{a}t. 709, CEA-Saclay, F-91191 Gif-sur-Yvette Cedex, France\\
\and
Laboratoire Traitement et Communication de l'Information, CNRS (UMR 5141) and T\'{e}l\'{e}com ParisTech, 46 rue Barrault F-75634 Paris Cedex 13, France\\
\and
Laboratoire de Physique Subatomique et de Cosmologie, Universit\'{e} Joseph Fourier Grenoble I, CNRS/IN2P3, Institut National Polytechnique de Grenoble, 53 rue des Martyrs, 38026 Grenoble cedex, France\\
\and
Laboratoire de Physique Th\'{e}orique, Universit\'{e} Paris-Sud 11 \& CNRS, B\^{a}timent 210, 91405 Orsay, France\\
\and
Max-Planck-Institut f\"{u}r Astrophysik, Karl-Schwarzschild-Str. 1, 85741 Garching, Germany\\
\and
Max-Planck-Institut f\"{u}r Radioastronomie, Auf dem H\"{u}gel 69, 53121 Bonn, Germany\\
\and
MilliLab, VTT Technical Research Centre of Finland, Tietotie 3, Espoo, Finland\\
\and
Niels Bohr Institute, Blegdamsvej 17, Copenhagen, Denmark\\
\and
Observational Cosmology, Mail Stop 367-17, California Institute of Technology, Pasadena, CA, 91125, U.S.A.\\
\and
Optical Science Laboratory, University College London, Gower Street, London, U.K.\\
\and
SISSA, Astrophysics Sector, via Bonomea 265, 34136, Trieste, Italy\\
\and
School of Physics and Astronomy, Cardiff University, Queens Buildings, The Parade, Cardiff, CF24 3AA, U.K.\\
\and
Space Sciences Laboratory, University of California, Berkeley, California, U.S.A.\\
\and
Special Astrophysical Observatory, Russian Academy of Sciences, Nizhnij Arkhyz, Zelenchukskiy region, Karachai-Cherkessian Republic, 369167, Russia\\
\and
UPMC Univ Paris 06, UMR7095, 98 bis Boulevard Arago, F-75014, Paris, France\\
\and
Universidad de Chile, Casilla 36-D, Santiago, Chile\\
\and
Universit\'{e} de Toulouse, UPS-OMP, IRAP, F-31028 Toulouse cedex 4, France\\
\and
University of Granada, Departamento de F\'{\i}sica Te\'{o}rica y del Cosmos, Facultad de Ciencias, Granada, Spain\\
% ...
}

%% file: main_table_hii_ame.tex
\begin{landscape}
\begin{table}[tmb]
\scriptsize
\begingroup
\caption{List of 98 candidate AME regions. Names are given by the central Galactic coordinates. The detection method used for each source is indicated by the superscript letter after the name. Fitted parameters are based on the aperture photometry (see Sect.~\ref{sec:aperflux}). Regions that show significant ($>5\sigma$) AME have boldface names, except for those with a high potential UC\hii\ contribution ($f_{\rm max}^{\rm UC\hii}>0.25$). Values for $\nu_{\rm sp}$ and $f_{\rm max}^{\rm UC\hii}$ are only given if $\sigma_{\rm AME}>2$. In four cases, $A_{\rm sp}$ was fixed at zero (see Sect.~\ref{sec:model_fitting}) and are marked with a dagger.  Note that EM$_{\rm ff}$ is an effective Emission Measure integrated over the aperture.}
\label{tab:main_list}
\vskip -6mm
\setbox\tablebox=\vbox{
\newdimen\digitwidth
\setbox0=\hbox{\rm 0}
\digitwidth=\wd0
\catcode`*=\active
\def*{\kern\digitwidth}
\newdimen\signwidth
\setbox0=\hbox{+}
\signwidth=\wd0
\catcode`!=\active
\def!{\kern\signwidth}
\newdimen\pointwidth
\setbox0=\hbox{.}
\pointwidth=\wd0
\catcode`?=\active
\def?{\kern\pointwidth}
\halign{\hbox to 1.1in{#\leaderfil}\tabskip=1.0em&
\hfil$#$\hfil&
\hfil$#$\hfil&
\hfil$#$\hfil&
\hfil$#$\hfil&
\hfil$#$\hfil&
\hfil$#$\hfil&
\hfil$#$\hfil&
\hfil$#$\hfil&
\hfil$#$\hfil&
\hfil$#$\hfil&
\hfil$#$\hfil&
\hfil$#$\hfil&
\hfil$#$\hfil&
\hfil$#$\hfil&
\vtop{\hsize=5cm\hangafter=1\hangindent=1pt\noindent\strut#\strut\par}\hfil\tabskip=0pt\cr
\noalign{\doubleline}
\omit&\theta&\omit\hfil EM$_{\rm ff}$\hfil&\tau_{250}&T_{\rm d}&\beta_{\rm d}&\Delta T_{\rm CMB}&A_{\rm sp}&\sigma_{\rm AME}&\nu_{\rm sp}&f_{\rm max}^{\rm UC\hii}&G_{0}&S_{\rm resid}^{\rm 28.4}&\frac{S_{\rm resid}^{28.4}}{S_{100\mu{\rm m}}}&\chi^2_r&\omit\hfil Notes\hfil\cr
\omit\hfil Name\hfil&[\deg]&\omit\hfil[cm$^{-6}\,$pc]\hfil&[\times10^5]&\omit\hfil[K]\hfil&&[\mu$K$_{\rm CMB}]&[10^{20}$\,cm$^{-2}]&&[$GHz$]&&&[$Jy$]&[\times 10^{4}]\cr
\noalign{\vskip 3pt\hrule\vskip 5pt}
\bf G004.24+18.09$^{a}$& >2& **1.0\pm1.8& * 1.0\pm 0.1& 17.1\pm0.3& 1.70\pm0.05& *\llap{$-$}16\pm1& *\mathbf{ 6.2\pm 0.4}& \mathbf{15.6}& 29.0\pm0.9& 0.00& *0.88\pm0.09& *5.3\pm0.7& 16.2\pm3.4& 1.35& LDN137/141\cr
\bf G005.40+36.50$^{a}$& 1.5\pm 1.5& **0.9\pm1.5& * 2.5\pm 0.3& 14.0\pm0.3& 1.99\pm0.06& *40\pm2& *\mathbf{ 2.1\pm 0.1}& \mathbf{18.7}& 25.9\pm2.3& 0.00& *0.27\pm0.04& *1.7\pm0.3& *8.0\pm1.9& 1.09& LDN134\cr
 G008.51$-$00.31\tablefootmark{a}& >2& *1338\pm643& 30.0\pm 4.7& 19.1\pm0.7& 2.02\pm0.08& **198\pm220& * 54\pm 28& *1.9& **\ldots& \ldots& *1.7\pm0.3& *46\pm54& *1.3\pm1.5& 0.16& Synch. Faint SNR8.3$-$0.0; 1.2\,Jy $@$\,1\,GHz\cr
 G010.19$-$00.32\tablefootmark{a}& 1.2\pm 0.5& *2067\pm502& 37.8\pm 5.1& 19.7\pm0.6& 2.05\pm0.07& **335\pm189& * 80\pm 23& *3.4& 26.0\pm2.1& 1.01& *2.0\pm0.4& *67\pm42& *1.1\pm0.7& 0.25& Kes62. Synch. SNR9.9$-$0.8; 6.7\,Jy $@$\,1\,GHz\cr
 G012.80$-$00.19\tablefootmark{a}& 1.7\pm 0.7& *2414\pm473& * 46.4\pm 13.5& 19.5\pm1.2& 2.03\pm0.10& **344\pm230& * 85\pm 31& *2.7& 26.0\pm2.2& 0.47& *1.9\pm0.7& *71\pm39& *1.0\pm0.6& 0.14& W33. Synch. Several weak SNRs\cr
 G015.06$-$00.69\tablefootmark{a}& 0.4\pm 0.2& *7472\pm506& 29.5\pm 3.4& 20.6\pm0.5& 2.10\pm0.11& **533\pm251& * 77\pm 40& *1.9& **\ldots& \ldots& *2.6\pm0.4& *60\pm44& *0.9\pm0.7& 0.22& M17\cr
\bf G017.00+00.85$^{a}$& 0.6\pm 0.3& **869\pm130& * 22.3\pm 10.8& 18.2\pm1.6& 2.08\pm0.18& ***75\pm109& *\mathbf{ 68\pm 12}& *\mathbf{5.3}& 35.0\pm6.2& 0.03& *1.3\pm0.7& *58\pm18& *3.2\pm1.3& 0.87& M16, W37\cr
\bf G023.47+08.19$^{a}$& 1.5\pm 1.2& **1.6\pm1.0& * 4.8\pm 0.5& 13.9\pm0.2& 1.97\pm0.07& *33\pm4& *\mathbf{ 4.2\pm 0.5}& *\mathbf{8.5}& 23.0\pm5.7& 0.00& *0.26\pm0.02& *3.2\pm0.6& *9.8\pm2.5& 1.38& LDN462\cr
 G028.79+03.49\tablefootmark{a}& 0.3\pm 0.1& *512\pm38& * 3.3\pm 0.2& 22.9\pm0.3& 1.49\pm0.05& *\llap{$-$}100\pm27& * 0.6\pm 3.0& *0.2& **\ldots& \ldots& *5.1\pm0.4& *\llap{$-$}0.7\pm3.8& *\llap{$-$}0.1\pm0.4& 1.77& W40\cr
 G030.77$-$00.03\tablefootmark{a}& 1.1\pm 0.2& *4730\pm255& 46.2\pm 3.3& 21.4\pm0.3& 1.93\pm0.06& **603\pm180& 194\pm 22& *8.5& 28.5\pm0.4& 0.37& *3.4\pm0.3& 166\pm29& *1.5\pm0.3& 0.76& W43\cr
 G035.20$-$01.74\tablefootmark{a}& >2& **713\pm129& 13.0\pm 1.8& 19.0\pm0.5& 1.74\pm0.12& ****4\pm101& * 21\pm 11& *2.0& **\ldots& \ldots& *1.6\pm0.3& *16\pm14& *1.4\pm1.2& 0.06& W48\cr
 G037.79$-$00.11\tablefootmark{b}& 1.3\pm 0.3& *1768\pm348& 27.0\pm 8.5& 19.7\pm1.3& 1.90\pm0.10& **184\pm157& * 61\pm 18& *3.4& 25.5\pm2.8& 0.35& *2.0\pm0.8& *49\pm29& *1.3\pm0.8& 0.36& W47. Synch. Several weak SNRs\cr
 G040.52+02.53\tablefootmark{a}& 0.6\pm 0.3& *231\pm37& * 2.4\pm 0.5& 20.6\pm1.0& 1.67\pm0.15& ***9\pm37& * 1.0\pm 4.1& *0.2& **\ldots& \ldots& *2.7\pm0.8& *1.1\pm4.9& *0.3\pm1.3& 0.09& W45\cr
 G043.20$-$00.10\tablefootmark{a}& 1.1\pm 0.3& *1433\pm100& 18.4\pm 5.1& 19.4\pm1.1& 1.88\pm0.09& *187\pm74& 50.5\pm 9.4& *5.4& 26.0\pm2.2& 1.23& *1.9\pm0.6& *39\pm12& *1.8\pm0.6& 0.37& W49\cr
 G045.47+00.06\tablefootmark{a}& 1.8\pm 0.6& *748\pm70& * 9.1\pm 0.6& 21.4\pm0.4& 1.63\pm0.07& **97\pm67& 39.1\pm 6.6& *5.9& 27.3\pm0.8& 0.80& *3.3\pm0.3& 33.4\pm9.5& *2.2\pm0.7& 0.75& NRAO601 - SNR in SIMBAD\cr
 G053.63+00.19\tablefootmark{a}& 1.9\pm 0.7& *625\pm49& * 5.9\pm 0.6& 20.4\pm0.4& 1.70\pm0.07& **44\pm38& 17.0\pm 4.6& *3.7& 27.0\pm1.1& 0.57& *2.5\pm0.3& 14.2\pm6.1& *1.6\pm0.7& 0.30& W52\cr
 G059.42$-$00.21\tablefootmark{a}& 1.1\pm 0.5& *294\pm27& * 6.8\pm 2.3& 18.3\pm1.2& 1.92\pm0.12& **\llap{$-$}26\pm22& 18.0\pm 2.6& *7.0& 26.8\pm1.3& 0.37& *1.3\pm0.5& 14.6\pm3.9& *2.6\pm0.8& 0.71& W55 \cr
 G061.47+00.11\tablefootmark{a}& 1.2\pm 0.8& *328\pm23& * 2.4\pm 0.8& 21.8\pm1.6& 1.59\pm0.12& ***4\pm22& * 4.7\pm 2.5& *1.9& **\ldots& \ldots& *3.7\pm1.7& *3.9\pm3.8& *0.8\pm0.7& 0.58& HII LBN061.50+00.29. SH2-88 \cr
 G062.98+00.05\tablefootmark{a}& 1.2\pm 1.5& *201\pm14& * 3.5\pm 1.1& 18.9\pm1.1& 1.82\pm0.10& **38\pm12& * 9.8\pm 1.3& *7.5& 25.7\pm2.5& 0.25& *1.6\pm0.6& *8.1\pm2.0& *2.3\pm0.7& 1.45& S89\cr
 G068.16+01.02\tablefootmark{a}& 0.8\pm 0.5& *187\pm13& * 0.9\pm 0.2& 20.6\pm0.9& 1.89\pm0.20& **40\pm12& *$0\dag$& *\ldots& **\ldots& \ldots& *2.7\pm0.7& *0.0\pm2.0& *0.0\pm1.2& 0.22& S98\cr
 G071.59+02.85\tablefootmark{a}& 1.0\pm 0.4& *286\pm51& * 2.0\pm 1.2& 19.9\pm2.5& 1.75\pm0.24& ***\llap{$-$}5\pm34& * 9.2\pm 5.0& *1.8& **\ldots& \ldots& *2.2\pm1.6& *7.8\pm6.3& *3.0\pm2.5& 0.06& S101\cr
 G075.81+00.39\tablefootmark{a}& 0.7\pm 0.8& *423\pm50& * 4.4\pm 0.7& 19.3\pm0.7& 1.72\pm0.08& *118\pm36& 12.4\pm 4.9& *2.5& 26.8\pm1.3& 1.07& *1.8\pm0.4& *9.0\pm6.1& *2.0\pm1.4& 0.11& HII GAL075.84+00.40. SH2-105. Cyg 2N\cr
 G076.38$-$00.62\tablefootmark{a}& >2& *138\pm42& * 1.8\pm 0.8& 20.8\pm2.3& 2.20\pm0.36& **52\pm39& *$0\dag$& *\ldots& **\ldots& \ldots& *2.8\pm1.9& *2.2\pm8.1& *0.4\pm1.5& 0.35& S106\cr
 G081.59+00.01\tablefootmark{a}& 1.0\pm 0.6& *3726\pm304& * 7.1\pm 2.0& 24.1\pm1.6& 1.67\pm0.10& **140\pm139& * 32\pm 25& *1.3& **\ldots& \ldots& *6.9\pm2.7& *25\pm29& *0.8\pm1.0& 0.21& HII DR23. Near Cygnus X\cr
 G093.02+02.76\tablefootmark{a}& 0.8\pm 0.3& *614\pm49& * 2.9\pm 0.3& 20.4\pm0.4& 1.58\pm0.09& **\llap{$-$}23\pm35& * 5.9\pm 3.8& *1.6& **\ldots& \ldots& *2.5\pm0.3& *4.7\pm4.7& *1.1\pm1.2& 0.18& HII GAL093.06+2.81\cr
 G094.47$-$01.53\tablefootmark{a}& 0.8\pm 0.3& *270\pm33& * 0.6\pm 0.3& 20.9\pm2.6& 1.79\pm0.40& ***\llap{$-$}3\pm23& * 1.5\pm 2.4& *0.6& **\ldots& \ldots& *2.9\pm2.2& *1.3\pm3.0& *1.0\pm2.5& 0.07& HII LBN094.79$-$01.77. LDN1059\cr
 G098.00+01.47\tablefootmark{a}& 0.5\pm 0.1& *155\pm14& * 0.9\pm 0.3& 19.2\pm1.2& 1.73\pm0.12& **5\pm9& * 7.6\pm 1.2& *6.1& 35.1\pm6.3& 0.47& *1.8\pm0.7& *6.1\pm1.6& *6.3\pm1.9& 0.46& RNe GM1-12, DNe TGU H582\cr
 G099.60+03.70\tablefootmark{a}& >2& *544\pm24& * 0.6\pm 0.2& 21.5\pm1.4& 2.09\pm0.22& **\llap{$-$}41\pm16& * 1.4\pm 2.3& *0.6& **\ldots& \ldots& *3.4\pm1.4& *0.1\pm2.3& *0.1\pm1.3& 0.33& Tr 37 cluster. Large complex incl. LDN1111\cr
 G102.88$-$00.69\tablefootmark{a}& 0.6\pm 0.1& *372\pm27& * 1.3\pm 0.3& 20.7\pm0.9& 1.64\pm0.13& **\llap{$-$}60\pm17& * 5.4\pm 2.1& *2.5& 32.3\pm3.9& 0.02& *2.7\pm0.7& *4.6\pm2.7& *2.1\pm1.3& 0.38& HII G102.9$-$00.7. LDN1161/1163\cr
 G107.20+05.20\tablefootmark{a}& 1.0\pm 0.3& *258\pm19& * 3.0\pm 0.3& 20.2\pm0.4& 1.87\pm0.08& **59\pm17& 16.3\pm 1.6& *9.9& 29.2\pm1.1& 1.03& *2.4\pm0.3& 13.9\pm2.1& *2.7\pm0.5& 0.94& AME-G107.2+5.2\tablefootmark{1} \cr
 G109.01+00.00\tablefootmark{b}& 1.1\pm 1.9& **51\pm11& * 1.3\pm 0.4& 19.5\pm1.1& 1.81\pm0.16& *\llap{$-$}18\pm9& * 3.7\pm 2.4& *1.5& **\ldots& \ldots& *1.9\pm0.6& *2.3\pm1.9& *1.4\pm1.2& 0.10& DNe TGU H684. Near HII LBN108.80$-$00.98\cr
 G110.25+02.58\tablefootmark{ab}& 0.8\pm 0.1& *296\pm23& * 1.0\pm 0.3& 25.6\pm2.1& 1.52\pm0.10& **\llap{$-$}42\pm18& * 6.9\pm 2.0& *3.4& *40\pm10& 5.41& *9.7\pm4.7& *4.5\pm2.8& *0.8\pm0.5& 0.24& HII G110.2+02.5. LBN110.11+02.44 \cr
 G118.09+04.96\tablefootmark{a}& 0.8\pm 0.2& 1295\pm22& * 1.4\pm 0.1& 24.0\pm0.4& 1.81\pm0.09& **26\pm16& *$0\dag$& *\ldots& **\ldots& \ldots& *6.6\pm0.7& *\llap{$-$}1.1\pm3.5& *\llap{$-$}0.2\pm0.5& 0.49& Weak SNR NGC7822\cr
 G123.13$-$06.27\tablefootmark{a}& 0.4\pm 0.1& *182\pm13& * 0.7\pm 0.1& 20.5\pm0.4& 1.65\pm0.09& *11\pm6& * 2.0\pm 1.1& *1.8& **\ldots& \ldots& *2.6\pm0.3& *1.5\pm1.1& *1.4\pm1.1& 0.37& S184\cr
\bf G133.27+09.05$^{ab}$& 1.3\pm 0.5& *25\pm3& * 4.5\pm 0.3& 14.5\pm0.2& 1.83\pm0.06& **8\pm8& *\mathbf{ 3.4\pm 0.4}& *\mathbf{8.5}& 24.3\pm4.1& 0.08& *0.32\pm0.03& *2.7\pm0.9& *6.2\pm2.2& 0.13& LDN1358/1355/1357\cr
 G133.74+01.22\tablefootmark{a}& 0.6\pm 0.2& *2155\pm116& * 3.3\pm 0.8& 25.3\pm1.5& 1.67\pm0.08& **53\pm56& 14.8\pm 9.8& *1.5& **\ldots& \ldots& *9.2\pm3.3& *10\pm10& *0.5\pm0.5& 0.97& W3\cr
\bf G142.35+01.35$^{a}$& 1.9\pm 1.7& ***0\pm10& * 4.4\pm 1.3& 16.8\pm0.9& 1.76\pm0.10& **37\pm10& *\mathbf{ 6.7\pm 0.7}& *\mathbf{9.5}& 24.8\pm3.5& 0.07& *0.8\pm0.3& *5.3\pm2.1& *3.6\pm1.5& 0.45& DNe TGU H942\cr
 G151.62$-$00.28\tablefootmark{a}& 1.0\pm 0.7& *348\pm30& * 1.3\pm 0.1& 21.0\pm0.4& 1.38\pm0.08& ***8\pm23& * 3.6\pm 2.3& *1.5& **\ldots& \ldots& *3.0\pm0.3& *2.8\pm2.9& *1.5\pm1.6& 0.21& HII SH2-209\cr
 G158.40$-$20.60\tablefootmark{a}& >2& ***0\pm11& * 3.6\pm 0.2& 15.5\pm0.1& 1.74\pm0.06& **56\pm11& * 5.3\pm 0.3& 16.3& 26.6\pm1.6& 0.26& *0.49\pm0.03& *4.4\pm1.3& *6.9\pm2.2& 1.69& LDN1450/1452\cr
\bf G160.26$-$18.62$^{abc}$& 1.5\pm 0.7& *66\pm6& * 4.6\pm 0.3& 17.4\pm0.2& 1.92\pm0.06& *51\pm8& \mathbf{ 13.3\pm 0.8}& \mathbf{17.4}& 27.8\pm0.3& 0.04& *0.98\pm0.07& 12.0\pm1.2& *4.4\pm0.7& 0.70& Perseus AME-G160.26$-$18.62\tablefootmark{1}\cr
\bf G160.60$-$12.05$^{a}$& 1.4\pm 0.2& *379\pm19& * 0.5\pm 0.2& 21.7\pm1.6& 2.01\pm0.16& ***\llap{$-$}8\pm10& *\mathbf{ 8.3\pm 1.6}& *\mathbf{5.1}& *50\pm17& 0.00& *3.7\pm1.7& *3.8\pm1.9& *2.6\pm1.3& 0.33& NGC1499 (California nebula)\cr
 G166.44$-$24.08\tablefootmark{ac}& 1.4\pm 1.2& *11.2\pm1.5& * 0.5\pm 0.0& 22.7\pm0.4& 1.77\pm0.06& *14\pm1& * 0.9\pm 0.2& *4.3& 20.8\pm8.5& 0.00& *4.7\pm0.6& *0.5\pm0.2& *0.3\pm0.2& 0.58& Pleiades\tablefootmark{2}\cr
 G173.56$-$01.76\tablefootmark{a}& 0.5\pm 0.1& *401\pm32& * 0.4\pm 0.2& 22.7\pm2.1& 1.68\pm0.17& **\llap{$-$}17\pm15& * 2.1\pm 2.6& *0.8& **\ldots& \ldots& *4.8\pm2.7& *1.9\pm2.9& *1.6\pm2.5& 0.39& NGC1893\cr
 G173.62+02.79\tablefootmark{abc}& 0.7\pm 0.2& *194\pm16& * 3.9\pm 0.2& 18.7\pm0.2& 1.64\pm0.04& ***\llap{$-$}6\pm12& * 7.9\pm 1.4& *5.6& 26.2\pm1.9& 1.22& *1.47\pm0.09& *6.5\pm1.9& *2.2\pm0.7& 0.36& S235, AME-G173.6+2.8\tablefootmark{1} \cr
\bf G180.80+04.30$^{a}$& 1.2\pm 0.6& **0.7\pm1.8& * 0.7\pm 0.1& 17.7\pm0.5& 1.73\pm0.12& *\llap{$-$}14\pm7& *\mathbf{ 2.9\pm 0.2}& \mathbf{13.2}& 25.5\pm2.7& 0.00& *1.08\pm0.17& *2.6\pm0.5& *6.9\pm1.6& 1.13& LDN1557 \cr
 G182.36+00.22\tablefootmark{ab}& 1.0\pm 0.5& **3.6\pm1.5& * 1.9\pm 0.6& 16.3\pm1.0& 1.78\pm0.12& *64\pm6& * 2.4\pm 0.5& *4.5& 26.9\pm1.2& 0.36& *0.7\pm0.2& *2.0\pm1.1& *3.6\pm2.0& 0.14& HII LBN182.30+00.07 \& RNe/DNe\cr
 G190.00+00.46\tablefootmark{a}& 0.5\pm 0.1& *278\pm19& * 3.5\pm 0.2& 19.4\pm0.2& 1.67\pm0.05& **46\pm15& 11.5\pm 1.6& *7.4& 27.7\pm0.3& 0.58& *1.83\pm0.12& *9.7\pm2.1& *2.6\pm0.7& 0.91& NGC2174/2175\cr
\bf G192.34$-$11.37$^{a}$& 1.5\pm 0.9& **8\pm6& * 1.4\pm 0.4& 17.3\pm1.0& 1.93\pm0.12& *25\pm5& *\mathbf{ 9.1\pm 0.7}& \mathbf{12.3}& 30.2\pm2.0& 0.00& *0.9\pm0.3& *7.6\pm1.0& 11.7\pm2.3& 2.46& LDN1582/1584\cr
 G192.60$-$00.06\tablefootmark{a}& 1.1\pm 0.6& *65\pm8& * 1.0\pm 0.3& 20.8\pm1.5& 1.83\pm0.14& ***\llap{$-$}7\pm10& * 3.8\pm 0.9& *4.3& 27.2\pm0.9& 0.50& *2.9\pm1.3& *3.2\pm1.2& *1.6\pm0.7& 0.12& S255\cr
\noalign{\vskip 5pt\hrule\vskip 3pt}}}
\endPlancktablewide                 
\endgroup
\end{table}
\end{landscape}

\addtocounter{table}{-1}
\begin{landscape}
\begin{table}[tmb]
\scriptsize
\begingroup
\caption{\textbf{Continued.}}
\vskip -6mm
\setbox\tablebox=\vbox{
\newdimen\digitwidth
\setbox0=\hbox{\rm 0}
\digitwidth=\wd0
\catcode`*=\active
\def*{\kern\digitwidth}
\newdimen\signwidth
\setbox0=\hbox{+}
\signwidth=\wd0
\catcode`!=\active
\def!{\kern\signwidth}
\newdimen\pointwidth
\setbox0=\hbox{.}
\pointwidth=\wd0
\catcode`?=\active
\def?{\kern\pointwidth}
\halign{\hbox to 1.1in{#\leaderfil}\tabskip=1.0em&
\hfil$#$\hfil&
\hfil$#$\hfil&
\hfil$#$\hfil&
\hfil$#$\hfil&
\hfil$#$\hfil&
\hfil$#$\hfil&
\hfil$#$\hfil&
\hfil$#$\hfil&
\hfil$#$\hfil&
\hfil$#$\hfil&
\hfil$#$\hfil&
\hfil$#$\hfil&
\hfil$#$\hfil&
\hfil$#$\hfil&
\vtop{\hsize=5cm\hangafter=1\hangindent=1pt\noindent\strut#\strut\par}\hfil\tabskip=0pt\cr
\noalign{\doubleline}
\omit&\theta&\omit\hfil EM$_{\rm ff}$\hfil&\tau_{250}&T_{\rm d}&\beta_{\rm d}&\Delta T_{\rm CMB}&A_{\rm sp}&\sigma_{\rm AME}&\nu_{\rm sp}&f_{\rm max}^{\rm UC\hii}&G_{0}&S_{\rm resid}^{\rm 28.4}&\frac{S_{\rm resid}^{28.4}}{S_{100\mu{\rm m}}}&\chi^2_r&\omit\hfil Notes\hfil\cr
\omit\hfil Name\hfil&[\deg]&\omit\hfil[cm$^{-6}\,$pc]\hfil&[\times10^5]&\omit\hfil[K]\hfil&&[\mu$K$_{\rm CMB}]&[10^{20}$\,cm$^{-2}]&&[$GHz$]&&&[$Jy$]&[\times 10^{4}]\cr
\noalign{\vskip 3pt\hrule\vskip 5pt}
\bf G201.62+01.63$^{a}$& 0.8\pm 0.4& *161\pm16& * 1.4\pm 0.2& 18.0\pm0.6& 2.11\pm0.17& **90\pm12& *\mathbf{ 9.6\pm 1.3}& *\mathbf{7.4}& 27.4\pm0.7& 0.06& *1.2\pm0.2& *7.9\pm1.5& *6.3\pm1.5& 0.88& LPH96 201.6+1.6. LDN1608/1609\tablefootmark{3} \cr
\bf G203.24+02.08$^{a}$& 1.6\pm 0.9& 103\pm9& * 3.8\pm 1.0& 16.9\pm0.9& 1.77\pm0.09& **19\pm10& *\mathbf{ 7.8\pm 0.9}& *\mathbf{8.3}& 30.3\pm2.1& 0.02& *0.8\pm0.2& *6.6\pm1.9& *4.3\pm1.4& 0.55& HII G203.2+02.1. LDN1613. Cone nebula.\cr
\bf G204.70$-$11.80$^{c}$& >2& *23\pm6& * 2.1\pm 0.2& 16.5\pm0.3& 1.84\pm0.06& **7\pm7& *\mathbf{ 6.6\pm 0.9}& *\mathbf{7.4}& 30.1\pm1.9& 0.00& *0.70\pm0.09& *5.8\pm1.2& *9.5\pm2.5& 1.23& LDN1622\cr
 G208.80$-$02.65\tablefootmark{a}& 0.9\pm 0.6& 124\pm8& * 0.8\pm 0.2& 19.0\pm0.8& 1.64\pm0.13& **12\pm10& * 1.8\pm 0.9& *2.0& **\ldots& \ldots& *1.7\pm0.4& *1.6\pm1.5& *2.2\pm2.1& 0.26& S280\cr
 G209.01$-$19.38\tablefootmark{a}& 0.3\pm 0.1& *4278\pm226& 14.8\pm 1.2& 21.2\pm0.3& 2.13\pm0.09& **371\pm102& * 32\pm 18& *1.8& **\ldots& \ldots& *3.1\pm0.3& *24\pm19& *0.5\pm0.4& 1.13& M42 (Orion nebula)\cr
\bf G211.98$-$01.17$^{a}$& 0.8\pm 0.2& *164\pm10& * 0.8\pm 0.1& 20.4\pm0.4& 1.65\pm0.09& *59\pm8& *\mathbf{ 4.8\pm 0.9}& *\mathbf{5.5}& 27.3\pm0.8& 0.14& *2.5\pm0.3& *3.9\pm1.1& *3.4\pm1.1& 0.94& HII LBN211.65$-$01.51\cr
 G213.71$-$12.60\tablefootmark{ab}& 0.6\pm 0.1& *68\pm5& * 2.7\pm 0.2& 20.1\pm0.2& 1.73\pm0.05& *13\pm8& * 7.6\pm 0.5& 14.9& 27.0\pm1.1& 3.93& *2.31\pm0.17& *6.5\pm1.0& *1.4\pm0.3& 1.45& RNe NGC2170. Mon R2\cr
 G218.05$-$00.38\tablefootmark{ab}& 1.2\pm 1.2& *10.3\pm1.4& * 0.8\pm 0.3& 19.9\pm1.4& 1.51\pm0.12& *\llap{$-$}13\pm5& * 3.7\pm 0.4& *9.4& 25.9\pm2.3& 0.52& *2.2\pm0.9& *3.3\pm1.1& *4.1\pm1.5& 2.60& Mon R2. LDN1650\cr
\bf G219.18$-$08.93$^{a}$& 0.5\pm 0.5& **0.6\pm1.1& * 1.7\pm 0.1& 16.2\pm0.2& 1.86\pm0.04& **0\pm2& *\mathbf{ 2.0\pm 0.1}& \mathbf{18.9}& 23.3\pm5.2& 0.00& *0.64\pm0.04& *1.5\pm0.2& *2.9\pm0.6& 1.64& RNe LBN1015, DNe TGU H1544/1546\cr
\bf G231.83$-$02.00$^{a}$& 1.4\pm 1.0& *13.7\pm1.0& * 1.7\pm 0.2& 16.8\pm0.4& 1.50\pm0.07& *36\pm6& *\mathbf{ 3.4\pm 0.3}& \mathbf{11.4}& 26.5\pm1.6& 0.03& *0.79\pm0.12& *2.9\pm1.1& *6.5\pm2.7& 0.60& DNe TGU H1593\cr
\bf G234.20$-$00.20$^{ab}$& 0.8\pm 0.7& *87\pm5& * 2.5\pm 0.2& 18.5\pm0.3& 1.63\pm0.07& *\llap{$-$}46\pm8& *\mathbf{ 8.2\pm 0.6}& \mathbf{14.5}& 26.3\pm1.8& 0.05& *1.38\pm0.14& *6.8\pm1.3& *4.3\pm1.0& 1.25& HII LBN1050\cr
\bf G239.40$-$04.70$^{a}$& 1.6\pm 0.5& *80\pm4& * 1.4\pm 0.1& 17.3\pm0.4& 1.89\pm0.14& *18\pm8& *\mathbf{ 5.2\pm 0.5}& *\mathbf{9.9}& 30.7\pm2.5& 0.00& *0.93\pm0.14& *4.5\pm0.9& *6.2\pm1.5& 0.95& LDN1667. HII LBN1059. V{*}\,VY\,Cma\cr
 G243.16+00.42\tablefootmark{a}& 0.3\pm 0.1& *171\pm10& * 0.4\pm 0.1& 21.5\pm1.1& 1.83\pm0.16& **\llap{$-$}2\pm6& * 3.6\pm 0.8& *4.2& 29.5\pm1.4& 0.00& *3.4\pm1.0& *3.0\pm1.0& *2.8\pm1.0& 0.56& NGC2467\cr
\bf G247.60$-$12.40$^{ab}$& 1.2\pm 0.9& *10.3\pm1.4& * 0.5\pm 0.0& 18.8\pm0.5& 1.59\pm0.13& **\llap{$-$}5\pm4& *\mathbf{ 3.6\pm 0.3}& \mathbf{14.0}& 27.3\pm0.8& 0.00& *1.6\pm0.2& *3.0\pm0.6& *8.3\pm1.9& 0.79& DNe TGU H1630\cr
 G253.80$-$00.20\tablefootmark{ab}& 0.9\pm 0.5& *227\pm13& * 2.2\pm 0.2& 19.3\pm0.3& 1.63\pm0.07& **61\pm13& 10.2\pm 1.1& *9.5& 28.7\pm0.6& 0.31& *1.81\pm0.15& *8.8\pm1.7& *4.3\pm1.0& 0.57& Gum 10. RNe BRAN124/126\cr
\bf G259.30$-$13.50$^{a}$& 1.1\pm 0.6& *33.0\pm2.0& * 0.3\pm 0.0& 19.8\pm0.4& 1.68\pm0.07& **7\pm3& *\mathbf{ 2.9\pm 0.3}& \mathbf{10.5}& 27.9\pm0.2& 0.00& *2.1\pm0.3& *2.7\pm0.4& *7.1\pm1.5& 1.10& BRAN34. DNe TGU H1667\cr
 G260.50+00.40\tablefootmark{a}& >2& *243\pm16& * 2.9\pm 0.4& 18.7\pm0.6& 1.70\pm0.07& **58\pm14& * 6.4\pm 1.9& *3.3& 23.9\pm4.6& 0.10& *1.5\pm0.3& *4.8\pm2.4& *2.1\pm1.1& 0.15& Vela-D mol cloud. BRAN166\cr
 G265.15+01.45\tablefootmark{a}& 0.5\pm 0.5& *360\pm41& * 2.5\pm 0.3& 21.7\pm0.8& 1.48\pm0.16& **\llap{$-$}125\pm100& * 8.2\pm 9.0& *0.9& **\ldots& \ldots& *3.7\pm0.8& *1.8\pm7.3& *0.3\pm1.4& 0.12& RCW36\cr
 G267.95$-$01.06\tablefootmark{a}& 0.3\pm 0.1& *3170\pm165& * 3.6\pm 0.9& 24.0\pm1.4& 1.98\pm0.09& *195\pm67& * 24\pm 13& *1.9& **\ldots& \ldots& *6.6\pm2.3& *19\pm14& *0.9\pm0.7& 0.70& RCW38\cr
 G270.27+00.84\tablefootmark{a}& 0.9\pm 0.6& *106\pm24& * 4.7\pm 0.5& 18.0\pm0.4& 1.71\pm0.10& **\llap{$-$}32\pm28& * 7.5\pm 3.4& *2.2& 24.3\pm4.1& 0.34& *1.18\pm0.16& *6.3\pm4.6& *2.2\pm1.7& 0.04& RCW41 \cr
 G274.01$-$01.15\tablefootmark{a}& 0.1\pm 0.1& *332\pm18& * 0.4\pm 0.0& 25.7\pm1.0& 1.65\pm0.17& **\llap{$-$}37\pm10& * 2.7\pm 1.6& *1.6& **\ldots& \ldots& 10.1\pm2.4& *1.8\pm2.0& *0.7\pm0.7& 0.41& HII Gum26. RNe BRAN261/258\cr
 G282.02$-$01.16\tablefootmark{a}& 1.1\pm 0.6& 1726\pm82& * 6.6\pm 0.6& 23.1\pm0.5& 1.47\pm0.05& ***8\pm61& 29.8\pm 6.8& *4.4& 28.9\pm0.8& 0.56& *5.3\pm0.7& 23.5\pm8.8& *1.3\pm0.5& 0.43& RCW46\cr
 G284.30$-$00.30\tablefootmark{ab}& 0.7\pm 0.2& *4534\pm249& * 6.7\pm 0.6& 24.2\pm0.6& 1.71\pm0.10& **215\pm135& * 47\pm 20& *2.4& 31.9\pm3.6& 0.64& *7.0\pm1.0& *35\pm22& *1.0\pm0.7& 0.49& RCW49, NGC3247\tablefootmark{4}\cr
 G287.48$-$00.63\tablefootmark{a}& 0.7\pm 0.1& *8492\pm322& * 6.3\pm 0.4& 25.8\pm0.4& 1.89\pm0.10& **350\pm134& * 52\pm 27& *2.0& **\ldots& \ldots& 10.2\pm0.9& *35\pm30& *0.6\pm0.5& 0.84& RCW53, NGC3372 (Carina nebula)\cr
 G289.80$-$01.15\tablefootmark{a}& 1.1\pm 1.4& *1250\pm111& * 2.2\pm 1.0& 25.2\pm2.9& 1.42\pm0.15& **\llap{$-$}76\pm96& *$0\dag$& *\ldots& **\ldots& \ldots& *8.9\pm6.2& **\llap{$-$}8\pm23& *\llap{$-$}0.8\pm2.3& 0.12& Gum35, RCW54\cr
 G291.63$-$00.52\tablefootmark{a}& 0.5\pm 0.1& *4456\pm215& * 3.7\pm 0.3& 26.2\pm0.5& 1.74\pm0.08& *193\pm94& * 34\pm 17& *1.9& **\ldots& \ldots& 11.2\pm1.3& *24\pm19& *0.8\pm0.7& 0.51& RCW57, NGC3603/3576\cr
\bf G293.35$-$24.47$^{b}$& >2& *15.2\pm0.8& * 0.4\pm 0.1& 18.1\pm0.5& 1.50\pm0.07& *30\pm3& *\mathbf{ 1.4\pm 0.2}& *\mathbf{8.9}& 23.7\pm4.8& 0.00& *1.2\pm0.2& *1.1\pm0.3& *4.7\pm1.7& 0.41& DNe TGU H1811/1810/1815\cr
 G294.98$-$01.71\tablefootmark{a}& 0.6\pm 0.3& 1033\pm66& * 2.3\pm 1.0& 23.3\pm2.4& 1.62\pm0.16& **62\pm50& 12.6\pm 6.2& *2.0& 33.6\pm5.1& 0.63& *5.6\pm3.4& *9.1\pm8.7& *1.1\pm1.1& 0.10& RCW62\cr
 G298.60$-$00.20\tablefootmark{a}& 0.8\pm 0.1& *2169\pm103& * 6.8\pm 1.6& 22.0\pm1.2& 1.73\pm0.07& *137\pm59& 28.1\pm 8.8& *3.2& 26.6\pm1.6& 0.82& *4.0\pm1.3& *22\pm10& *1.2\pm0.6& 0.23& Several DNe. HII GAL298.56$-$00.11\cr
 G305.27+00.15\tablefootmark{a}& 0.6\pm 0.1& *3090\pm145& 17.4\pm 1.5& 21.1\pm0.4& 2.08\pm0.08& *417\pm83& * 78\pm 12& *6.3& 29.1\pm1.0& 0.92& *3.1\pm0.3& *65\pm14& *1.2\pm0.3& 0.74& RCW74\cr
 G311.94+00.12\tablefootmark{a}& >2& *2655\pm174& 21.3\pm 1.9& 21.7\pm0.4& 1.71\pm0.05& ***25\pm121& * 67\pm 15& *4.2& 25.4\pm2.8& 0.51& *3.6\pm0.4& *53\pm19& *1.1\pm0.4& 0.18& Several DNe. HII GAL311.89+00.10\cr
 G317.51$-$00.11\tablefootmark{b}& >2& *2226\pm141& 19.0\pm 1.6& 20.7\pm0.4& 1.78\pm0.06& *178\pm95& * 56\pm 12& *4.6& 27.7\pm0.4& 0.67& *2.8\pm0.3& *47\pm15& *1.4\pm0.5& 0.15& Several DNe. HII GAL317.60$-$00.36\cr
\bf G318.49$-$04.28$^{a}$& 1.0\pm 0.3& *102\pm17& * 2.9\pm 0.3& 17.2\pm0.3& 1.99\pm0.11& **46\pm15& *\mathbf{ 8.0\pm 1.5}& *\mathbf{5.2}& 30.8\pm2.6& 0.00& *0.92\pm0.10& *6.6\pm2.1& *4.0\pm1.4& 0.58& RNe GN15.11.0. DNe TGU H1978 \cr
 G320.27$-$00.27\tablefootmark{ab}& 1.5\pm 1.2& 1349\pm91& 10.5\pm 1.0& 22.6\pm0.5& 1.66\pm0.05& *134\pm64& 53.3\pm 7.9& *6.8& 27.3\pm0.8& 0.69& *4.6\pm0.6& *44.3\pm10.0& *1.6\pm0.4& 0.63& HII GAL320.23$-$00.29. Several DNe. \cr
 G322.15+00.61\tablefootmark{a}& >2& **36\pm56& * 3.9\pm 1.1& 18.7\pm1.1& 1.92\pm0.26& **33\pm56& 16.0\pm 6.5& *2.5& 32.5\pm4.2& 0.26& *1.5\pm0.6& *13\pm10& *3.5\pm3.1& 0.09& RCW92\cr
 G327.30$-$00.50\tablefootmark{a}& 1.3\pm 0.8& *2882\pm212& 18.1\pm 7.5& 20.2\pm1.7& 2.08\pm0.17& **190\pm127& * 46\pm 19& *2.4& 29.3\pm1.2& 0.68& *2.3\pm1.2& *40\pm23& *1.1\pm0.7& 0.32& HII GAL327.30$-$00.60. Several DNe \cr
 G336.90+00.00\tablefootmark{a}& 1.5\pm 0.5& *4543\pm308& 29.6\pm 2.8& 22.7\pm0.5& 1.82\pm0.06& ***\llap{$-$}91\pm182& * 79\pm 27& *2.9& 27.6\pm0.5& 1.01& *4.8\pm0.7& *67\pm33& *0.7\pm0.3& 0.14& Kes39. Weak SNR G337.0$-$00.1; 1.5\,Jy $@$\,1\,GHz\cr
 G343.48$-$00.04\tablefootmark{a}& >2& *1491\pm177& * 35.4\pm 11.5& 18.6\pm1.2& 1.94\pm0.11& ***61\pm117& * 54\pm 16& *3.4& 24.9\pm3.3& 0.55& *1.4\pm0.6& *43\pm21& *1.3\pm0.7& 0.29& HII GAL343.4$-$00.0. Several DNe.\cr
\bf G344.75+23.97$^{b}$& >2& *12.1\pm0.8& * 0.3\pm 0.0& 20.3\pm0.4& 1.86\pm0.07& **\llap{$-$}7\pm2& *\mathbf{ 1.4\pm 0.2}& *\mathbf{6.0}& 26.9\pm1.2& 0.00& *2.4\pm0.3& *1.2\pm0.3& *2.1\pm0.6& 1.45& MBM 121/122. Synch.\cr
 G345.40$-$00.94\tablefootmark{a}& >2& 66\pm113& * 5.7\pm 3.8& 19.3\pm2.8& 2.38\pm0.49& **221\pm110& * 48\pm 10& *4.6& 33.2\pm4.8& 0.86& *1.8\pm1.6& *43\pm17& *3.9\pm2.1& 0.24& RCW117\cr
 G348.73$-$00.75\tablefootmark{a}& 0.7\pm 0.5& *1956\pm176& * 4.2\pm 1.1& 22.9\pm2.1& 2.08\pm0.41& ***\llap{$-$}24\pm122& * 10\pm 15& *0.7& **\ldots& \ldots& *5.0\pm2.8& **7\pm19& *0.4\pm1.0& 0.11& RCW122\cr
 G351.29+00.68\tablefootmark{a}& 0.5\pm 0.2& *3071\pm239& 13.1\pm 1.8& 22.3\pm0.7& 2.04\pm0.12& **290\pm169& * 25\pm 23& *1.1& **\ldots& \ldots& *4.3\pm0.9& *17\pm30& *0.3\pm0.6& 0.18& Extension to the west of AME-G353.05+16.90\cr
\bf G351.31+17.28$^{a}$& >2& *161\pm12& * 0.9\pm 0.4& 25.4\pm2.6& 1.69\pm0.15& ***4\pm13& *\mathbf{ 7.5\pm 1.4}& *\mathbf{5.3}& 28.5\pm0.4& 0.01& *9.4\pm5.7& *6.5\pm2.2& *1.1\pm0.4& 0.24& HII LBN1105/1104. Several DNe \cr
 G351.65$-$01.23\tablefootmark{a}& 1.8\pm 1.7& **443\pm222& 11.8\pm 4.2& 19.7\pm1.4& 2.20\pm0.26& **291\pm150& * 49\pm 24& *2.1& 28.5\pm0.4& 1.38& *2.0\pm0.9& *41\pm31& *1.9\pm1.5& 0.05& Kes52\cr
\bf G353.05+16.90$^{bc}$& 0.9\pm 0.5& *11\pm2& * 6.5\pm 1.0& 21.0\pm0.7& 1.78\pm0.05& *121\pm13& \mathbf{ 22.2\pm 0.7}& \mathbf{29.8}& 30.5\pm2.3& 0.23& *3.0\pm0.6& 20.0\pm1.9& *1.5\pm0.2& 0.80& $\rho$~Ophiuchi AME-G353.05+16.90\tablefootmark{1}\cr
 G353.16+00.74\tablefootmark{a}& 0.4\pm 0.1& *5725\pm307& 16.4\pm 5.2& 21.7\pm1.5& 2.13\pm0.12& **576\pm155& * 85\pm 26& *3.2& 32.5\pm4.1& 0.28& *3.6\pm1.5& *71\pm31& *1.2\pm0.6& 0.40& HII GAL353.19+0.67. HII NGC6357\cr
\bf G353.97+15.79$^{a}$& >2& *31\pm6& * 4.7\pm 1.5& 18.7\pm1.2& 1.69\pm0.10& *\llap{$-$}23\pm9& \mathbf{ 10.8\pm 1.0}& \mathbf{10.9}& 26.6\pm1.5& 0.11& *1.5\pm0.6& *9.0\pm2.9& *2.3\pm0.8& 0.72& In Ophiuchus\cr
 G355.44+00.11\tablefootmark{a}& 1.9\pm 0.6& *1444\pm478& * 41.6\pm 13.1& 18.3\pm1.1& 2.12\pm0.12& **338\pm198& * 69\pm 23& *2.9& 26.1\pm2.0& 0.22& *1.3\pm0.5& *56\pm39& *1.4\pm1.0& 0.25& Synch. SNR G355.6$-$0.0; 3\,Jy $@$\,1\,GHz\cr
\bf G355.63+20.52$^{a}$& 1.7\pm 0.8& *12\pm2& * 2.2\pm 0.2& 16.1\pm0.3& 2.01\pm0.10& *60\pm6& *\mathbf{ 7.1\pm 0.5}& \mathbf{13.3}& 26.9\pm1.2& 0.00& *0.61\pm0.06& *5.9\pm0.7& *9.0\pm1.8& 1.70& In Ophiuchus\cr
\noalign{\vskip 5pt\hrule\vskip 3pt}}}
\endPlancktablewide                 
\tablefoot{
\tablefootmark{a}{Detected in bandmerged catalogue.}
\tablefootmark{b}{Detected in component subtracted map.}
\tablefootmark{c}{Previously known source from the literature.}
\tablefootmark{1}{\cite{planck2011-7.2}}
\tablefootmark{2}{\cite{Genova-Santos2011}}
\tablefootmark{3}{\cite{Finkbeiner2002,Dickinson2006}}
\tablefootmark{4}{\cite{Dickinson2007}}
}
\endgroup
\end{table}
\end{landscape}